\documentstyle[preprint,prb,aps]{revtex}

\begin{document}
\draft
\input{psfig}
\title{Anisotropic Spin Hamiltonians due to Spin--Orbit \\
and Coulomb
Exchange Interactions}

\author{T. Yildirim\cite{NIST} and A. B. Harris}
\address{Department of Physics, University of Pennsylvania,
Philadelphia, Pennsylvania 19104-6369}
\author{Amnon Aharony and O. Entin--Wohlman}
\address{School of Physics and Astronomy, Raymond and Beverly
Sackler Faculty of Exact Sciences, \\
Tel Aviv University, Tel Aviv 69978, Israel}
\date{\today}
\maketitle
 
\begin{abstract}
Here we correct, extend, and clarify results concerning the spin
Hamiltonian ${\cal H}_S$ used to describe the ground manifold
of Hubbard models for magnetic insulators
in the presence of spin--orbit interactions.
Most of our explicit results are for a tetragonal lattice as applied
to some of the copper oxide lamellar systems and are obtained within
the approximation that ${\cal H}_S$ consists of a sum of 
nearest--neighbor bond
Hamiltonians.  We consider both a ``generic'' model in
which hopping takes place from one copper ion to another and
a ``real'' model in which holes can hop from a copper ion to
an intervening oxygen $2p$ band.  Both models include
orbitally--dependent direct and exchange Coulomb interactions
involving two orbitals.
Our analytic results have been confirmed by numerical
diagonalizations for two holes occupying any of the $3d$
states and, if applicable, the oxygen $2p$ states.
An extension of the perturbative scheme used by Moriya is
used to obtain analytic results for ${\cal H}_S$ up to order
${\bf t}^2$ (${\bf t}$ is the matrix of hopping coefficients)
for arbitrary crystal symmetry for both the ``generic'' and
``real'' models.  With only direct orbitally--independent
Coulomb interactions, our results reduce to Moriya's apart from
some minor modifications.  For the tetragonal case, we show
to all orders in ${\bf t}$ and $\lambda$, the spin--orbit
coupling constant, that ${\cal H}_S$ is isotropic in the absence of
Coulomb exchange terms and assuming only nearest--neighbor hopping.
In the presence of Coulomb exchange, scaled by $K$, the anisotropy in
${\cal H}_S$ is biaxial and is shown to be of order $Kt^2\lambda^2$.
Even when ${\bf K}=0$, for systems of sufficiently low symmetry, the
anisotropy in ${\cal H}_S$ is proportional to $t^6\lambda^2$ when
the direct on--site Coulomb interaction $U$ is independent of the
orbitals involved and of order $t^2\lambda^2$ otherwise.  These
latter results apply to the orthorhombic phase of La$_2$CuO$_4$.

\end{abstract}

\section{INTRODUCTION}
\medskip

A long standing problem which has attracted much interest recently
concerns the mechanism whereby spin--orbit interactions give rise to
magnetic anisotropy in magnetic insulators.  This subject, which
was extensively investigated three decades
ago,[\onlinecite{TMPR,DSV}]
has recently been the object of renewed attention due to interest in
the lamellar copper oxide systems.[\onlinecite{LACUO}] The first of these to
be extensively investigated, La$_2$CuO$_4$, has a small orthorhombic
distortion away from a tetragonal structure [\onlinecite{ORTHO}] and
the above mechanism was shown [\onlinecite{Coffey,Coffey2,SEA}]
to give rise to an anisotropic
exchange, including that of the antisymmetric Dzyaloshinskii--Moriya
type.  In that system there are two anisotropy
energies.[\onlinecite{LACUOA,CL}]
One of these, the out--of--plane anisotropy, is of the form
$\alpha N_z^2$, where $N_z$ is the $z$--component of the staggered
magnetization, the $z$--axis is taken to be perpendicular to the
copper oxide plane, and $\alpha$ is an anisotropy constant.
This energy causes the spins to lie in the basal
plane.  There is also an in--plane anisotropy energy which selects
the orientation of the spins within the basal plane.  Until recently
the discussions of the origins of anisotropy were confined to the
orthorhombic structure.  However, more recently a family of copper
oxide materials of similar structure, but which are actually tetragonal,
have been studied [\onlinecite{TETRA1,TETRA2}] and found to have roughly the
same out--of--plane anisotropy as La$_2$CuO$_4$.  The earlier studies
[\onlinecite{Coffey,Coffey2,SEA}] did not predict any anisotropy in
the tetragonal limit.  Accordingly, a
reanalysis of anisotropy for the tetragonal systems ought to show
a common origin of the out--of--plane anisotropy which does not
rely on the orthorhombic distortion.  That is the main purpose of this
paper.  However, in the course of this work, we have found that
a number of general questions concerning both the results and the
methodology required some clarification, which this paper is intended
to provide.

A microscopic basis for superexchange between magnetic ions was first given
almost forty years ago by Anderson.[\onlinecite{PWA}]
In the language of a Hubbard model, [\onlinecite{HUB}]
his calculation started
from an orbitally nondegenerate band in which there is one electron per site
in the limit of large Coulomb interaction $U$ whenever two electrons occupy
the same site.  If the kinetic energy is completely neglected,
each electron (or hole) may be characterized by its spin.
When kinetic energy (described by hopping) is included perturbatively,
one finds a spin Hamiltonian, which in low order perturbation theory can
be expressed as the sum of contributions ${\cal H}(i,j)$ from each bond
$(i,j)$.  This spin Hamiltonian describes the perturbative removal of
degeneracy.  In higher order in the hopping, one encounters contributions
to the spin Hamiltonian from plaquettes (at order $t^4/U^3$, where $t$ is
a hopping matrix element) and eventually from even higher order clusters.
Ignoring higher--order contributions, Anderson obtained an isotropic
exchange interaction between nearest--neighbor spins,
\begin{equation}
\label{HIJEQ}
{\cal H}(i,j) = J(i,j) {\bf S}(i) \cdot {\bf S}(j) \ ,
\end{equation}
where $J(i,j)=4 t_{ij}^2 /U$ and $t_{ij}$ is the hopping matrix
element between sites $i$ and $j$.

Soon afterwards Moriya [\onlinecite{TMPR}] used Anderson's formalism
to study the effect of spin--orbit interactions on superexchange
between magnetic ions.  He showed that for sufficiently low symmetry
the most general effective spin Hamiltonian for two spin 1/2 magnetic
ions, such as Cu$^{++}$, is of the form
\begin{equation}
\label{JDMEQ}
{\cal H}(i,j) = J(i,j) {\bf S}(i)\cdot {\bf S}(j) + 
{\bf D}(i,j) \cdot {\bf S}(i)\times{\bf S}(j) +
{\bf S}(i) \cdot {\bf M}(i,j) \cdot {\bf S}(j)  \ ,
\end{equation}
where ${\bf M} (i,j)$ is a symmetric $3 \times 3$ tensor.
The first term represents the isotropic symmetric exchange. The second
and third terms represent the antisymmetric and symmetric anisotropies,
respectively.  Moriya's results were obtained to second order in the
hopping perturbation, but in principle provided a framework in which
the spin--orbit interaction could be included to arbitrary order.
Convenient explicit results were given to lowest nontrivial order in
the spin--orbit coupling constant, $\lambda$.

Much more recently, Thio {\it et al.} [\onlinecite{ORTHO}] found that
La$_2$CuO$_4$ is described by Eq. (2).  Consequently Coffey and
co--workers, [\onlinecite{Coffey}] invoked this
Hamiltonian to describe the CuO planes in the cuprates. They  found that
${\bf D}(i,j)$ can not be the same for all bonds $<ij>$, as was assumed by
a number of previous authors. The form of the ${\bf D}(i,j)$ is determined
by the symmetry properties of the crystal structure. The first attempt
at a microscopic calculation of the vectors ${\bf D}(i,j)$ was made
by Coffey, Rice, and Zhang [\onlinecite{Coffey2}] in the framework of
the Moriya theory of the anisotropic superexchange interactions.
Within this theory, ${\bf D}(i,j)$ is of order $\lambda$, whereas
${\bf M}(i,j)$ is of order $\lambda^2$.  Therefore, many authors neglected
${\bf M}$.  Naively, one expected a gap in the spin--wave spectrum due
to anisotropy, and this is what one finds when ${\bf M}(i,j)$ is
neglected.  Subsequently, Shekhtman {\it et al.} (SEA)
[\onlinecite{SEA}] have shown that ${\bf M}(i,j)$ can {\it never}
be neglected.  Most interestingly, when ${\bf M}(i,j)$ is included,
they found a hidden symmetry in ${\cal H}(i,j)$, as a result of which,
inclusion of spin--orbit interactions did not reduce the degeneracy
of the ground state of the pair of spins $(i,j)$. [\onlinecite{TAK}]
Their result was that ${\cal H}(i,j)$ could be written
in the following form:
\begin{equation}
\label{PRL}
{\cal H}(i,j) = \left( J -  {D^2\over 4J} \right) {\bf S} (i)
\cdot {\bf S} (j) + {\bf D}(i,j) \cdot {\bf S}(i) \times {\bf S}(j) +
{\bf S}(i) \cdot { {\bf D} \otimes {\bf D} \over 2J } \cdot {\bf S}(j) \ ,
\end{equation}
where the vector ${\bf D}(i,j)$ is bond--dependent and
$[ {\bf A} \otimes {\bf B}]_{\mu \nu } = A_\mu B_\nu$.
As SEA show, the result (\ref{PRL}) indicates that
although the pair interaction is not of the isotropic form of Eq.
(\ref{HIJEQ}), it is rotationally invariant and hence the energy level
spectrum of the pair interactions consists of a singlet and a triplet,
just as it would in the absence of spin--orbit interactions.  In previous
work the terms in ${\bf D}(i,j)$ and those in
$M_{\mu,\nu}(i,j) \equiv D_\mu(i,j) D_\nu (i,j)/(2J)$ were not treated on
an equal footing, and therefore this hidden symmetry was never noticed.
Furthermore, SEA showed that even though each individual bond might
have this hidden symmetry, the crystal as a whole could have anisotropy
because of the frustration caused by the competition between exchange
interactions of different bonds. [\onlinecite{TAK}]  In particular, for
La$_2$CuO$_4$ they found that the anisotropy was a result of
this frustration.

All the work cited so far relied on the idea, introduced by
Moriya, that the effect of spin--orbit interactions could be
taken into account by a gauge transformation on the hopping
between sites.  As used by Moriya to obtain results up to order
$t^2/U$, this formulation is correct and convenient.
However, this formulation
does not form a correct basis for calculations to higher order
in $t/U$.  Thus, as we shall see, the hidden symmetry of SEA,
although maintained at order $t^2$ for constant $U$, is
broken at order $t^6$ for constant $U$ or at order
$t^2$ for nonconstant $U$.  (Here constant $U$ means that
the Coulomb interaction between holes in two orbitals
does not depend on which orbitals are involved.) In addition,
the calculations of Shekhtman {\it et al.} [\onlinecite{SAE}] or
Bonesteel [\onlinecite{NEB}] for the anisotropy of the cuprates
were based on terms requiring the existence of a distortion from
tetragonal symmetry.  However, the  easy plane anisotropy is
observed [\onlinecite{TETRA1,TETRA2}] to have 
similar magnitudes in both the orthorhombic and tetragonal cuprates
isostructural to La$_{2}$CuO$_{4}$.  The main reason for the failure
of the previous calculations to give anisotropy for the tetragonal
cuprates was the fact that these calculations neglected
the Coulomb exchange interaction.  From the results of Barriquand
and Sawatzky (BS)[\onlinecite{GS}] one can see that they
partially included such interactions.  However, it remained unclear
which aspects of the BS results would persist when the calculation
was pursued more systematically.  In fact, in Ref. \onlinecite{PRL2}
it was shown that for tetragonal symmetry Coulomb exchange
interactions played a crucial role in determining the anisotropy.

In view of the above history, the following points remained to be
clarified and are addressed in the present paper.  1)  One
should generalize Moriya's results for ${\cal H}(i,j)$ to
the case of nonconstant $U$.  Having done that, we find that
when reduced to the case of constant $U$, our present results
differ in a small way from those of Moriya, who overlooked some 
$\lambda$-dependent contributions to the magnitude of
the isotropic exchange interaction, $J$.  We also give general results
for superexchange interactions, i. e. for the case when
the copper ions are separated by an intervening oxygen ion.
However, the results are given in a general form which can equally
apply to systems of ions other than Cu, as long as their ground
state is orbitally nondegenerate.  2) Since earlier calculations for
the cuprates omitted hopping between excited states of the
Cu ions, we have reanalyzed the
role of symmetry at arbitrary order in the matrix elements
$t_{{\rm Cu-Cu}} \equiv t$ which describe the effective
hopping between copper ions. We find that in the absence of
Coulomb exchange interactions [i. e.  for ${\bf K}=0$ in Eq.
(\ref{HAM}), below], one recovers isotropic exchange for a
simplified ``generic'' model which describes the complete 3d
band for copper ions on a simple tetragonal Bravais lattice.
This isotropy is the result of the high symmetry
of the crystal field levels and the resulting high symmetry
of the hopping matrix elements.  This result
shows that in the absence of Coulomb exchange terms,
one retains isotropy in ${\cal H}(i,j)$ to all orders in both
$\lambda$ and $t_{ij}/U$
and thus that inclusion of Coulomb exchange interactions is
essential to obtain anisotropic exchange interactions in the tetragonal
case.  3)
For a tetragonal lattice we find that this accidental isotropy in
${\cal H}(i,j)$ is removed at order $t^2 \lambda^2 K$ when Coulomb exchange
interactions are allowed and we give detailed expressions 
for the exchange anisotropy in terms of the hopping matrix
elements and the matrix elements of the Coulomb interaction.
4)  For a crystal with arbitrarily low symmetry
(i. e. when the crystal field states have no special symmetry),
we expect to (and do) find a removal of degeneracy of the
spin triplet.  This breaking of rotational invariance occurs at
order $t^6$ for the case of constant $U$ and at order $t^2$
when $U$ is nonconstant.  These results modify the conclusion
given in Ref. \onlinecite{SEA}.  5) In contrast to all previous
work, we also found an in--plane anisotropy originating from
the anisotropy of the spin--wave zero--point energy.

Most of the above results have been obtained analytically, both
for the ``generic'' model (with only Cu ions) and for the
``real'' model (in which the Cu ions are separated by oxygen ions).
Furthermore, we have corroborated our results by comparing
them to results obtained by numerically diagonalizing the
Hamiltonian which describes all possible states of two
holes on one bond.  For the ``generic'' model, there are 20
single--particle orbitals, 10 on each copper ion, so that
in all there are 190 two--hole states.  For the ``real'' model
there are six additional $2p$ states on the oxygen ion, so there
are 325 two--holes states in all.  In the tetragonal case, where
we know that the exchange interaction matrix $J_{\mu \nu} (i,j)$
[see Eq. (\ref{HJMN}), below] is diagonal, its values may be
deduced from the values of the energy splittings of the ground
manifold, as is discussed in Appendix A.

Briefly, this paper is organized as follows.  In Sec. II, we first
introduce ``generic'' and ``real'' Hamiltonians and then discuss
the perturbative framework we use to calculate the spin Hamiltonian
${\cal H}_S$.  The actual perturbative calculations of ${\cal H}_S$
are described in Sec. III, although many of the details are
relegated to Appendices.  Here we give expressions for $J(i,j)$,
${\bf D}(i,j)$, and ${\bf M}(i,j)$ up to order $t^2\lambda^2
\Delta {\cal H}_c$ for the ``generic'' model and analogously
for the ``real'' model, where $\Delta {\cal H}_c$ represents
Coulomb interactions beyond the approximation in which the
Coulomb exchange ${\bf K}$ is zero and ${\bf U}$ is constant.
The case of tetragonal site symmetry is discussed in Sec. IV
for both the ``generic'' and ``real'' models.  There we prove
a theorem, valid to all orders in the hopping matrix elements
and spin--orbit coupling, which says that for nearest--neighbor
hopping, the complete spin Hamiltonian is isotropic when Coulomb
exchange is absent.  Here we display explicitly the leading
contribution to the anisotropic exchange when Coulomb exchange
is treated perturbatively.  In Sec. V we discuss the experimental
consequences of these results.  In particular we estimate the
anisotropies and spin--wave gaps which our work would predict.
In Sec. VI we study the case of arbitrarily low symmetry for the
``generic'' model and show that the anisotropy in $J_{\mu \nu}(i,j)$
is of order $t^6 \lambda^2$ for the case of constant $U$
and of order $t^2 \lambda^2$ when $U$ is not constant.
Finally in Sec. VII we summarize the conclusions of this work.
A brief summary of our major conclusions has been
given previously.[\onlinecite{PRL1,PRL2}]

\medskip
\section{HUBBARD HAMILTONIAN AND SYMMETRY OF EXCHANGE}
\medskip

\subsection{Generic Model}

In this section we introduce a general Hamiltonian,
versions of which will be studied in this paper.
We start from the following generic model,[\onlinecite{ERRATUM}]
which captures the symmetries of the cuprates.  For holes which
reside only on the Cu ions, this model is given by

\begin{eqnarray}
\label{HAM}
{\cal H}  &= &
\sum_{i,\alpha,\sigma} \epsilon_{i\alpha} d_{i\alpha \sigma}^{\dag}
d_{i\alpha \sigma} + \sum_{\rm holes, h}  \lambda  {\bf L}({h}) \cdot
{\bf S}({h}) \nonumber \\
&+& \sum_{\stackrel{\alpha,\beta,\sigma}{i\ne j}}
t_{i\alpha,j\beta} ( d_{i\alpha \sigma}^{\dag}d_{j\beta \sigma}
+ d_{j\beta \sigma}^{\dag}d_{i\alpha \sigma} )
\nonumber \\
& + &  \frac{1}{2}\sum_{\stackrel{i,\alpha,\alpha'}{\sigma,s}}
U_{i\alpha,i\alpha'} d_{i\alpha \sigma}^{\dag} d_{i\alpha' s}^{\dag}
d_{i\alpha' s} d_{i\alpha \sigma}
\nonumber \\
& + &  \frac{1}{2}\sum_{\stackrel{i,\alpha \neq \alpha'}{\sigma,s}}
K_{i\alpha,i\alpha'} d_{i\alpha \sigma}^{\dag} d_{i\alpha' s}^{\dag}
d_{i\alpha s} d_{i\alpha' \sigma}
\nonumber \\
& + &  \frac{1}{2}\sum_{\stackrel{\alpha,\beta,\sigma,s}{i\ne j}}
V_{i\alpha,j\beta} ( d_{i\alpha \sigma}^{\dag} d_{j\beta s}^{\dag}
d_{j\beta s} d_{i\alpha \sigma}
+ d_{j\beta \sigma}^{\dag} d_{i\alpha s}^{\dag} d_{i\alpha s}
d_{j\beta \sigma} ) \nonumber \\
& + &  \frac{1}{2}\sum_{\stackrel{\alpha,\beta,\sigma,s}{i\ne j}}
N_{i\alpha,j\beta} ( d_{i\alpha \sigma}^{\dag} d_{j\beta s}^{\dag}
d_{i\alpha s} d_{j\beta \sigma}
+ d_{j\beta \sigma}^{\dag} d_{i\alpha s}^{\dag} d_{j\beta s}
d_{i\alpha \sigma} ) \ .
\end{eqnarray}
Here $d_{i \alpha \sigma}^{\dag}$ creates a hole in the $\alpha$th spatial
orbital, whose single--particle energy is $\epsilon_{i\alpha}$,
with $z$--component of spin $\sigma$ on the Cu ion at site $i$.
In general we allow hopping with matrix elements $t_{i \alpha ,j \beta}$
between the $\alpha$ orbital on site $i$ and the $\beta$ orbital on
site $j$.
This Hamiltonian also includes direct Coulomb interactions
between electrons on the same site (scaled by ${\bf U}$) and
on different sites (scaled by ${\bf V}$) and exchange Coulomb
interactions between electrons on the same site (scaled by ${\bf K}$)
and on different sites (scaled by ${\bf N}$).
Our numerical work indicates that when ${\bf t} \not= 0$,
the effects of ${\bf V}$ and ${\bf N}$ are not qualitatively
different from those of ${\bf U}$ and ${\bf K}$, respectively.
Since the latter are often dominant, we
shall neglect ${\bf V}$ and ${\bf N}$.
In principle one should also include Coulomb terms with four
states, $U_{\alpha \beta \gamma \delta} d_{i \alpha \sigma}^\dagger
d_{i \beta s}^\dagger d_{i \gamma s} d_{i \delta \sigma}$.
Here we follow most of the literature and start with the simpler
Eq. (\ref{HAM}), which involves only the Hartree--like terms
${\bf U}$ and the simple Coulomb exchange terms ${\bf K}$.
Equation (\ref{HAM}) can easily be extended to the ``real'' model
in which we include p states on the oxygen ions, with hopping
between them and the d states on the nearest--neighboring Cu ions.

\medskip
\subsection{Single Site Hamiltonian for Non--Interacting Holes}
\medskip

In this subsection we briefly discuss the basis states used in
the perturbative scheme described in the next subsection.
We first consider ions within a single--particle picture.  We
therefore start by considering the effects of the crystal field
Hamiltonian, ${\cal H}_x$, and the spin--orbit interaction,
${\cal H}_{\rm so}$.  The former is constructed so as to give the
observed ionic levels.
Including only such energies the single--particle Hamiltonian is
\begin{equation}
{\cal H}_{x} + {\cal H}_{\rm so} = 
\sum_{i\alpha\sigma} \epsilon_{i\alpha}d_{i\alpha\sigma}^{\dagger}
d_{i\alpha\sigma} +
\lambda\sum_{\stackrel{i\alpha\beta}{\sigma\sigma '}}
\Bigl( \omega_{i}(\alpha ,\beta) \Bigr)_{\sigma\sigma '}
d_{i\alpha\sigma}^{\dagger}d_{i\beta\sigma'} \ ,
\end{equation}
where
\begin{equation}
\Bigl( \omega_{i}(\alpha ,\beta)\Bigr)_{\sigma\sigma '}
\equiv \frac{1}{2}\sum_{\mu}
<i\alpha \mid L_{\mu}\mid i\beta >
\Bigl(\sigma_{\mu}\Bigr)_{\sigma\sigma '},\label{I4}
\end{equation}
in which $<i\alpha \mid L_{\mu}\mid i\beta >\equiv L^\mu_{\alpha \beta}$
is the matrix element
of the $\mu $--component of the orbital angular momentum between
the two single--particle states, and
$\sigma_{\mu}$ is the Pauli matrix.  We shall often present
results for $i$--independent matrix elements of ${\bf L}$.

For many purposes it is convenient to diagonalize the single--particle,
single--site Hamiltonian ${\cal H}_x+{\cal H}_{\rm so}$.
We may choose the wave functions $\mid i\alpha >$ to be real,
in which case the matrix elements $L_{\alpha\beta}^{\mu}$ are
purely imaginary.  As a result, every single--particle energy of
${\cal H}_x+{\cal H}_{\rm so}$
is at least doubly degenerate. That is, the two linearly--independent wave
functions which are related to one another by time reversal,
\begin{eqnarray}
\psi_{a}&=&\sum_{\alpha }(y_{\alpha a}\mid \alpha \uparrow>
+z_{\alpha a}\mid \alpha\downarrow>),\nonumber\\
\phi_{a}&=&\sum_{\alpha}(-z_{\alpha a}^{\ast}\mid\alpha\uparrow>+y_{\alpha
a}^{\ast}\mid\alpha\downarrow>),\label{I6}
\end{eqnarray}
belong to the same energy. We use Greek indices to label the crystal
field states in the absence of spin--orbit interactions and Roman ones
for the eigenstates of ${\cal H}_x +{\cal H}_{\rm so}$. The
latter can be characterized
by pseudospin quantum numbers, $\sigma =\pm 1$, and are associated
with the creation operators $c_{ia\sigma}^{\dagger}$.  These
operators are related to the $d_{i\alpha\sigma}^{\dagger}$'s via
\begin{equation}
\label{UNITARY}
d_{i\alpha\sigma}^\dagger = \sum_{a\sigma_{1}}(m_{\alpha
a}^{i})^\ast_{\sigma\sigma_{1}}c_{ia\sigma_{1}}^\dagger ,
\end{equation}
where the unitary matrix ${\bf m}_{\alpha a}^i$ is
\begin{equation}
\label{MMAT}
{\bf m}_{\alpha a}^{i}\equiv \left(\begin{array}{cc}
y_{\alpha a}^{i} \ \  - (z_{\alpha a}^{i})^\ast \\
z_{\alpha a}^{i} \ \  (y_{\alpha a}^{i})^\ast
\end{array}\right) 
\equiv  u_{\alpha a}^{i}{\bf I} + i {\bf v}_{\alpha a}^{i}\cdot
{\vec {\sigma}} \ ,
\label{I8}
\end{equation}
where $u_{\alpha a}^{i}$ is a real scalar, ${\bf v}_{\alpha a}^{i}$ 
is a real vector and
${\bf I}$ is the $2\times 2$ unit matrix. This leads to
\begin{equation}
{\cal H}_{x}+{\cal H}_{\rm so}\equiv\sum_{ia\sigma}
E_{i a}c_{ia\sigma}^{\dagger}c_{ia\sigma},\label{I9}
\end{equation}
where
\begin{equation}
\Bigl\{\sum_{\alpha} \epsilon_{i\alpha}({\bf m}_{\alpha
a}^{i})^{\dagger} {\bf m}_{\alpha b}^{i}+\frac{\lambda}{2}
\sum_{\alpha\beta}({\bf m}_{\alpha a}^{i})^{\dagger}\omega (\alpha ,\beta)
{\bf m}_{\beta b}^{i}\Bigr\}_{\sigma_{1}\sigma_{2}} =
E_{ia}\delta_{ab}\delta_{\sigma_{1}\sigma_{2}}
\ .\label{I10}
\end{equation}
(Here the dagger operation on ${\bf m}_{\alpha a}^i$ operates
only in terms of the two by two matrices as in Eq. (\ref{MMAT})
and is not to be applied to the scripts $i$, $\alpha$, or $a$.) 
The transformation ${\bf m}_{\alpha a}$ that diagonalizes the
single--particle Hamiltonian is in general different for each site.
Consequently, the single--site energies may depend on the site index.
However, in certain situations, for example, in the presence of
the tetragonal to orthorhombic distortion in La$_2$CuO$_4$,
it is possible to define the transformation such that the
single--particle energies are site independent.  This will be the
case for some of the explicit calculations which are presented
below for the cuprates.

\medskip
\subsection{Formulation of Perturbation Theory}
\medskip

For Cu$^{++}$ ions in a
$d^9$ configuration we are dealing with an ionic ground state
having one $3d$ hole whose spin is arbitrary.  When we include
the oxygen ions in the model, those ions have filled $2p$
bands in their ground state.  In either case, in the absence of
hopping, i. e. for ${\bf t}=0$, the many--electron ground state manifold
is one in which one hole of arbitrary spin resides on each copper
ion.  The energy levels within this ground manifold, when the
remaining terms in the Hamiltonian, especially hopping, are
considered, are the object of our study.

When hopping is introduced as a perturbation, the splitting of
the hitherto degenerate ground state manifold can be described
by a spin Hamiltonian, ${\cal H}_S$.  In view of time reversal
invariance ${\cal H}_S$ will consist of two--spin interactions (between
nearest and further neighbors), four--spin interactions,
and so forth.  In the present paper most of our
results will be for the nearest--neighbor two--spin coupling
constants, except for the general theorem of Sec. IV, which
makes no assumptions about the specific form of ${\cal H}_S$.
If we only consider two spin interactions between nearest--neighboring
spins, we effectively write
\begin{equation}
\label{SUMHIJ}
{\cal H}_S = \sum_{\langle ij \rangle} {\cal H}(i,j) \ ,
\end{equation}
where $\langle ij \rangle$ indicates a sum over pairs of
nearest--neighboring sites and for spins 1/2
\begin{equation}
\label{HJMN}
{\cal H}(i,j) = \sum_{\mu \nu } J_{\mu \nu }(i,j)  S_\mu (i) S_\nu (j) \ ,
\end{equation}
where $\mu$ and $\nu$ label Cartesian components.  We refer to the
case when $J_{\mu, \nu } (i,j) = J(i,j) \delta_{\mu , \nu }$, where
$\delta$ is the Kronecker delta function, as isotropic exchange.
[To avoid confusion between the two kinds of exchange,
the terms in Eq. (\ref{HAM}) proportional to ${\bf K}$
are referred to as Coulomb exchange.]  Appendix A contains a
discussion of the possible anisotropies in ${\cal H}(i,j)$.

The major objective of this paper is to discuss the symmetry of
the matrix ${\bf J}(i,j)$ and develop  perturbative expressions
for it on the basis of the generic Hamiltonian of Eq. (\ref{HAM})
and its generalization to include the intervening oxygen ions. 
>From our point of view the most important early work was
that of Moriya,[\onlinecite{TMPR}] who studied a simplified version
of the above model.  The most significant simplifications
necessary to obtain Moriya's main result were
to neglect the Coulomb exchange, ${\bf K}$, and to assume constant
$\bf U$, i. e. to assume that $U_{i\alpha , i \beta}$ did not depend
on either the site index $i$ or
the orbital indices $\alpha$ and $\beta$.  In particular,
when $U_{i\alpha, i \beta}$ is independent of $\alpha$ and $\beta$,
the wave functions for the two--hole states are Slater determinants
of the one--hole states as obtained by the canonical transformation
of Eq. (\ref{UNITARY}).  In other words, in this very special case,
the exact eigenstates of the Hamiltonian
${\cal H}_x+{\cal H}_{\rm so}$ also diagonalize the Coulomb
interaction, ${\cal H}_c$.  In terms of these new single--particle
states the transformed hopping Hamiltonian now assumes the form
\begin{mathletters}
\label{EQ14}
\begin{equation}
\label{I16a}
{\cal H}_{\rm hop}=\sum_{i,j}T_{ij},
\end{equation}
where
\begin{equation}
T_{ij}=\sum_{\stackrel{ab}{\sigma\sigma
'}}\Bigl(\tilde{t}_{ab}^{ij}\Bigr)_{\sigma\sigma'}
c_{ia\sigma}^{\dagger} c_{jb\sigma '} \label{I16b}
\end{equation}
represents hops from site $j$ to site $i$,
and $(\tilde{\bf t}_{ab}^{ij})$ is the $2\times 2$ matrix
\begin{equation}
\label{I16c}
\Bigl(\tilde{\bf t}_{ab}^{ij}\Bigr)\equiv\sum_{\alpha\beta}
t_{i\alpha,j\beta}({\bf m}_{\alpha a}^{i})^{\dagger}
{\bf m}_{\beta b}^{j} \equiv A_{ab}^{ij}{\bf I}+i{\bf B}_{ab}^{ij}
\cdot {\vec \sigma} \ ,
\end{equation}
in which $A_{ab}^{ij}$ $({\bf B}_{ab}^{ij})$ is a real scalar
(vector), that can be found using Eq. (\ref{I8}) and the
representation in which $t_{i\alpha ,j\beta}$ is real.
\end{mathletters}
By hermiticity these coefficients obey
\begin{equation}
\label{HERM}
A_{ab}^{ij} = A_{ba}^{ji} \ , \ \ \ \ \
{\bf B}_{ab}^{ij} = - {\bf B}_{ba}^{ji} \ .
\end{equation}
Actually (and this seems to have caused much subsequent confusion),
Moriya did not write down Eqs. (\ref{I16a})--(\ref{I16c}).
Instead, in a further simplification, he truncated ${\bf t}$ to
include only hopping between the ${\bf t}=0$ ground states.
Even for his calculations at order $t^2$, this
simplification is slightly incorrect.  However,
we should emphasize that this truncation is totally inappropriate
for a discussion of effects of order higher than $t^2$, since
hopping between excited states then comes into play.  Also,
when ${\bf U}$ is not constant, hopping between exact eigenstates of
${\cal H}_x+{\cal H}_{\rm so} + {\cal H}_c$ is no longer a
single--particle interaction.  To see this note that there are matrix
elements between an initial state, in which both holes are in their
ground states on different ions, and a final state in which, for
instance, both holes are in excited states of one ion.  Such a process
explicitly relies on the fact that the two--hole states are not
simply obtained from single--hole states.  Thus, in this case,
when ``final--state interactions'' are present, the hopping
perturbation involves four electron operators.

Accordingly, to study the case when ${\bf U}$ is not constant
and when Coulomb exchange is not neglected, we write
${\cal H}_c= {\cal H}_{c0} + \Delta {\cal H}_c$, where
\begin{equation}
{\cal H}_{c0}=\frac{1}{2}U_{0}
\sum_{\stackrel{i\alpha\beta}{\sigma\sigma '}}
d_{i\alpha\sigma}^{\dagger}d_{i\beta\sigma '}^{\dagger}
d_{i\beta\sigma'} d_{i\alpha\sigma} \equiv\frac{1}{2}U_{0}
\sum_{\stackrel{iab}{\sigma\sigma '}}c_{ia\sigma}^{\dagger}
c_{ib\sigma '}^{\dagger}c_{ib\sigma '}c_{ia\sigma},\label{I12}
\end{equation}
and the additional Coulomb terms resulting from nonconstant ${\bf U}$
and ${\bf K}$ take the form
\begin{equation}
\Delta{\cal H}_{c}=\frac{1}{2}\sum_{i}\sum_{\stackrel{aba'b'}
{\sigma\sigma'\sigma_{1}\sigma_{1}'}} \Bigl(
\Delta \tilde{U}_{\sigma\sigma '\sigma_{1}\sigma_{1}'}(i;abb'a')
+ \tilde{K}_{\sigma\sigma '\sigma_{1}\sigma_{1}'}
(i;abb'a')\Bigr)c_{ia\sigma}^{\dagger}
c_{ib\sigma '}^{\dagger}c_{ib'\sigma_{1}}
c_{ia'\sigma_{1}'},\label{I13}
\end{equation}
with\begin{mathletters}
\label{I14}
\begin{equation}
\Delta\tilde{U}_{\sigma\sigma '\sigma_{1}\sigma_{1}'}(i;abb'a')\equiv
\sum_{\alpha\alpha '}\Delta U_{\alpha\alpha '}^i \Bigl(
({\bf m}_{\alpha a}^{i})^{\dagger}
{\bf m}_{\alpha a'}^{i}\Bigr)_{\sigma\sigma_{1}'}
\Bigl(({\bf m}_{\alpha 'b}^{i})^{\dagger}
{\bf m}_{\alpha 'b'}^{i}
\Bigr)_{\sigma '\sigma_{1}},\label{I14a}
\end{equation}
\begin{equation}
\tilde{K}_{\sigma\sigma '\sigma_{1}\sigma_{1}'}(i;abb'a')\equiv
\sum_{\alpha\alpha '}K_{\alpha\alpha '}^i
\Bigl(({\bf m}_{\alpha a}^{i})^{\dagger} {\bf m}_{\alpha'a'}^{i}
\Bigr)_{\sigma\sigma_{1}'} \Bigl(({\bf m}_{\alpha 'b}^{i})^{\dagger}
{\bf m}_{\alpha b'}^{i} \Bigr)_{\sigma '\sigma_{1}} \ . \label{I14b}
\end{equation}
\end{mathletters}
Expressions for $\Delta {\bf U}$ and ${\bf K}$ for tetragonal
crystal--field states in terms of Racah
parameters are given in Appendix B.

In the following, we will calculate the effective
spin Hamiltonian using perturbation theory in which we take
the unperturbed Hamiltonian to be
\begin{equation}
\label{UNPERT}
{\cal H}_0 =  {\cal H}_x + {\cal H}_{\rm so} + {\cal H}_{c0} =
\sum_{i a \sigma} E_{i a} c_{ia \sigma}^\dagger c_{ia \sigma}
+ \frac{1}{2}U_{0} \sum_{\stackrel{iab}{\sigma\sigma '}}
c_{ia\sigma}^{\dagger}
c_{ib\sigma '}^{\dagger}c_{ib\sigma '}c_{ia\sigma}
\end{equation}
and the perturbation to be
\begin{equation}
V = {\cal H}_{\rm hop} + \Delta {\cal H}_c \ ,
\end{equation}
where these quantities are given in Eqs. (\ref{EQ14}) and (\ref{I13}).

\medskip
\section{Perturbative Contributions to ${\cal H}(
\protect{\small {i,j}})$}
\medskip

\subsection{Contributions of Order ${\bf t}^2$}

The lowest order contributions to ${\cal H}(i,j)$
are second order in ${\bf t}$.  At this order in ${\bf t}$
in the absence of
the Coulombic perturbation $\Delta {\cal H}_c$, we can use
the result of Eq. (\ref{PSI2}) in Appendix C to evaluate
\begin{equation}
{\cal H}^{(2)}(i,j) = - 2 \langle \psi_0' \mid T_{ij}
{1 \over {\cal H}_0 } T_{ji} \mid \psi_0 \rangle \ , 
\end{equation}
where the factor of 2 accounts for the similar term when the
hopping is in the reverse direction.  Here
$\mid \psi_0 \rangle$ and $\mid \psi_0' \rangle$ are
states in the ground manifold with one hole per site, and
the superscript (2) indicates a result which is second order in
${\bf t}$.  In using the result in Appendix C we must truncate the
matrix element so that it remains within this manifold.  Also,
in evaluating this expression it is convenient to use
the identity
\begin{equation}
\label{IDENT}
c_{i0\sigma}^{\dagger}c_{i0\sigma '}\equiv \Bigl[ \frac{1}{2}
+{\bf S}(i) \cdot\vec{\sigma}\Bigr]_{\sigma '\sigma}
\end{equation} 
whereby we obtain the result
\begin{eqnarray}
\label{III6}
{\cal H}^{(2)}(i,j)&=& - \sum_{b} \Biggl( {\rm Tr} \Bigl\{
\tilde{\bf t}_{0b}^{ij} \tilde{\bf t}_{b0}^{ji}
\Bigl[\frac{1}{2} + {\bf S}(i) \cdot\vec{\sigma} \Bigr]\Bigr\}
/(U_{0}+E_{jb}) +  ( i \leftrightarrow j ) \Biggr) \nonumber\\
&+& 2 {\rm Tr} \Bigl\{\tilde{\bf t}_{00}^{ij}\Bigl[ \frac{1}{2}
+ {\bf S}(j) \cdot\vec{\sigma}\Bigr]\tilde{\bf t}_{00}^{ji} \Bigl[
\frac{1}{2}+ {\bf S}(i) \cdot\vec{\sigma}\Bigr]\Bigr\} /U_0 \ ,
\end{eqnarray}
where the traces are over the $2\times 2$ matrices in $\sigma$--space
and $( i \leftrightarrow j )$ denotes the sum of all previous
terms with $i$ and $j$ interchanged.

The first term in (\ref{III6}), which only involves hopping of
a single hole (from site $i$ to $j$ and back), is easily shown to
be independent of the spins at $i$ and $j$. [This follows directly
from the identities of Eq. (\ref{HERM}), or more simply from
time reversal invariance.] Therefore, this term contributes a
spin--independent constant, and does not affect the splitting of
the ground state.  Similarly, the terms coming from the factors
of $1/2$ inside the square brackets in the second term also give
constants. To order $t^{2}$ we have thus arrived at an effective
magnetic Hamiltonian of the form of Eq. (\ref{SUMHIJ}), with
\begin{equation}
{\cal
H}^{(2)}(i,j)=\frac{2}{U_{0}}{\rm Tr} \Bigl\{
\tilde{\bf t}_{00}^{ij} \Bigl( {\bf S}(j) \cdot\vec{\sigma}\Bigr)
\tilde{\bf t}_{00}^{ji}
\Bigl( {\bf S}(i) \cdot\vec{\sigma}\Bigr)\Bigr\}.\label{III7}
\end{equation}
In view of Eq. (\ref{I16c}), this becomes
\begin{equation}
{\cal H}^{(2)}
(i,j)=\frac{2}{U_{0}}{\rm Tr} \Bigl\{\Bigl(A_{00}^{ij}+
i{\bf B}_{00}^{ij}\cdot\vec{\sigma}\Bigr)
\Bigl( {\bf S}(j) \cdot\vec{\sigma}\Bigr)
\Bigl(A_{00}^{ji}+i{\bf B}_{00}^{ji}\cdot\vec{\sigma}\Bigr)
\Bigl({\bf S}(i) \cdot\vec{\sigma}\Bigr)\Bigr\}.\label{III8}
\end{equation}
The symmetry of this form is further discussed in Appendix D,
where we show that in fact ${\cal H}^{(2)}(i,j)$ is of the
isotropic form of Eq. (1).

\subsection{Contribution of Order ${\bf t}^2 \Delta {\cal H}_c$}

To calculate the contributions of the Coulomb terms of Eq.
(\ref{I13}) to the magnetic exchange we need to carry out third
order perturbation theory.  By taking two factors of the hopping
matrix element we generate terms of order $\tilde {\bf t}^2$.
We must include an additional factor of $\Delta {\cal H}_c$.
This factor is only relevant in the intermediate state
when there are two holes on the same site.  The relevant matrix
element for third--order perturbation theory is written in
Eq. (\ref{PSI4}) of Appendix C.  In using this result
it is convenient to use the identity of Eq. (\ref{IDENT}).
Then we obtain the correction to the energy at second order in
$\tilde {\bf t}$ including perturbatively the leading Coulombic
contributions (which we indicate by the superscript ``(2,c)''):
\begin{eqnarray}
{\cal H}^{(2,c)}(i,j)=&-&\Biggl(
\sum_{\stackrel{\sigma\sigma_{1}\sigma_{2} \sigma_3}{ss'}}
\sum_{ab}\frac{1}{(U_{0}+E_{ia})}\frac{1}{(U_{0}+E_{ib})}
\Bigl(\tilde{t}_{b0}^{ij} \Bigr)_
{\sigma_{2}\sigma_{1}}\Bigl(\tilde{t}_{0a}^{ji}\Bigr)_{\sigma_{3}s}
\nonumber \\ &&
\times \Bigl[\Delta\tilde{U}_{ss'\sigma_{2}\sigma}(i;a0b0)+
\tilde{K}_{ss'\sigma_{2}\sigma}(i;a0b0)
-\Delta\tilde{U}_{ss'\sigma\sigma_{2}}(i;a00b)-
\tilde{K}_{ss'\sigma\sigma_{2}}(i;a00b)\Bigr]
\nonumber \\ &&
\times \Bigl(\frac{1}{2}+ {\bf S}(i) \cdot
\vec{\sigma}\Bigr)_{\sigma s'}\Bigl(\frac{1}{2}+ {\bf S}(j) \cdot\vec
{\sigma}\Bigr)_{\sigma_{1}\sigma_{3}}
+ ( i \leftrightarrow j ) \Biggr) \ ,
\end{eqnarray}
where we have used the property
$\Delta\tilde{U}_{\sigma\sigma'\sigma_{1}\sigma_{1}'}(i;abb'a')
=\Delta\tilde{U}_{\sigma'\sigma\sigma_{1}'\sigma_{1}}(i;baa'b')$, 
$\tilde{K}_{\sigma\sigma'\sigma_{1}\sigma_{1}'}
(i;abb'a')=\tilde{K}_{\sigma'\sigma\sigma_{1}'\sigma_{1}}(i;baa'b')$.
(In writing the above result we set $E_{i,0}=E_{j,0}=0$ for
simplicity.) In order to carry out the spin summations, we insert
here the explicit expressions for $\Delta\tilde{U}$ and $\tilde{K}$,
Eqs. (\ref{I14}). This leads to
\begin{eqnarray}
\label{H2C}
{\cal H}^{(2,c)}(i,j)&=&- \sum_{\alpha\alpha'} \Biggl[ 
\Biggl( \Delta U_{\alpha\alpha'} \Bigl[
{\rm Tr} \Bigl\{ \Bigl( \frac{1}{2}+ {\bf S}(i)
\cdot\vec{\sigma}\Bigr)
\Bigl({\bf x}_{\alpha'\alpha'}^{ji}\Bigr)^{\dagger}
\Bigl(\frac{1}{2}+{\bf S}(j)
\cdot\vec{\sigma}\Bigr){\bf x}_{\alpha\alpha}^{ji}\Bigr\}\nonumber\\
&-& {\rm Tr} \Bigl\{\Bigl(\frac{1}{2}+ {\bf S}(i) \cdot\vec{\sigma}
\Bigr) ({\bf m}_{\alpha '0}^{i})^{\dagger}
{\bf m}_{\alpha '0}^{i}\Bigr\}
{\rm Tr} \Bigl\{\Bigl(\frac{1}{2}+ {\bf S}(j) \cdot
\vec{\sigma}\Bigr){\bf w}_{\alpha\alpha}^{ji}\Bigr\}\Bigr]\nonumber\\
&+&K_{\alpha\alpha '}\Bigl[ {\rm Tr} \Bigl\{
\Bigl(\frac{1}{2}+ {\bf S}_{i}\cdot\vec{\sigma}
\Bigl)({\bf x}_{\alpha\alpha '}^{ji}\Bigr)^{\dagger}
\Bigl(\frac{1}{2}+ {\bf S}(j)
\cdot\vec{\sigma}\Bigr){\bf x}_{\alpha\alpha '}^{ji}\Bigr\}\nonumber\\
&-& {\rm Tr} \Bigl\{\Bigl(\frac{1}{2}+ {\bf S}(i)
\cdot\vec{\sigma} \Bigr) ({\bf m}_{\alpha '0}^{i})^{\dagger}
{\bf m}_{\alpha 0}^{i}
\Bigr\} {\rm Tr} \Bigl\{\Bigl(\frac{1}{2}+ {\bf S}(j)
\cdot\vec{\sigma}\Bigl){\bf w}_{\alpha\alpha '}^{ji}\Bigr\}\Bigr]\Biggr)
+(i\leftrightarrow j)\Biggr] ,\label{III24}
\end{eqnarray}
where
\begin{mathletters}
\label{III25}
\begin{equation}
{\bf x}_{\alpha\alpha'}^{ji}=\sum_{a}\frac{1}{U_{0}+E_{ia}}
\tilde{\bf t}_{0a}^{ji}({\bf m}_{\alpha a}^{i})^{\dagger}
{\bf m}_{\alpha '0}^{i} \equiv X_{\alpha\alpha '}^{ji} {\bf I}
+i{\bf Y}_{\alpha\alpha '}^{ji} \cdot\vec{\sigma}, \label{III25a}
\end{equation}
and
\begin{equation}
{\bf w}_{\alpha\alpha '}^{ji}=\sum_{ab}\frac{1}{(U_{0}+E_{ia})}
\frac{1}{(U_{0}+E_{ib})} \tilde{\bf t}_{0a}^{ji}(
{\bf m}_{\alpha a}^{i})^\dagger
{\bf m}_{\alpha'b}^{i}\tilde{\bf t}_{b0}^{ij} \equiv
W_{\alpha\alpha '}^{ji} {\bf I}
+i{\bf Z}_{\alpha\alpha '}^{ji}\cdot\vec{\sigma}, 
\label{III25b}
\end{equation}
in which $X_{\alpha \alpha'}^{ji}$ and $W_{\alpha \alpha'}^{ji}$ are
real scalars and ${\bf Y}_{\alpha \alpha'}^{ji}$ and
${\bf Z}_{\alpha \alpha'}^{ji}$
are real vectors. It is straightforward to verify that in Eq.
(\ref{III24}) the terms which involve two traces, as well as
those coming from the factors of $1/2$ and involving one spin
variable, do not contribute to spin--dependence in the spin
Hamiltonian.  This follows by noting that (a)
$({\bf m}_{\alpha'0}^{i})^{\dagger} {\bf m}_{\alpha '0}^{i}$
and ${\bf w}_{\alpha\alpha}^{ji}$ are proportional to the unit matrix,
and (b) one can interchange $\alpha$ and $\alpha'$ in the sums.
\end{mathletters}

The full effective magnetic Hamiltonian, to order ${\bf t}^{2}$, is
obtained by combining ${\cal H}^{(2)}(i,j)$, Eq. (\ref{III7}),
with ${\cal H}^{(2,c)}(i,j)$, Eq. (\ref{III24}).
The result has the form of Eq. (\ref{JDMEQ}), with

\begin{mathletters}
\label{III26}
\begin{eqnarray}
J(i,j)&=&\frac{2}{U_{0}}{\rm Tr} \{\Bigl(
\tilde {\bf t}_{00}^{ji}\Bigr)\Bigl(
\tilde{\bf t}_{00}^{ij}\Bigr)\}
-\sum_{\alpha\alpha '}\Delta U_{\alpha\alpha '}
{\rm Tr} \{\Bigl( {\bf x}_{\alpha '\alpha '}^{ji}\Bigr)^{\dagger}
\Bigl({\bf x}_{\alpha\alpha}^{ji}\Bigr)^{\dagger}+
\Bigl({\bf x}_{\alpha '\alpha '}^{ij}\Bigr)^{\dagger}\Bigl(
{\bf x}_{\alpha\alpha}^{ij}\Bigr)^{\dagger}\}\nonumber\\
&-&\sum_{\alpha\alpha '}K_{\alpha\alpha '}{\rm Tr} \{\Bigl(
{\bf x}_{\alpha\alpha '}^{ji}\Bigr)^{\dagger}\Bigl(
{\bf x}_{\alpha\alpha '}^{ji}\Bigr)^{\dagger}+
\Bigl( {\bf x}_{\alpha\alpha '} ^{ij}\Bigr)^{\dagger}
\Bigl( {\bf x}_{\alpha\alpha '}^{ij}\Bigr)^{\dagger}\},\label{III26a}
\end{eqnarray}
\begin{eqnarray}
{\bf D}(i,j)&=&-\frac{i}{U_{0}}\Bigl({\rm Tr}
\{\tilde{\bf t}_{00}^{ji}\}{\rm Tr} \{\tilde{\bf t}_{00}^{ij}
\vec{\sigma}\}-(i\leftrightarrow j)\Bigr)\nonumber\\
&+&\frac{i}{2}\sum_{\alpha\alpha'}\Delta U_{\alpha\alpha '}
\Biggl\{ \Biggl[ {\rm Tr} \{ {\bf x}_{\alpha\alpha}^{ji}\}
{\rm Tr} \{\Bigl( {\bf x}_{\alpha '\alpha'}^{ji}\Bigr)^{\dagger}
\vec{\sigma}\} - {\rm Tr} \{\Bigl( {\bf x}_{\alpha '\alpha '}^{ji}
\Bigr)^{\dagger}\} {\rm Tr} \{ {\bf x}_{\alpha\alpha}^{ji}
\vec{\sigma}\}\Bigr) \Biggr] - (i\leftrightarrow j)
\Biggr\} \nonumber\\
&+&\frac{i}{2}\sum_{\alpha\alpha'} K_{\alpha\alpha '} \Biggl\{
\Biggl[ {\rm Tr} \{ {\bf x}_{\alpha\alpha '}^{ji}\}{\rm Tr} \{\Bigl(
{\bf x}_{\alpha \alpha'}^{ji}\Bigr)^{\dagger}\vec{\sigma}\}
- {\rm Tr} \{\Bigl( {\bf x}_{\alpha \alpha '}^{ji}\Bigr)^{\dagger} \}
{\rm Tr} \{ {\bf x}_{\alpha\alpha '}^{ji}\vec{\sigma}\}\Biggr]
-(i\leftrightarrow j)\Bigr] \Biggr\} \ ,\label{III26b}
\end{eqnarray}
\begin{eqnarray}
{\bf M}(i,j) &=&\frac{1}{U_{0}}\Bigl( {\rm Tr}
\{\tilde{\bf t}_{00}^{ji}\vec{\sigma}\}\otimes {\rm Tr}
\{\tilde{\bf t}_{00}^{ij}\vec{\sigma}\}+(i\leftrightarrow j)\Bigl)\nonumber\\
&-&\frac{1}{2}\sum_{\alpha\alpha '}
\Delta U_{\alpha\alpha '}\Biggl\{ \Biggl[ {\rm Tr} \{
{\bf x}_{\alpha\alpha}^{ji}\vec {\sigma}\}\otimes 
{\rm Tr} \{\Bigl( {\bf x}_{\alpha '\alpha '}^{ji}\Bigr)^{\dagger}
\vec{\sigma}\} + {\rm Tr} \{\Bigl(
{\bf x}_{\alpha '\alpha '}^{ji}\Bigr)^{\dagger} \vec{\sigma}\}\otimes
{\rm Tr} \{ {\bf x}_{\alpha\alpha}^{ji}\vec{\sigma}\}\Biggr]
+ (i\leftrightarrow j) \Biggr\} \nonumber\\
&-&\frac{1}{2}\sum_{\alpha\alpha '}
K_{\alpha\alpha '}\Biggl\{ \Biggl[ {\rm Tr} \{
{\bf x}_{\alpha\alpha '}^{ji}\vec
{\sigma}\}\otimes {\rm Tr} \{\Bigl(
{\bf x}_{\alpha \alpha '}^{ji}\Bigr)^{\dagger}
\vec{\sigma}\}+ {\rm Tr} \{\Bigl(
{\bf x}_{\alpha \alpha '}^{ji}\Bigr)^{\dagger}
\vec{\sigma}\}\otimes {\rm Tr} \{
{\bf x}_{\alpha\alpha '}^{ji}\vec{\sigma}\}\Biggr] +
(i\leftrightarrow j) \Biggr\} \nonumber\\
&+&\sum_{\alpha\alpha '}K_{\alpha\alpha '}\Bigl[{\rm Tr} \{(
{\bf m}_{\alpha '
0}^{i})^{\dagger} {\bf m}_{\alpha 0}^{i}\vec{\sigma}\}\otimes {\rm Tr}
\{{\bf w}_{\alpha\alpha'}^{ji}\vec{\sigma}\}
+(i\leftrightarrow j)\Bigr] \ . \label{III26c}
\end{eqnarray}
One notes that when the contributions of $\Delta U_{\alpha\alpha '}$ and
$K_{\alpha\alpha '}$ are ignored, Eqs. (\ref{III26})
reproduce Eq. (\ref{PRL}), with
${\bf D} (i,j) =-i[{\rm Tr} \{\tilde{\bf t}_{00}^{ji}\}
{\rm Tr} \{\tilde{\bf t}_{00}^{ij}\vec{\sigma}\} /U_{0}
-(i\leftrightarrow j)]$, and
$J=2{\rm Tr} \{\tilde{\bf t}_{00}^{ji}\tilde{\bf t}_{00}^{ij}\}/U_{0}$.
\end{mathletters}
The results (\ref{III26}) hold for general site symmetry, and to
all orders in the spin--orbit coupling. They become particularly
simple in the special case of tetragonal symmetry, as is
discussed in Sec. IV.  In Eq. (\ref{III26a}) we see that even
when ${\bf U}$ is a constant and ${\bf K}=0$, $J(i,j)$
does depend on $\lambda$. Moriya's expression for $J(i,j)$ is only
correct to zeroth order in $\lambda$.

\subsection{The copper--oxygen--copper bond}

Here we derive the effective magnetic Hamiltonian of the copper
spins for the bond Cu--O--Cu. The spin--orbit interaction on the
oxygen is much smaller than that on the copper,[\onlinecite{VV}]
and therefore may be neglected. Then the microscopic
Hamiltonian (\ref{HAM}) is modified as follows. First, the kinetic
energy now represents hopping between the oxygen and the copper ions.
That is, in place of Eqs. (\ref{EQ14}) we now have
\begin{equation}
\label{HOP1} 
{\cal H}_{\rm hop}= \sum_{iq} T_{qi} + {\rm h. \ c.} \ ,
\end{equation}
where
\begin{equation}
\label{HOP2}
T_{qi} = \sum_{an} \sum_{\sigma \sigma'} \Bigl(
\bar t_{na}^{qi}\Bigr)_{\sigma\sigma '}
p_{qn\sigma}^{\dagger}c_{ia\sigma '} \ ,
\end{equation}
in which $p_{qn\sigma}$ ($p_{qn\sigma}^{\dagger}$) are the destruction
(creation) operators for a hole on one of the states ($n$) of the
$q$th oxygen, and
\begin{equation}
\label{HOP3}
\Bigl( \bar t_{na}^{qi}\Bigr)_{\sigma\sigma '}
=\sum_{\alpha}t_{n\alpha}^{qi} \Bigl(m_{\alpha a}^{i}
\Bigr)_{\sigma\sigma '} \ .
\end{equation}
Here $t^{qi}_{n\alpha}$ describes hopping from the $\alpha$th
orbital on the $i$th copper ion to the $n$th orbital on the $q$th 
oxygen ion.  The matrix element
$\Bigl(\bar {t}^{qi}_{n \alpha} \Bigr)_{\sigma \sigma'}$ describes
hopping between the copper states [see Eq. (\ref{UNITARY})] which
diagonalize ${\cal H}_x+{\cal H}_{\rm so}$ and the $q$th oxygen ion.

Second, we add to the Hamiltonian the on--site single--particle
energies and the Coulomb interactions on the oxygen. These terms
are written in the form
\begin{eqnarray}
\label{III28}
{\cal H}_{p}&=&\sum_{qn\sigma}\epsilon_{n}p_{qn\sigma}^{\dagger}
p_{qn\sigma}+\frac{1} {2}\sum_{qnn'}\sum_{\sigma\sigma '}U_{p}^{(q)}
p_{qn\sigma}^{\dagger}p_{qn'\sigma '}^{\dagger}
p_{qn'\sigma '}p_{qn\sigma}\nonumber\\ &+&\frac{1}{2}
\sum_{qnn'}\sum_{\sigma\sigma '}\Delta U_{nn'}^{(q)}
p_{qn\sigma}^{\dagger}p_{qn'\sigma '}^{\dagger}p_{qn'\sigma '}
p_{qn\sigma}+ \frac{1}{2}
\sum_{qnn'}\sum_{\sigma\sigma '}K_{nn'}^{(q)}
p_{qn\sigma}^{\dagger}p_{qn'\sigma '}^{\dagger}p_{qn\sigma'}
p_{qn'\sigma} \nonumber \\ & \equiv &
{\cal H}_0^{(p)} + \Delta {\cal H}_c^{(p)} \ ,
\end{eqnarray}
where ${\cal H}_0^{(p)}$ is the first line of this equation
and $\Delta {\cal H}_c^{(p)}$ is the second line.
Thus, the total Hamiltonian for this case is taken to be
${\cal H}_0 + V$, where
\begin{equation}
{\cal H}_0 =  {\cal H}_x + {\cal H}_{\rm so} + {\cal H}_{c0}
+ {\cal H}_0^{(p)} 
\end{equation}
and $V$, which we treat perturbatively, is
\begin{equation}
V = {\cal H}_{\rm hop} + \Delta {\cal H}_c + \Delta {\cal H}_c^{(p)} \ .
\end{equation}

In the above, the index $q$ distinguishes between oxygens on the bond
along the $x$-- and $y$--directions from the copper ion in question.
However, the perturbation expansion gives results in the form of
contributions summed over all pairs of single bonds between
nearest--neighboring copper ions $i$ and $j$.
Then, the index $q$ is fixed once
the values of $i$ and $j$ are specified, as one sees from Fig. 1.
Accordingly, we henceforth omit the index $q$, so that, for instance,
${\bar {\bf t}}^{qi}_{n a} \rightarrow {\bar {\bf t}}^{i}_{na}$,
${\bar {\bf t}}^{iq}_{an} \rightarrow {\bar {\bf t}}^{i}_{an}$,
$T_{qi} \rightarrow T_i$, and
$p_{qn\sigma}^\dagger \rightarrow p_{n \sigma}^\dagger$.

We now turn to the perturbation expansion, from which we obtain
the magnetic Hamiltonian of the copper spins.  It is clear that the
lowest order contribution to the effective interaction between two
copper spins is of order ${\bf t}^{4}$. There are two
possible channels in this order, which we denote by $a$ and $b$.
In channel $a$, the hole is transferred from one of the coppers to the
oxygen, then to the second copper, and then back to the first copper
via the oxygen. Hence in this channel there are two holes on the
{\it copper} in the intermediate state. In channel $b$, the hole is
transferred from one of the Cu ions to the oxygen, and then a second
hole is taken from the second copper to the same oxygen. Afterwards
the two holes return to the coppers, i.e., back to the ground
state in which there is one hole on each Cu ion. Thus in channel $b$
there are two holes on the {\it oxygen} in the intermediate state.
When the terms coming from the Coulomb interactions
$\Delta U_{nn'}$ and $K_{nn'}$ for the oxygen ions
are included, then their effect will appear only in
channel $b$, in which the two holes have a state 
where both are on the oxygen.

It turns out that for channel $a$ all our previous expressions,
derived for the Cu--Cu bond, hold with the replacement
\begin{equation}
\label{III29}
\tilde{\bf t}_{ab}^{ij}=\sum_{n}\frac{\bar {\bf t}_{an}^{i}
\bar {\bf t}_{nb}^{j}} {\epsilon_{n}} \ .
\end{equation}
We show this explicitly in Appendix E for the $\bar {\bf t}^{4}$ process.
Similar arguments hold for the processes of order
${\bar {\bf t}}^{4}\Delta\tilde{\bf U}$ and
${\bar {\bf t}}^{4}\tilde{\bf K}$, where $\Delta\tilde{\bf U}$
and $\tilde{\bf K}$ are the Coulomb interactions on the copper ion.
(Note that for this channel  $\Delta \tilde{U}$ and
$\tilde{K}$ represent the Coulomb interactions on the copper).

It thus remains to investigate the perturbation expansion in channel
$b$.  Applying once to $\psi_0$ the term in the Hamiltonian of Eq.
(\ref{HOP1}), which describes hopping from the copper ions to
the intervening oxygen ion, one obtains
\begin{eqnarray}
\label{III30}
\mid\psi_{1}> & \equiv &  \left( T_i + T_j \right) c_{i0\sigma}^\dagger
c_{j0\sigma_1}^\dagger c_{j0\sigma_1} c_{i0\sigma}  \mid \psi_0 \rangle
\nonumber \\ &=&
\sum_{n}\sum_{\sigma\sigma_{1}\sigma_{2}}\Bigl[\Bigl( {\bar t}_{n0}^i
\Bigr)_{\sigma_{2}\sigma} p_{n\sigma_{2}}^{\dagger}
c_{j0\sigma_{1}}^{\dagger}
+ \Bigl( {\bar t}_{n0}^j \Bigr)_{\sigma_{2}\sigma_{1}}
c_{i0\sigma}^{\dagger} p_{n\sigma_{2}}^{\dagger}\Bigr]
c_{j0\sigma_{1}}c_{i0\sigma}\mid\psi_{0} \rangle \ ,
\end{eqnarray}
which represents virtual states with energy $\epsilon_{n}$. In the
next order, the second hole is put on the same oxygen. This leads to 
\begin{eqnarray}
\label{III31}
| \psi_{2b} \rangle & \equiv & \left( {1 \over {\cal H}_0} T_j
{1 \over {\cal H}_0} T_i + {1 \over {\cal H}_0} T_i
{1 \over {\cal H}_0} T_j \right)
c_{i0\sigma}^\dagger c_{j0\sigma_1}^\dagger c_{j0 \sigma_1}
c_{i0\sigma} | \psi_0 \rangle \nonumber \\
&=& \sum_{nn'}\sum_{\sigma\sigma_{1}}\sum_{\sigma_{2}\sigma_{3}}
\Bigl(\frac{1}{\epsilon_{n}}+\frac{1}{\epsilon_{n'}}\Bigr)\frac{1}
{\epsilon_{n}+\epsilon_{n'}+U_{p}}
\Bigl({\bar t_{n0}^i}\Bigr)_{\sigma_{2}\sigma}
\Bigl( {\bar t_{n'0}^j}\Bigr)_{\sigma_{3}
\sigma_{1}}p_{n\sigma_{2}}^{\dagger}
p_{n'\sigma_{3}}^{\dagger}c_{j0\sigma_{1}}
c_{io\sigma}\mid\psi_{0}\rangle \ .
\end{eqnarray}
In order to return to the ground state two 
more powers of the hopping are needed. This gives
\begin{eqnarray}
\label{III32}
&& {\cal H}^{(2b)} (i,j)=-\sum_{nn'} \Bigl(\frac{1}{\epsilon_{n}}
+\frac{1}{\epsilon_{n'}}\Bigr)^{2}\frac{1}{\epsilon_{n}
+\epsilon_{n'} +U_{p}}\Bigl[ {\rm Tr} \Bigl\{\Bigl(\frac{1}{2}
+\vec{\sigma}\cdot {\bf S}(i) \Bigl){\bar {\bf t}_{0n}^i}
{\bar {\bf t}_{n0}^i}\Bigr\} \nonumber \\ &\times &
{\rm Tr} \Bigl\{\Bigl(\frac{1}{2}
+\vec{\sigma}\cdot {\bf S}(j) \Bigr)
{\bar {\bf t}_{0n'}^j} {\bar {\bf t}_{n'0}^j}\Bigr\}
-{\rm Tr} \Bigl\{\Bigl(\frac{1}{2}
+\vec{\sigma}\cdot {\bf S}(i) \Bigr)
{\bar {\bf t}_{0n'}^i} {\bar {\bf t}_{n'0}^j}\Bigl( \frac{1}{2}
+ \vec{\sigma} \cdot {\bf S}(j)\Bigr) {\bar {\bf t}_{0n}^j}
{\bar {\bf t}_{n0}^i}\Bigr\}\Bigr]\ ,
\end{eqnarray}
where we used the identity Eq. (\ref{IDENT}). (We labeled this
contribution with a superscript 2 because even though it is fourth
order in the ${\bf {\bar t}}$'s, it is really a second--order process
in terms of a
renormalized Cu--Cu hopping interaction. The superscript ``b''
indicates a contribution from channel b.) In a similar way to the
arguments given after Eq. (\ref{III6}), one can convince oneself
that the first term in Eq. (\ref{III32}) as well as the terms coming
from the factors of $1/2$ in the second term do not contribute to
the spin Hamiltonian. Thus, to order ${\bf {\bar t}}^{4}$, the
contribution of channel $b$ is
\begin{equation}
\label{III33}
{\cal H}^{(2b)}(i,j)=\sum_{nn'}\Bigl(\frac{1}{\epsilon_n}
+\frac{1}{\epsilon_{n'}}\Bigr)^{2}
\frac{1}{\epsilon_{n}+\epsilon_{n'}+U_{p}}
{\rm Tr} \Bigl\{\vec{\sigma}\cdot {\bf S}(i)
{\bar {\bf t}}_{0n'}^{i}
{\bar {\bf t}}_{n'0}^{j}\vec{\sigma}\cdot {\bf S}(j)
{\bar {\bf t}}_{0n}^{j} {\bar {\bf t}}_{n0}^{i}\Bigr\}\ .
\end{equation}

Next we calculate the effect of the Coulomb terms $\Delta U_{nn'}$
and $K_{nn'}$ of Eq. (\ref{III28}). To this end we apply them to the
state $|\psi_{2b}\rangle$ of Eq. (\ref{III31}). The result is
\begin{eqnarray}
- {1 \over {\cal H}_0} \Biggl( \Delta {\bf U} + {\bf K} \Biggr) |
\psi_{2b} \rangle &=&
-\sum_{nn'}\sum_{\sigma\sigma_{1}}\sum_{\sigma_{2}\sigma_{3}}
\Bigl(\frac{1} {\epsilon_{n}}+\frac{1}{\epsilon_{n'}}\Bigr)
\Bigl(\frac{1}{\epsilon_{n}+\epsilon_{n'}+U_{p}}\Bigr)^{2}
\Bigl( {\bar t}_{n0}^{i}\Bigr)_{\sigma_{2}\sigma}
\Bigl( {\bar t}_{n'0}^{j} \Bigr)_{\sigma_{3}\sigma_{1}}\nonumber\\
&\Bigl[& \Delta U_{nn'}p_{n\sigma_{2}}^{\dagger}
p_{n'\sigma_{3}}^{\dagger} +K_{nn'}
p_{n'\sigma_{2}}^{\dagger}p_{n\sigma_{3}}^{\dagger}
\Bigr]c_{j0\sigma_{1}}
c_{i0\sigma}\mid\psi_{0}>.\label{III34}
\end{eqnarray}
Finally we apply two factors of the hopping which bring the holes
back to the ground state.  This leads to
\begin{eqnarray}
{\cal H}^{(2b,c)}(i,j)=\sum_{nn'}\Bigl[\Bigl(
\frac{1}{\epsilon_{n}}+\frac{1}{\epsilon_{n'}}\Bigr)
\frac{1}{\epsilon_{n}+\epsilon_{n'}+U_{p}}\Bigr]^{2}
\Bigl[&-&\Delta U_{nn'}{\rm Tr} \Bigl\{\vec{\sigma}\cdot {\bf S}(i)
{\bar {\bf t}_{0n'}^i} {\bar {\bf t}_{n'0}^j}
\vec{\sigma}\cdot {\bf S}(j) {\bar {\bf t}_{0n}^j}
{\bar {\bf t}_{n0}^i}\Bigr\}\nonumber\\
+K_{nn'}{\rm Tr} \Bigl\{\vec{\sigma}\cdot {\bf S}(i)
{\bar {\bf t}_{0n'}^i} {\bar {\bf t}_{n0}^i}\Bigr\}{\rm Tr}
\Bigl\{\vec{\sigma}
\cdot {\bf S}(j) {\bar {\bf t}_{0n}^j} {\bar {\bf t}_{n'0}^j}\Bigr\}
&-&K_{nn'}{\rm Tr} \Bigl\{\vec{\sigma}\cdot { \bf S}(i)
{\bar {\bf t}_{0n}^i} {\bar {\bf t}_{n'0}^j}\vec{\sigma}\cdot {\bf S}(j)
{\bar {\bf t}_{0n'}^j} {\bar {\bf t}_{n0}^i}\Bigr\}\Bigr].\label{III35}
\end{eqnarray}

Combining Eqs. (\ref{III33}) and (\ref{III35}) we obtain the
magnetic interaction arising from
channel $b$ in the form of Eq. (\ref{JDMEQ}), with
\begin{mathletters}
\label{III36}
\begin{eqnarray}
\label{III36a}
J^{(b)}(i,j) &=&\sum_{nn'}\Bigl(\frac{1}{\epsilon_{n}}
+\frac{1}{\epsilon_{n'}}\Bigr)^{2} \frac{1}{\epsilon_{n}
+\epsilon_{n'}+U_{p}}\Bigl(1-\frac{\Delta U_{nn'}}{\epsilon_{n}
+ \epsilon_{n'}+U_{p}}\Bigr){\rm Tr} \Bigl\{ {\bar {\bf t}}_{0n'}^{i}
{\bar {\bf t}}_{n'0}^{j} \Bigl( {\bar {\bf t}}_{0n}^{j}
{\bar {\bf t}}_{n0}^{i}\Bigr) ^{\dagger}\Bigr\}\nonumber\\
&-&\sum_{nn'}\Bigl(\frac{1}{\epsilon_{n}}+\frac{1}{\epsilon_{n'}}
\Bigr)^{2} \frac{K_{nn'}}{(\epsilon_{n}+\epsilon_{n'}
+U_{p})^{2}}{\rm Tr} \Bigl\{ {\bar {\bf t}}_{0n}^{i}
{\bar {\bf t}}_{n'0}^{j}\Bigl( {\bar {\bf t}}_{0n'}^{j}
{\bar {\bf t}}_{n0}^{i}\Bigr)^{\dagger}\Bigr\}\ ,
\end{eqnarray}
\begin{eqnarray}
\label{III36b}
{\bf D}^{(b)} (i,j) &=&-\frac{i}{2}\sum_{nn'}\Bigl(\frac{1}{\epsilon_{n}}
+\frac{1}{\epsilon_{n'}}\Bigr)
^{2}\frac{1}{\epsilon_{n}+\epsilon_{n'}+U_{p}}\Biggl[
\Bigl(1-\frac{\Delta U_{nn'}}{\epsilon_{n}+\epsilon_{n'}+U_{p}}
\Bigr)\nonumber\\
&\Bigl(& {\rm Tr} \Bigl\{ {\bar {\bf t}}_{0n}^{j} 
{\bar {\bf t}}_{n0}^{i}\Bigr\}
{\rm Tr} \Bigl\{ {\bar {\bf t}}_{0n'}^{i}
{\bar {\bf t}}_{n'0}^{j}\vec{\sigma}\Bigr\}
-{\rm Tr} \Bigl\{ {\bar {\bf t}}_{0n'}^{i}
{\bar {\bf t}}_{n'0}^{j}\Bigr\}
{\rm Tr} \Bigl\{ {\bar {\bf t}}_{0n}^{j}
{\bar {\bf t}}_{n0}^{i}\vec{\sigma}\Bigr\}\Bigr)\nonumber\\
&& -\frac{K_{nn'}}{\epsilon_{n}+\epsilon_{n'}+U_{p}} \Bigl(
{\rm Tr} \Bigl\{ {\bar {\bf t}}_{0n'}^{j} {\bar {\bf t}}_{n0}^{i}
\Bigr\}{\rm Tr} \Bigl\{ {\bar {\bf t}}_{0n}^{i}
{\bar {\bf t}}_{n'0}^{j}\vec{\sigma}\Bigr\}-
{\rm Tr} \Bigl\{ {\bar {\bf t}}_{0n}^{i}
{\bar {\bf t}}_{n'0}^{j}\Bigr\}
{\rm Tr} \Bigl\{ {\bar {\bf t}}_{0n'}^{j} {\bar {\bf t}}_{n0}^{i}
\vec{\sigma}\Bigr\}\Bigl)\Biggr],
\end{eqnarray} 
\begin{eqnarray}
{\bf M}^{(b)}(i,j) &=&\frac{1}{2}\sum_{nn'}\Bigl(\frac{1}{\epsilon_{n}}
+\frac{1}{\epsilon_{n'}}\Bigr)^{2}
\frac{1}{\epsilon_{n}+\epsilon_{n'}+U_{p}}
\Biggl[\Bigl(1-\frac{\Delta U_{nn'}}
{\epsilon_{n}+\epsilon_{n'}+U_{p}}\Bigr)\nonumber\\
&\Bigl(& {\rm Tr} \Bigl\{ {\bar {\bf t}}_{0n'}^{i}
{\bar {\bf t}}_{n'0}^{j}\vec{\sigma}\Bigr\}\otimes
{\rm Tr} \Bigl\{ {\bar {\bf t}}_{0n}^{j}
{\bar {\bf t}}_{n0}^{i}\vec{\sigma}\Bigr\}
+{\rm Tr} \Bigl\{ {\bar {\bf t}}_{0n}^{j} {\bar {\bf t}}_{n0}^{i}
\vec{\sigma}\Bigr\}\otimes {\rm Tr} \Bigl\{ {\bar {\bf t}}_{0n'}^{i}
{\bar {\bf t}}_{n'0}^{j}\vec{\sigma}\Bigr\}\Bigr)\nonumber\\
&& +\frac{K_{nn'}}{\epsilon_{n}+\epsilon_{n'}
+U_{p}} \Bigl( 2 {\rm Tr} \Bigl\{ {\bar {\bf t}}_{0n'}^{i}
{\bar {\bf t}}_{n0}^{i}
\vec{\sigma}\Bigr\}\otimes {\rm Tr} \Bigl\{ {\bar {\bf t}}_{0n}^{j}
{\bar {\bf t}}_{n'0}^{j}\vec{\sigma}\Bigr\}\nonumber\\
&-&{\rm Tr} \Bigl\{ {\bar {\bf t}}_{0n}^{i}
{\bar {\bf t}}_{n'0}^{j}\vec{\sigma}\Bigr\}\otimes
{\rm Tr} \Bigl\{ {\bar {\bf t}}_{0n'}^{j}
{\bar {\bf t}}_{n0}^{i}\vec{\sigma}\Bigr\}
-{\rm Tr} \Bigl\{ {\bar {\bf t}}_{0n'}^{j} {\bar {\bf t}}_{n0}^{i}
\vec{\sigma}\Bigr\}\otimes
{\rm Tr} \Bigl\{ {\bar {\bf t}}_{on}^{i} {\bar {\bf t}}_{n'0}^{j}
\vec{\sigma}\Bigr\}\Bigl)\Biggr].\label{III36c}
\end{eqnarray}
\end{mathletters}

The full magnetic Hamiltonian for the copper spins of the Cu--O--Cu
bond is obtained by combining the results of Eq. (\ref{III36}) for
channel $b$, with those for channel $a$ given by Eqs. (\ref{III26}),
in conjunction with the identification of Eq. (\ref{III29}).  These
results generalize those of Refs. \onlinecite{SAE,NEB} and
\onlinecite{JAPSPR}, which were obtained in the absence of the Coulomb
terms $\Delta {\bf U}$ and ${\bf K}$.

\begin{center}
\section{TETRAGONAL SYMMETRY}
\end{center}

This section consists of three subsections. In Sec. A, we apply
a canonical transformation to show that without Coulomb exchange
interactions the effective spin Hamiltonian is isotropic at all
orders of $t$ and $\lambda$. In Sec. B, we use this theorem to
isolate the most important contribution to the anisotropy, namely
that involving the Coulomb exchange energy.  In Sec. C we
corroborate our analytical results of perturbation theory by
numerical solutions for single--bond clusters: Cu--Cu and Cu--O--Cu.

\begin{center}
\subsection{Canonical Transformation}
\end{center}

We start by proving our strongest result, namely that the spin
Hamiltonian ${\cal H}_S$ arising from the generic model is
isotropic for a wide class of models in the absence of
Coulomb exchange interactions.  In particular, this result
holds for a commonly used model of the cuprates, in which
interionic Coulomb interactions, ${\bf V}$ and ${\bf M}$ in
Eq. (\ref{HAM}), and the Coulomb exchange terms ${\bf K}$
are neglected, hopping is between nearest--neighboring Cu ions,
and the site symmetry is tetragonal.
Strictly speaking, the only use we make of site symmetry is
that it has to be high enough so that the $3d$ spatial orbitals
which diagonalize the crystal field Hamiltonian are
$\psi_0({\bf r}) \sim x^2 -y^2$, $\psi_1 ({\bf r}) \sim 3z^2-r^2$,
$\psi_x ({\bf r}) \sim yz$, $\psi_y ({\bf r}) \sim xz$, and
$\psi_z  ({\bf r}) \sim xy$.  Here the $z$ axis coincides with the
tetragonal $c$--axis and the $x$-- and $y$--axes coincide with the
nearest--neighbor directions in the plane perpendicular to the
$c$--axis, as shown in Fig. 1.  These symmetry labels are chosen
so that $\psi_\alpha ({\bf r})$ transforms (under the operations of
tetragonal symmetry) like $L_\alpha$ for $\alpha = x, y, z$
and $\psi_0({\bf r})$ and $\psi_1({\bf r})$ transform like scalars.

An important observation is that the orbital angular momentum
operator ${\bf L}$ has matrix elements only between
states of specific symmetry.  For instance, $L_x$ connects
$\psi_x ({\bf r})$ only to the states $\psi_0({\bf r})$ and
$\psi_1({\bf r})$ and it connects $\psi_y({\bf r})$ to
$\psi_z( {\bf r})$ and vice versa.  Similar statements can be
made about the other components of ${\bf L}$.  We now
introduce a transformation in spin space (to pseudospin) such
that the spin--orbit interaction is diagonal with respect to
pseudospin.
For that purpose we introduce pseudospin $\vec \mu$ as follows
\begin{equation}
\label{CANON}
| \alpha , \mu >  = \psi_\alpha ({\bf r})
\sum_\eta [\sigma_\alpha]_{\mu , \eta}^* \phi_\eta \ \equiv
f_{\alpha , \mu }^{\dag} |{\rm vac} \rangle  \ ,
\end{equation}
where $\phi_\eta$ is a spin function for spin ``up'' if
$\eta=1/2$ and for spin ``down'' if $\eta=-1/2$, and
$| {\rm vac}\rangle$ denotes the vacuum state.  Here $\sigma_\alpha$
for $\alpha = x, \ y, \ z$ are the Pauli matrices, and
$\sigma_0=\sigma_1 = {\bf I}$ is the unit matrix.
As discussed in Appendix F, the above transformation is such that
the spin--orbit interaction is diagonal in pseudospin:
\begin{equation}
{\cal H}_{\rm so} = \lambda \sum_{k =i,j} \sum_{\alpha , \beta , \mu}
W(k, \alpha, \beta) f_{k \alpha \mu}^{\dag} f_{k \beta\mu} \ ,
\end{equation}
where $W(k, \alpha , \beta )$ is a spin--orbit matrix element.
[The transformation of Eq. (\ref{CANON}) should not be confused
with Eq. (\ref{UNITARY}).  The latter involves an exact diagonalization
and requires a knowledge of all the parameters.  In contrast,
the transformation of Eq. (\ref{CANON}) is independent of the
interaction parameters.  It merely brings the
Hamiltonian into block diagonal form in which there are
two identical blocks, one for $\mu=1$ and one for $\mu=-1$.]
Because the unitary transformation of Eq. (\ref{CANON}) does
not mix spatial states, it does not affect the form of the
Coulomb interactions scaled by ${\bf U}$.  Furthermore, in view of
the lattice symmetry hopping can only involve holes moving from one
site to a neighboring site {\it without changing their symmetry}.
Thus holes in a state $\psi_\alpha$ on one ion, where $\alpha= x, y$,
or $z$, can only hop to states of the same $\alpha$ on a
nearest--neighboring ion.  Likewise holes in states $\psi_0({\bf r})$
or $\psi_1 ({\bf r})$ on one ion can only hop to states
$\psi_0({\bf r})$ or $\psi_1 ({\bf r})$ on an adjacent ion.
Since states $\alpha$ and $\beta$ which are connected by
hopping must be states of the {\it same} symmetry, we have

\begin{equation}
\label{HOPSYM}
T_{ij} = \sum_{\alpha,\beta,\mu}
t_{i\alpha,j\beta} d_{i\alpha \mu}^{\dag} d_{j\beta \mu} =
\sum_{\alpha,\beta,\mu,\rho, \tau}
t_{i\alpha,j\beta} [\sigma_\alpha]_{\tau \mu} f_{i\alpha \tau}^{\dag}
[\sigma_\beta]_{\rho \mu}^* f_{j\beta \rho} =
\sum_{\alpha,\beta,\rho }
t_{i\alpha,j\beta} f_{i\alpha \rho}^{\dag} f_{j\beta \rho} \ . 
\end{equation}
In other words, the total Hamiltonian (for ${\bf K}=0$)
can be written in the form
\begin{eqnarray}
\label{ROTINV}
{\cal H}  &= &
\sum_k \sum_{\alpha,\mu} \epsilon_{k\alpha} f_{k\alpha \mu}^{\dag}
f_{k\alpha \mu} + \lambda \sum_k  \sum_{\alpha , \beta , \mu}
W(k, \alpha, \beta) f_{k \alpha \mu}^{\dag} f_{k \beta\mu}
+ \sum_{i,j, \alpha,\beta,\mu}
t_{i\alpha,j\beta} f_{i\alpha \mu}^{\dag}f_{j\beta \mu}
\nonumber \\
& + &  \frac{1}{2}\sum_k \sum_{\alpha,\alpha',\mu,\mu'}
U_{k\alpha,k\alpha'} f_{k\alpha \mu}^{\dag} f_{k\alpha' \mu'}^{\dag}
f_{k\alpha' \mu'} f_{k\alpha \mu} .
\end{eqnarray}
Thus this Hamiltonian can be written [\onlinecite{WRITE}] in terms
of the quantities
\begin{equation}
Q_{\alpha \beta}(i,j) \equiv \sum_\mu f_{i \alpha \mu}^{\dag}
f_{j \beta \mu}
\end{equation}
which themselves are invariant under rotations in pseudospin space.
Therefore ${\cal H}$ is invariant under rotations in pseudospin
space.  To construct the effective spin Hamiltonian
${\cal H}(i,j)$ involves using degenerate perturbation theory
to eliminate the excited states.  Accordingly, it is clear that
the resulting spin Hamiltonian will be rotationally invariant in
pseudospin space.  Since we have defined pseudospin so that in the
ground state (in which all the holes are in the state
$\psi_0$) pseudospin and real spin are identical,
it follows that for the tetragonal case with no Coulomb exchange
interactions the spin Hamiltonian is also rotationally invariant.
This theorem indicates that even though ${\cal H}_S$ may
include further neighbor two--spin interactions, four--spin
interactions, etc., it is nevertheless rotationally invariant,
so that the spin--wave spectrum can not have a gap at zero
wave vector in the absence of Coulomb exchange terms.

For the case of the nearest--neighbor exchange interaction we can
make some further explicit statements.  For tetragonal symmetry,
$J_{\mu \nu} (i,j)$ must also be diagonal (with its principal
axes along the tetragonal axes).
Thus, ${\cal H}(i,j)$ is an isotropic Heisenberg model.

The above theorem can be generalized to include the intervening oxygen
ions.  Here we consider a Hubbard model which includes the
three 2$p$ spatial orbitals.  Now we introduce different
unitary transformations for oxygen ions on $y$--directed
and $x$--directed bonds (see Fig. 1).  For
those on $y$--directed bonds we set $\psi_0 ({\bf r}) =|2p_y\rangle$,
$\psi_x({\bf r})= |2p_z \rangle$, and $\psi_z({\bf r})=|2p_x \rangle$.
We then introduce states $|\alpha , \mu \rangle$ by
\begin{equation}
\label{OXYTR}
|\alpha , \mu \rangle = \psi_\alpha ({\bf r}) \sum_\eta
[\sigma_\alpha ]_{\mu , \eta }^* \phi_\eta (\sigma) \ ,
\end{equation}
similar to Eq. (\ref{CANON}) which was used for the Cu d states.
We need to examine how the hopping and spin--orbit interactions
are affected by this transformation.  Note that hopping
along the $y$--direction can only take place between 
Cu states like $x^2-y^2$ and oxygen $2p_y$ states.  These
are both associated with symmetry 0 or 1.  Likewise an oxygen
$2p_z$ orbital can only hop to a copper $yz$ state, both
of which have symmetry label $x$.  Also an oxygen $2p_x$
orbital can only hop to a copper $xy$ state, both of which
have symmetry label $z$.  Thus with this labeling of
states, hopping occurs only between states of the same
symmetry label and the canonical transformation has no effect
on the hopping, just as in Eq. (\ref{HOPSYM}).  One can
verify that the spin--orbit interaction on the oxygen ions
does conserve pseudospin.  Oxygen ions on the $x$--directed
bonds are treated analogously.  For them we write
$\psi_0 ({\bf r}) =|2p_x\rangle$, $\psi_y({\bf r})= |2p_z \rangle$,
and $\psi_z({\bf r})= |2p_y \rangle$
and we again use Eq. (\ref{OXYTR}).  Then, we conclude
that the Hamiltonian of the entire lattice can be
expressed in terms of the quantities
$Q_{\alpha \beta} (i,j)$.  Thus the theorem holds with
intervening oxygen ions: in the absence of Coulomb
exchange, this model gives no anisotropy in ${\cal H}_S$
for a tetragonal lattice.

The fact that this conclusion is demonstrated to all orders in
perturbation theory represents an important new result.  The
low order perturbation result of BS is in accord
with this theorem.  As mentioned there, this conclusion modifies
the conventional wisdom that the anisotropy in the exchange
interaction is trivially related to the anisotropy of the $g$ tensor.
Finally, we emphasize that this theorem depends crucially on the fact
that the eigenstates of the crystal field are those of tetragonal
site symmetry and that hopping is only between nearest neighbors.
In addition, the theorem is only valid for Coulomb terms which
have the Hartree form, i.e., those which only involve
two orbitals, as in Eq. (\ref{HAM}).  More complicated Coulomb
terms (involving more than two orbitals) and Coulomb exchange terms
[involving ${\bf K}$ in Eq. (\ref{HAM})] will generate anisotropy.

\medskip
\begin{center}
\subsection{Anisotropy}
\end{center}
\medskip

Anisotropy can occur via various mechanisms.  One such mechanism
is to introduce Coulomb exchange interactions, as done implicitly
by BS.  In the context of the above discussion we note that exchange
interactions compete with spin--orbit interactions in the following
sense.  With only the former interactions the eigenstates of a single
ion are states of total real spin 1 or 0.  With no exchange but with
spin--orbit interactions, the eigenstates of a single ion are
states of total pseudospin 1 or 0.  In both cases, our numerical
evaluation of the energy levels gave singlets and triplets as
this argument requires.  However, the wave functions are different,
of course.  When both interactions are present, the degeneracies are
removed because wave functions can
not be simultaneous eigenfunctions of both real spin and pseudospin.
>From our calculation, treating hopping, spin--orbit,
$\Delta U$, and exchange interactions as perturbations,
we reach the following conclusions.
For the ``generic'' model, anisotropic exchange appears at order
${\bf t}^2 \lambda^2 {\bf K}$.  An efficient way to perform
this calculation is to use the hopping matrix elements,
$\tilde {\bf t}$, of Eq. (\ref{I16b}) and work to order
$\tilde {\bf t}^2 {\bf K}$, as we did in Sec. III.
The perturbation expansion yields the general expression, given in
Eq. (\ref{H2C}).  In Appendix G we analyze this expression for
tetragonal symmetry and find that it agrees with the result given
previously [\onlinecite{PRL2}], namely
\begin{eqnarray}
\label{JMU}
J_{\mu \mu}^{\rm anis}  & = &
- 2  \lambda^{2}  \left\{
 \frac{  \mid L_{0,\mu }^{\mu }\mid^{2} t_{0,1}^{2}K_{1,\mu }}
{  (\epsilon_{\mu } + \epsilon_{1}+U_{1,\mu })^{2}}
\left[ \frac{1}{ \epsilon_{\mu }} + \frac{1}
{\epsilon_{1}+U_{0,1}} \right]^{2}
\right.  \nonumber \\
&+&
\left.  \frac{ K_{0,\mu } }{(\epsilon_{\mu }+U_{0,\mu })^{2}}
\left | \frac{(t_{\mu ,\mu } - t_{0,0}) L_{0,\mu }^{\mu }}
{\epsilon_{\mu }} + \frac{t_{0,1}  L_{1,\mu }^{\mu }}
{\epsilon_{1} +U_{0,1}} \right |^{2} \right\} ,
\end{eqnarray}
where $L_{\alpha \beta}^\mu$ denotes the orbital angular
momentum matrix element, $\langle \alpha | L_\mu | \beta \rangle$,
$\mu$ assumes the values $x$, $y$, and $z$, and the superscript
``anis'' indicates that we have arbitrarily omitted isotropic
(i. e. $\mu$--independent) contributions.
The same expression is also derived directly from perturbation
theory in ${\bf t}$, $\lambda$, and ${\bf K}$ in Appendix H.

Now we briefly discuss the implications of the above result.
First of all, note that within tetragonal symmetry the result
does display the expected full anisotropy for a single bond,
under which $J_{xx}\equiv J_\parallel$, $J_{yy}\equiv J_\perp$,
and $J_{zz}$ are all different.  To get biaxiality
($J_\parallel \not= J_\perp$) requires either $t_{01}\not= 0$
or $t_{xx} \not= t_{yy}$, viz. Fig. 2.  Of course, in tetragonal
symmetry single--site quantities can not differentiate between the
$x$ ($\parallel$) and $y$ ($\perp$) directions.  To understand  why
$t_{01}\not=0$ introduces biaxiality, note that $t_{01}$
changes sign when the local $x$ coordinates are rotated into
the $y$ coordinates.  Also note that even in the limit when $U$
is considered to be very large, the result still does depend on
the hopping between excited levels through $t_{\mu \mu}$.
Finally, we remark that these expressions differ in several
respects from those of BS. This point is
discussed in Appendix I, where we give the results more explicitly. 

\begin{center}
\subsection{With Oxygens}
\end{center}

Turning now to the Cu--O--Cu bond in the tetragonal symmetry,
we again discuss separately the anisotropy resulting from channel
$a$ (the two holes occupy the same copper in the
intermediate state) and that coming from channel $b$ (the two
holes are on the oxygen in the intermediate state).

For channel $a$, we use the transformation of Eq. (\ref{III29}).
Using Eqs. (\ref{MMATEQ}) and Eq. (\ref{HOP3}) we obtain
\begin{equation}
\Bigl( {\tilde t}_{ab}^{ij} \Bigr)_{\sigma_1 \sigma_2} = \sum_{n
\alpha \beta} {1 \over \epsilon_n } t_{\alpha n}^i t_{n \beta}^j
{\bar m}_{\alpha a}^* {\bar m}_{\beta b} \Bigl( \sigma_\alpha^\dagger
\sigma_\beta \Bigr)_{\sigma_1 \sigma_2} \ ,
\end{equation}
where ${\bar m}_{\alpha a}$ are scalars [see Eq. (\ref{MMATEQ})].  For
tetragonal symmetry, $\alpha$ and $\beta$ belong to the same
symmetry class (e.g., $\alpha=\beta$ for $\alpha=x, y, $ or $z$, or
$\alpha$ and $\beta$ are 0 or 1).  Hence ${\tilde {\bf t}}_{ab}^{ij}$
becomes the unit matrix times a scalar given by
\begin{equation}
{\tilde t}_{ab}^{ij} = \sum_{\alpha \beta} t_{\alpha \beta}^{ij}
{\bar m}_{\alpha a }^* {\bar m}_{\beta b } \ , \ \ \ \ \ \ 
t_{\alpha \beta}^{ij} = \sum_n {1 \over \epsilon_n } t_{\alpha n}^i
t_{n \beta}^j \ ,
\end{equation}
in which $\alpha$ and $\beta$ refer to tetragonal d states.
The calculation can now proceed exactly as for the ``generic'' model
described in Appendix G (or H), with the effective hopping matrix
elements $t_{\alpha\beta}$ given by
\begin{eqnarray}
t_{00}&=&t_{0p_{x}}^{2}/\epsilon_{p_{x}},\ \
t_{01}=t_{0p_{x}}t_{p_{x}1}/\epsilon_{p_{x}},
\nonumber\\
t_{xx}&=&0,\ \ t_{yy}=t_{yp_{z}}^{2}/\epsilon_{p_{z}},\ \ 
t_{zz}=t_{zp_{y}}^{2}/\epsilon_{p_{y}},\label{IV4}
\end{eqnarray}
where $p_{x}$, $p_{y}$, and $p_{z}$ represent the p states on the
oxygen. Thus the contribution of channel $a$ 
to the symmetric anisotropy of the spin Hamiltonian for a bond along
the $x$--direction is reproduced by Eq. (\ref{JMU}), with the
replacements (\ref{IV4}).  Analogous expressions hold
for a Cu--O--Cu bond along the $y$--direction.

Now let us consider the magnetic anisotropy in channel $b$. Inspection
of Eqs. (\ref{III36b})
and (\ref{III36c}) shows that we need to examine the $2\times 2$
matrices ${\bar {\bf t}}_{0n} {\bar {\bf t}}_{n'0}$.
(We omit the site indices which are irrelevant for the tetragonal
symmetry). Using Eq. (\ref{HOP3}) and tetragonal symmetry, we write
\begin{equation}
{\bar {\bf t}}_{0n} {\bar {\bf t}}_{n'0}=
\sum_{\alpha\beta}t_{\alpha n}t_{n'\beta}
{\bf m}_{\alpha 0}^{\dagger} {\bf m}_{\beta 0}
= \sum_\alpha t_{\alpha n} t_{n' \alpha} {\bf m}_{\alpha 0}^\dagger
{\bf m}_{\alpha 0} \ . \label{IV5}
\end{equation}
Therefore, they are proportional to the unit
matrix. As a result, there is
no contribution to the magnetic anisotropy in channel $b$ in order
${\bf t}^{4}$. The reason is that in this channel (and to this order) 
the excited states on the copper are not visited at all.
Therefore, just as is
the case for the Cu--Cu bond when those states are ignored
[cf. Eq. (\ref{III7})] the magnetic Hamiltonian 
resulting from this channel is isotropic.

\medskip
\begin{center}
\subsection{Numerical Study}
\end{center}
\medskip

We have  checked our analytical results of perturbation theory
against results (shown in Fig. 3) obtained from exact
diagonalization for the four lowest levels out of the 190
possible two--hole states for a pair of Cu sites.  The relations
between the exchange constants and the four lowest levels 
are obtained in Appendix A.

For our numerical results shown in Fig. 3, we used the values of the
parameters listed in Table
I.[\onlinecite{ESKES,EANDS,MAM,HSC,JSG,CEM,ETS,ESF}]
The hopping matrix elements are related to $(pd\sigma )$ as follows:
$t_{0,p_{x}} = -\sqrt{3} t_{1,p_{x}} = \frac{\sqrt{3}}{2}(pd\sigma)$
and $t_{y,p_{z}} = t_{z,p_{y}} = (pd\pi)$, with
$(pd\pi ) \approx -\frac{1}{2} (pd\sigma)$.[\onlinecite{LFM}]
The expressions for $U_{\alpha,\beta}$ and $K_{\alpha,\beta}$
in terms of the Racah parameters were taken from Ref. \onlinecite{JSG}
and are listed in Appendix B.  We also checked that the
$J_{\mu \mu}^{\rm anis}$, shown in Fig. 3, agree to within about 10\%
with those obtained from the full 325 site Hamiltonian for the
Cu--O--Cu cluster.  Very crudely, as $A$ increases above 7 eV, the
$J_{\mu \mu}^{\rm anis}$ are inversely proportional
to $A^{-2}$ and are proportional to a linear combination of $B$ and $C$.
(When $B=C=0$, our theorem indicates that there is no anisotropy in
$J_{\mu\mu}$.) Thus our results are not highly sensitive to increasing
the value of $A$.  As $A$ is decreased below about 6 eV, perturbation
theory rapidly becomes increasingly inaccurate.  Now we discuss
briefly the numerical values of the Racah parameters.  We took the
values of $B$ and $C$ from Ref. 31.  Then, fixing the value of $A$
is equivalent to fixing the value of $U_0=A+4B+3C$.  Recently proposed
values for $U_0$ are 8.8, 8.8, 9.4, and 10.5 eV from Refs.
\onlinecite{ESKES}, \onlinecite{EANDS}, \onlinecite{MAM},
and \onlinecite{HSC}, respectively.  As a compromise, we took
$U_0=9.34$ or $A=7$ eV.  Our parameters yield an anisotropy in $J$
of order 0.03 meV and, as we shall see in the next section,
give an out--of--plane gap in the spin--wave spectrum within 10\% of the
experimental [\onlinecite{CL}] value 5 meV.

\begin{center}
\section{SPIN WAVE SPECTRUM OF THE EFFECTIVE SPIN HAMILTONIAN
IN TETRAGONAL SYMMETRY}
\end{center}

Given Eqs. (\ref{HJMN}) and (\ref{JMU}) for single bonds, the classical
ground state of the effective spin Hamiltonian is rotationally invariant
in the basal plane.  The out--of--plane anisotropy
$\alpha_{XY}\approx\Delta J/J_{0}$ is positive
(see Fig. 3), and therefore the spins order in that plane,
as is well established.
In the absence of spin wave fluctuations, the in--plane gap is zero.
However, the classical rotational invariance within the basal
plane is broken by the dependence of the spin--wave energies
on the angle $\theta$ between the staggered magnetization and the
crystal $x$--axis.  The purpose of this section is to study
this anisotropy and show that it leads to a nonvanishing
in--plane gap in the spin--wave spectrum. [\onlinecite{PRL1,PRL2}]

In order to show this, we start with the following general
Hamiltonian for the CuO$_{2}$ plane in a tetragonal system
\begin{equation}
\label{SW1}
H_{\rm eff} = \sum_{<ij>} H_{ij}  \ ,
\end{equation}
where for $<ij>$ along the $x$--direction, $H_{ij}$ is
\begin{equation}
\label{SW2}
H_{ij} =  J_{\parallel}S_{i}^{x}S_{j}^{x} + J_{\perp} S_{i}^{y}S_{j}^{y}
+ J_{z} S_{i}^{z} S_{j}^{z}
\end{equation}
and   for $<ij>$ along the $y$--direction $H_{ij}$ is
\begin{equation}
\label{SW3}
H_{ij} =  J_{\perp}S_{i}^{x}S_{j}^{x} + J_{\parallel} S_{i}^{y}S_{j}^{y}
+ J_{z} S_{i}^{z} S_{j}^{z} \ .
\end{equation}

We will now calculate the spin--wave spectrum of this Hamiltonian and
then the first quantum correction to the classical ground state energy.
We consider the case where the spins lie in the $xy$--plane and 
ordered antiferromagnetically ($J_{\parallel},J_{\perp}>J_{z} >0$). 
Assuming the staggered magnetization moment makes an angle
$\theta$ with the positive $x$--axis, we use the following 
transformation so that spins are parallel to the new $z$--axis:
\begin{equation}
\label{SW4}
{\bf S}_{i}   =
   \left(
\begin{array}{ccc}
0& -\sin \theta &
  \cos\theta \\
0& \cos \theta &
  \sin \theta \\
-1 & 0 & 0
\end{array}
\right)  {\bf S'}_{i} \ .
\end{equation}
Defining sublattice A to have up spins (in the rotated frame)
and sublattice B to have down spins (in the rotated frame), we have the
following bosonic spin representation:
\begin{equation}
\label{SW5}
S^{'x}_{i}  =
\sqrt{\frac{S}{2}} \; \left [ a_{i} +
a_{i}^{+} \right ] \ , \ \ \ \ 
S_{i}^{'y} = -i \sqrt{\frac{S}{2}} \; \left [ a_{i} -
 a_{i}^{+} \right ] \ , \ \ \ \
S_{i}^{'z} = S -  a_{i}^{+} a_{i}
\end{equation}
for sublattice A, and 
\begin{equation}
\label{SW6}
S^{'x}_{j}  = \sqrt{\frac{S}{2}} \; \left [ b_{j} +
b_{j}^{+} \right ] \ , \ \ \ \
S_{j}^{'y} = i \sqrt{\frac{S}{2}} \; \left [ b_{j} -
b_{j}^{+} \right ] \ , \ \ \ \
S_{j}^{'z} = - S +   b_{j}^{+} b_{j}
\end{equation}
for sublattice B.  For later convenience we consider the case of 
general spin, although in the end we set $S=1/2$.
Using Eqs. (\ref{SW4}--\ref{SW6}) we may write 
the effective spin Hamiltonian $H_{\rm eff}$ given in Eq. (\ref{SW1})
in momentum space as
\begin{equation}
\label{SW7}
H_{\rm eff} = E_{0} + 4 J_{\rm av} S\sum_{\bf q}  
\bigg[  a_{\bf q}^{\dag} a_{\bf q} + b_{\bf q}^{\dag} b_{\bf q} + \Biggl(
A_{\bf q} a_{\bf q}b_{-\bf q} + B_{\bf q} a_{\bf q} b_{\bf q}^{\dag}
 +  {\rm h. \ c.}
\Biggr)  \bigg] \ ,
\end{equation}
where ${\bf q}$ is summed over the first Brillouin zone of the magnetic
reciprocal lattice and
\begin{eqnarray}
\label{SW8}
E_{0} & = & - 2 J_{\rm av} N S^{2} \ , \hspace{1cm} J_{\rm av} =
\frac{1}{2}(J_{\parallel} + J_{\perp}) \ , \nonumber \\
A_{\bf q} & = & \frac{1}{4 J_{\rm av}}
\big[ J_{1} \cos(q_{x}a) + J_{2} \cos(q_{y}a) \big] \ , \nonumber \\ 
B_{\bf q} & = & - \frac{1}{4 J_{\rm av}} 
\big[ J_{3} \cos(q_{x}a) + J_{4} \cos(q_{y}a) \big] \ .
\end{eqnarray}
Here $N$ is the total number of spins and
\begin{eqnarray}
\label{SW9}
J_{1} = J_{\parallel} \sin^{2}\theta +
J_{\perp} \cos^{2}\theta + J_{z} \ , \nonumber \\
J_{2} = J_{\parallel} \cos^{2}\theta
+ J_{\perp} \sin^{2}\theta + J_{z} \ , \nonumber \\
J_{3} = J_{\parallel} \sin^{2}\theta
+ J_{\perp} \cos^{2}\theta - J_{z} \ , \nonumber \\
J_{4} = J_{\parallel} \cos^{2}\theta
+ J_{\perp} \sin^{2}\theta - J_{z} .
\end{eqnarray}
Henceforth we will set the lattice constant $a$ to unity.
Note that our conventions imply that $\sum_{\bf q} 1 = N/2$.
As one expects the classical ground state energy $E_{0}$ does not
depend on $\theta$ and thus we have complete degeneracy with
respect to $\theta$. However diagonalization of the Hamiltonian in
Eq. (\ref{SW7}) leads to the result
\begin{equation}
\label{SW10}
H_{\rm eff} = E_{0}' + \sum_{\bf q} \bigg\{ \omega_{+}({\bf q})
a_{\bf q}^{'\dag} a_{\bf q}^{'} +
\omega_{-}({\bf q}) b_{\bf q}^{'\dag} b_{\bf q}^{'} \bigg\} \ ,
\end{equation}
where the new ground state energy  $E_{0}'$ is now
\begin{equation}
\label{SW11}
E_{0}' =  - 2 (1 + \frac{1}{S} ) N J_{\rm av} S^{2} + \frac{1}{2} \sum_{\bf q}
\bigg\{\omega_{+}({\bf q}) + \omega_{-}({\bf q}) \bigg\}
\end{equation}
and thus does depend on $\theta$.  This dependence on $\theta$ arises
because the zero--point motion contribution (which is the sum
of spin--wave energies $\omega_{+}({\bf q}) + \omega_{-}({\bf q})$
over the Brillouin zone) depends on $\theta$.  The spin--wave energies are
\begin{eqnarray}
\label{SW12}
\omega_{+}({\bf q})& =&  4 J_{\rm av} S \sqrt{(1-B_{\bf q})^{2}-A_{\bf q}^{2}}
\nonumber \\
\omega_{-}({\bf q})& = & 4 J_{\rm av} S \sqrt{(1+B_{\bf q})^{2}-A_{\bf q}^{2}}
\ .
\end{eqnarray}
Note that when $J_{\parallel} = J_{\perp} = J_{z}$,
$B_{\bf q} $ is zero and thus we have two degenerate spin modes as usual.
When $J_{\parallel}, J_{\perp}$ and $J_{z}$ are
different, the two modes are no longer degenerate. 
This remains true when $ {\bf q} \rightarrow 0 $:
\begin{equation}
\label{SW13}
\omega_{+}(0) = 4  S \sqrt{ 2 J_{\rm av} (J_{\rm av} - J_{z})} \ ; \ \ \ \ \ \
\omega_{-}(0) = 0 \ .
\end{equation}
This result shows that we have only one gap in the noninteracting
spin wave picture even though the ground--state energy is anisotropic
and therefore selects [\onlinecite{CLH}] a value of $\theta$.

In Fig. 4 we plot the noninteracting spin--wave spectrum according to
Eq. (\ref{SW12}) along different directions in the Brillouin zone.
For illustrative purposes we arbitrarily chose values of the
$J$'s which correspond to much larger anisotropy than we have
for the cuprates.
An indication [\onlinecite{SPLIT}] in the spin--wave spectrum that
$\delta J \equiv J_\parallel - J_\perp$ is nonzero is the removal of
degeneracy between $\omega_+ ({\bf q})$ and $\omega_- ({\bf q})$
on the boundary of the Brillouin zone (where $q_x+q_y=\pi$).
This effect is illustrated in Fig. 4.  Even though noninteracting
spin--wave theory does not lead to two gaps at zero wave vector when
$J_\parallel \not= J_\perp$, one can obtain the second gap by
calculating the spin--wave spectrum including higher orders in $1/S$.
However, below we will estimate this in--plane gap without 
explicitly invoking spin--wave interactions.

For this purpose we study the quantum zero--point energy
(per spin) in detail.  It is given by
\begin{equation}
\label{SW14}
E_{Z} (\theta) = \frac{1}{2N} \sum_{\bf q}  
\bigg( \omega_{+} ({\bf q}) + \omega_{-} ({\bf q}) \bigg) \ .
\end{equation}
>From Eqs. (\ref{SW8}), (\ref{SW9}), and (\ref{SW12}) one can write
\begin{equation}
\label{SW15}
\omega_{\pm}( {\bf q} ) = 4 J_{\rm av} S \bigg[
( f \pm g) + ( h \pm k) \cos(2\theta)
\bigg(\frac{\delta J}{J_{\rm av}}\bigg) \bigg]^{1/2} \ ,
\end{equation}
where
\begin{eqnarray}
\label{SW16}
f &=& 1 - \frac{ J_{z} C_{+}^2}{ 4 J_{\rm av}}, \;\;\;\;\;\;\;\;\;\;\;\;\;
h = \frac{J_{z} C_{+} C_{-} }{8 J_{\rm av}} \nonumber\\
g & = & \frac{ (J_{\rm av}-J_{z})C_{+} } {2J_{\rm av}}, \;\;\;\;\;\;
k = -\frac{C_{-}}{4}  \; , \;\;\;\;\;\;   \delta J = J_{\parallel}
- J_{\perp}
\end{eqnarray}
with $C_{+},\; C_{-}$ given by
\begin{equation}
\label{SW17}
C_{\pm} = \cos(q_{x}a) \pm  \cos(q_{y}a) \ .
\end{equation}

To obtain the leading $\theta$--dependence of the mode energies,
we expand $\omega_{\pm}({\bf q} )$ up to second order in powers of 
$( \delta J/J_{\rm av})$:
\begin{equation}
\label{SW18}
\omega_\pm ({\bf q}) = 4 J_{\rm av} S (f\pm g)^{1/2} \bigg\{
1 + \frac{1}{2} \bigg(\frac{h\pm k}{ f \pm g} \bigg) \cos(2\theta)
\bigg( \frac{\delta J}{J_{\rm av}}\bigg) -
\frac{1}{8} \bigg(\frac{h\pm k}{ f \pm g} \bigg)^{2} \cos^{2}(2\theta)
\bigg( \frac{\delta J}{J_{\rm av}}\bigg)^{2} \bigg\}.
\end{equation}
By using this in Eq. (\ref{SW14}) we can obtain the leading
$\theta$--dependence of the quantum zero--point energy,
\begin{equation}
\label{SW21}
E_{Z}(\theta) =  2J_{\rm \rm av} S \Biggl[ C_{0}
+ C_{1} \cos(2\theta) {\delta J \over J_{\rm av} } - C_{2}
\cos^{2}(2\theta) \frac{ (\delta J)^{2}}{J_{\rm av}^2} \Biggr] \ ,  
\end{equation}
where the numerical constants are
\begin{eqnarray}
\label{SW20}
C_{0} &= \frac{1}{N} & \sum_{\bf q} \bigg\{
\sqrt{f+g} + \sqrt{f-g} \bigg\} \nonumber \\
C_{1} &= &\frac{1}{2N} \sum_{\bf q} \bigg\{
\frac{(h+k)}{\sqrt{f+g} } + \frac{(h-k)}{\sqrt{f-g} } 
\bigg\} = 0 + 0 = 0 \nonumber \\
C_{2} &=& \frac{1}{8N} \sum_{\bf q} \bigg\{
\frac{(h+k)^{2}}{(f+g)^{3/2} } + \frac{(h-k)^{2}}{(f-g)^{3/2} }
\bigg\} \ .
\end{eqnarray}
Note that  coefficients of odd powers of $\delta J$ vanish due to the
fact that these terms include odd power of $C_{-}$ which changes
sign under $q_{x} \leftrightarrow q_{y}$
while the other expressions are invariant under this operation.

In Fig. 5 we show $ E_{Z}(\theta)$  from Eq. (\ref{SW21}) and from 
the exact sum given in Eq. (\ref{SW14}) for
$J_{\parallel} = 1, J_{\perp} = 0.9$ 
and $J_{z} = 0.8$ for which $C_{0} = 0.44 + 0.39 = 0.83$
and $C_{2} = 2.95 \times  10^{-3} + 0.7 \times 10^{-2} 
\approx 1 \times  10^{-2}$, where the first and second numbers are
the contribution from out--of--plane  and in--plane modes,
respectively.  Note that the in--plane mode contributes almost twice
as much as the out--of--plane mode.  The agreement between the
exact and approximate results is excellent even though we have taken
$\delta J / J_{\rm av} \approx  0.1 $. Since in many real systems this 
ratio is extremely small, Eq. (\ref{SW21}) should give nearly
the exact value.  For $J_{\parallel} = J_{\perp} = J_{z} = J$ we have
\begin{equation}
C_{0} = 0.842 \ , \;\;\;\; C_{2} = 1 \times 10^{-2} \ .
\end{equation}
 
Note that the zero--point fluctuation energy favors the staggered
magnetization to point along a $[1,0]$ direction within the easy
[\onlinecite{TRANQ,BURLET}] plane.
Experiments [\onlinecite{RM}] indicate that this may be the case,
for YBa$_2$Cu$_3$O$_6$, where the dipolar energy does not
select a value of $\theta$, [\onlinecite{PRL1}]
although it is not easy to distinguish the direction of the
staggered magnetization in such systems. [\onlinecite{GEN}]
For other tetragonal cuprates, the magnetic structure in the
ground state is determined by the competition between $E_Z(\theta)$
and other anisotropies which result from inter--plane
interactions. [\onlinecite{PRL1}]

We are now ready to estimate the in--plane gap due to the
anisotropy of quantum zero point energy shown in Fig. 5.  To do
this we assume that the quantum zero--point energy is equivalent
to an effective Hamiltonian for general $S$ of the form
\begin{equation}
\label{SW22}
H_{\rm QZPE} = \sum_i
\gamma_{\rm in} S \bigg( {S_{i}^{x}}^{2} {S_{i}^{y}}^{2} / S^{4} \bigg) \ .
\end{equation}
Since we are interested in the region where ${\bf q} \approx 0$,
this effective interaction is probably adequate to approximate the
dependence of $E_{Z}(\theta)$ on $\theta_i$ even when
$\theta$ has a slow nonzero spatial variation.
Note that $H_{\rm QZPE}$ is of order $S$ because $E_{Z}(\theta)$ in
Eq. (\ref{SW14}) is of that order.  By comparing  the angular
dependence of Eq. (\ref{SW21}) and Eq. (\ref{SW22}), one obtains
\begin{equation}
\label{SW23}
\gamma_{\rm in} = 8 C_{2} \frac{(\delta J)^{2}}{J_{\rm av}}
\equiv 4J_{\rm av} \delta_{\rm in} \ .
\end{equation}
Transforming $S_{x}$ and $S_{y}$ into the local quantization axis by
using Eq. (\ref{SW4}) (with $\theta=0$) and Eqs. (\ref{SW5}--\ref{SW6}),
and keeping only the terms at order of $1/S^0=1$), we find that
\begin{equation}
\label{SW24}
H_{\rm QZPE} =   4NJ_{\rm av} \delta_{\rm in}  + 4J_{\rm av} \delta_{\rm in} 
\sum_{i} \bigg\{ a_{i}^{\dag} a_{i} + b_{i}^{\dag} b_{i} 
 -\frac{1}{2} \biggl[ a_{i}^{2} + b_{i}^{2} +
{\rm  h.c.} \biggr] \bigg\} \ .
\end{equation}
In momentum space, $H_{\rm QZPE}$ is
\begin{equation}
H_{\rm QZPE} = 4NJ_{\rm av} \delta_{\rm in} +  4J_{\rm av} \delta_{\rm in}
\sum_{\bf q} \bigg \{  a_{\bf q}^{\dag} a_{\bf q} + b_{\bf q}^{\dag}
b_{\bf q} -\frac{1}{2} \biggl[ a_{\bf q} a_{-\bf q} + b_{\bf q} b_{-\bf q}
+ {\rm h.c.} \biggr] \bigg \} \ .
\end{equation}
Hence the total Hamiltonian $H_{\rm tot} =  H_{\rm QZPE}+  H_{\rm eff}$, 
where $H_{\rm eff}$ is given in Eq. (\ref{SW7}), is
\begin{eqnarray}
H_{\rm tot}  =  E_{\rm tot} 
&+& 4J_{\rm av} S \Biggl[ \sum_{\bf q}  (1 + S^{-1} \delta_{\rm in})
\bigg ( a_{\bf q}^{\dag}a_{\bf q} + b_{\bf q}^{\dag} b_{\bf q} \bigg)
\nonumber \\ &-& \sum_{\bf q} \frac{1}{2} S^{-1} \delta_{\rm in}
\bigg (
a_{\bf q}^{\dag} a_{-\bf q}^{\dag} + a_{\bf q}a_{-\bf q}  
+ b_{\bf q}^{\dag} b_{-\bf q}^{\dag} + b_{\bf q}b_{-\bf q}  
\bigg ) \nonumber  \\ & + &
\sum_{\bf q} A_{\bf q} 
\bigg( a_{\bf q}^{\dag} b_{-\bf q}^{\dag}+a_{\bf q} b_{-\bf q} \bigg) +
\sum_{\bf q} B_{\bf q} \bigg( a_{\bf q} b_{\bf q}^{\dag}
+ a_{\bf q}^{\dag} b_{\bf q} \bigg) \Biggr] \ .
\end{eqnarray}
The spin--wave energies $\omega_{\pm} $ are given as
\begin{eqnarray}
\omega_{+}^{2} ({\bf q})  &=& \Biggl( 4J_{\rm av}S \Biggr)^2
\bigg( 1 - A_{\bf q} - B_{\bf q} \bigg)
\bigg( 1 + 2S^{-1} \delta_{\rm in} + A_{\bf q} -  B_{\bf q} \bigg)
\ , \nonumber \\
\omega_{-}^{2} ({\bf q}) &=& \Biggl( 4J_{\rm av}S \Biggr)^2
\bigg( 1 + A_{\bf q} + B_{\bf q} \bigg)
\bigg( 1 + 2 S^{-1} \delta_{\rm in} - A_{\bf q} + B_{\bf q} \bigg) \ .
\end{eqnarray}
To get the in--plane and out--of--plane gaps, we set ${\bf q} = 0$
in which case
\begin{eqnarray}
\label{GAP}
\omega_{+} (q=0) &= &4 J_{\rm av} S \sqrt{ 2 [1 - (J_{z}/J_{\rm av})]
[1 + S^{-1} \delta_{\rm in} ] } \ , \nonumber \\
\omega_{-} (q=0) & =& 4J_{\rm av} S  \sqrt{  2[1 + (J_z/J_{\rm av})]
S^{-1} \delta_{\rm in}} ,
\end{eqnarray}
where $\omega_{-}(q=0)$ is the in--plane--gap due to the
quantum zero point energy that we are looking for.
Note that $\omega \propto \sqrt S$, as was originally
found [\onlinecite{EFS}] in a similar situation where
the gap is due to quantum
zero--point effects.  Thus we see that the noninteracting result,
plotted in Fig. 4, which gives one gapless mode, needs to
be modified as we have just done.

We now give a numerical evaluation of the gaps $\omega_\pm (q=0)$.
For that purpose we approximate the result of Eq. (\ref{GAP}) as
\begin{eqnarray}
\omega_+ (q=0) &=& 4S \sqrt{ 2J_{\rm av}(J_{\rm av}-J_z) } \ ,
\nonumber \\ \omega_{-} (q=0) &=&  8J_{\rm av} S
\sqrt{ \delta_{\rm in}/S} = 8 \delta J \sqrt {(2S)C_2}
\approx 0.8 \mid J_\parallel - J_\perp \mid  \ ,
\end{eqnarray}
where we used Eq. (\ref{SW23}). To evaluate $\omega_+(q=0)$ we use
the experimental value [\onlinecite{CL},\onlinecite{325}]
$J_{\rm av}=130$ meV and take
$J_{\rm av}-J_z = (J_\perp-J_z) - (J_\perp-J_\parallel)2=30 \ \mu$eV
from Fig. 3, in which case
\begin{equation}
\omega_+(q=0) = 2S (5.9 {\rm meV}) \ .
\end{equation}
Measurements [\onlinecite {TETRA2}] show that zero--point fluctuations
reduce $2S$ to about 0.8.  Using this value, we get
$\omega_+(q=0)= 4.7$ meV, which compares favorably with the
experimental [\onlinecite{CL}] value of 5 meV.  From the data shown
in Fig. 3 we see that the in--plane gap, $\omega_-(q=0)$ should
be about 25 $\mu$eV.  It would be interesting to observe this via
an infra--red absorption experiment.  Because the theoretical estimate
of the frequency range is uncertain, it might be useful to locate the
mode at high magnetic field and follow it back to zero applied field.

\medskip
\begin{center}
\section{THE LOWEST SYMMETRY MODEL}
\end{center}
\medskip

In Sec. IV we showed that for the model of Eq. (\ref{HAM})
in tetragonal symmetry the anisotropy
vanishes in the absence of Coulomb exchange.
This was due to the fact that we have only
hopping between orbitals of the same symmetry. 
However the theorem breaks down when we have nonzero
hopping between orbitals of different symmetry. Thus in this
section we consider a system with lower symmetry to show that
we can have anisotropy without Coulomb exchange.

Here we again consider the effective spin Hamiltonian
for two copper ions, but now we do not assume any particular
symmetry.  Thus the orbitals localized on the two Cu ions
which diagonalize ${\cal H}_x$ are no longer the same and
will be some arbitrary
linear combinations of $  x^2-y^2, 3z^2-r^2 , xy,yz, zx$,
respectively.  We write these orbitals as
\begin{equation}
\tilde{d}_{i,\alpha}^{\dag} =  \sum_{\beta} R^{-1}_{\alpha,\beta}
(i) d_{\beta}^{\dag} ,
\end{equation}
where  $R_{\alpha,\beta} (i)$ is the matrix element
of the orthogonal matrix which gives the new states in terms
of the undistorted d--orbitals for the $i$th copper ion.

Within these orbitals the hopping matrix elements now are
\begin{equation}
\tilde{t}_{i\alpha,j\beta} =  \sum_{\gamma,\eta}
R_{\alpha,\gamma}(i) R_{\beta,\eta}(j) t_{\gamma,\eta} ,
\end{equation}
where $t_{\gamma,\eta}$ is the usual overlap integral
between the undistorted d--orbitals, listed above.
Similarly, the matrix elements of angular momentum in this
new basis are
\begin{equation}
{\bf\tilde{L}}_{i\alpha,i\beta} =  \sum_{\gamma,\eta}
R_{\alpha,\gamma}(i) R_{\beta,\eta}(i) {\bf L}_{\gamma,\eta} \ .
\end{equation}

\medskip
\begin{center}
\subsection{Numerical Study}
\end{center}
\medskip

We now present our numerical results for the effective spin
Hamiltonian when we use these new hopping and angular momentum
matrices for two arbitrarily chosen matrices ${\bf R}(i)$ and
${\bf R}(j)$.  In Fig. 6 we show the anisotropy (energy
differences between triplet states) for two different situations;
(1) the on--site Coulomb repulsive interactions depend on the
orbitals ($U_{\alpha,\beta}\neq U$) (2) Orbital--independent
(constant)
Coulomb interactions: $U_{\alpha,\beta}=U$.  As in the tetragonal
case (but now with no Coulomb exchange interaction), we have
full anisotropy for both cases.  By fitting the numerical
results shown in Fig. 6 we showed that anisotropy is proportional
to $t^{2}\lambda^{2}$ for nonconstant ${\bf U}$ and to
$t^{6}\lambda^{2}$ for constant ${\bf U}$.  In the next section
we give an analytic proof that for constant ${\bf U}$ the
anisotropy vanishes up to order ${\bf t}^4$.

\medskip
\begin{center}
\subsection{Order $\tilde t^4$ Results for Constant $U$ and ${\bf K}=0$}
\end{center}
\medskip

With only $nn$ hopping on the square lattice, there are no
contributions at order $\tilde{t}^{3}$. The calculations at order
$\tilde{t}^{4}$ will generate two types of
contributions:[\onlinecite{HUBEX}] one a four--spin interaction,
the other two--spin interactions either between nearest neighbors
or between next--nearest neighbors.
The first type is generated when a hole hops around a closed loop,
i. e., from site 1 to 2, the hole which had been earlier on site
2 hops to 3 and so on, until the hole from site 4 hops to 1.
The second type of interaction is generated both by closed
loop processes and by various arrangements of four hops involving
two or three sites.

In this paper we are mainly concerned with the evaluation of the
$nn$ pair exchange interactions.  In particular we have concentrated
mostly on the anisotropy of these interactins due to spin--orbit
interactions.  To study the contributions to this anisotropy
from repeated hopping within a single bond to order
$\tilde{t}^{4}$, we use Eq. (\ref{PSI2}) of Appendix C and apply to it
two more hopping terms, $T_{ij}{\cal H}_0^{-1}T_{ji}{\cal H}_0^{-1}$,
ending at a ground state. After some algebra we obtain
\begin{eqnarray}
\label{III21}
\Delta{\cal H}(i,j) &=& {\rm Tr}
\Bigl\{\Bigl(\vec{\sigma}\cdot {\bf S}_{j}\Bigr)
\Bigl(\sum_{ab}\tilde{t}_{0b}^{ji}
\tilde{t}_{ba}^{ij}\tilde{t}_{a0}^{ji}\Bigr)
\Bigl(\vec{\sigma}\cdot {\bf S}_{i}\Bigr)
\Bigl(\tilde{t}_{00}^{ij}\Bigr)\Bigr\}
\nonumber\\
&\times &\frac{2}{U_{0}}\Bigl[\frac{1}{E_{ja}(U_{0}+E_{ib})}
+\frac{1}{E_{ib} (U_{0}+E_{ja})}
-\frac{1}{(U_{0}+E_{ja})(U_{0}+E_{ib})}\Bigr] 
+ \Biggl( i \leftrightarrow j \Biggr)  \ .
\end{eqnarray}

Combining Eqs. (\ref{III7}) and (\ref{III21}), we end up with Eq.
(\ref{III9}), in which both $(A_{1}+i {\bf B}_{1}\cdot\vec{\sigma})$
and $(A_{2}+i {\bf B}_{2}\cdot\vec{\sigma})$ are of the form
$\tilde{t}_{00}^{ij}+O(\tilde{t}^{3})$. Since at order $\tilde{t}^{2}$
we had ${\bf D}_{2}=0$, Eq. (\ref{III10}) now yields
${\bf D}_{2}=O(\tilde{t}^{4})$, and thus the energy splitting of the
triplet due to ${\bf D}_{2}$ is of order [cf. Eqs.  (\ref{III11})
and (\ref{III15})]
${\bf D}_{2}^{2}/A_{1}A_{2}=O(\tilde{t}^{6})$, irrespective of the
details of Eq. (\ref{III21}). Thus, the $nn$ magnetic exchange
interaction becomes anisotropic only at order $\tilde{t}^{6}$,
and this is correct to all orders in the spin--orbit coupling
$\lambda $ and for all lattice symmetries. This result is
indeed confirmed by our single--bond numerical diagonalization,
as we showed in Fig. 6.

We end this discussion with two comments. First, note that the
separation of $(A_{i}+i {\bf B}_{i}\cdot\vec{\sigma})$ into a sum
of terms of orders $\tilde{t}$ and $\tilde{t}^{3}$ was only
possible because the sums over $a$ and $b$ in Eq.  (\ref{III21})
all appeared within one matrix [which appears between
$(\vec{\sigma}\cdot {\bf S}_{j})$ and
$(\vec{\sigma}\cdot {\bf S}_{i})$]. This would not have been
possible if we had contributions of the kind
$\sum_{a} {\rm Tr} \{(\vec{\sigma}\cdot {\bf S}_{j})
T_{1}^{a}(\vec{\sigma} \cdot {\bf S}_{i})T_{2}^{a}\}$,
representing interference between different hopping paths.
Such contributions arise at order $\tilde{t}^{6}$, and generate
further anisotropy. [Without them, the analysis of Appendix D
indicates that the energy levels would be two singlets and a
doublet].  Second, note that the symmetry contained in Eq.
(\ref{III8}) would persist to all orders,
had we ignored excited states, allowing only $\tilde{t}_{00}^{ij}$. 

\medskip
\begin{center}
\section{DISCUSSION AND CONCLUSIONS}
\end{center}
\medskip

\subsection{Discussion}

It is clear that the role of spin--orbit interactions in causing
anisotropy in the exchange interaction is an interesting and
subtle one.  In particular, there has been much controversy concerning
the way the exchange interaction $J_{\mu\nu}$ depends on the
crystal symmetry and under what conditions one expects to find a
gap in the spin--wave spectrum.  For a long time after Moriya's
seminal paper it was thought that one could neglect ${\bf M}$
in Eq. (\ref{JDMEQ}) and that the spin Hamiltonian for a single bond
would be anisotropic if the Dzyaloshinskii vector ${\bf D}$ were
nonvanishing.  It was then observed by Kaplan [\onlinecite{TAK}] and by
SEA, [\onlinecite{SEA}] that although ${\bf M}$ is of order $\lambda^2$
and ${\bf D}$ is of order $\lambda$, one must nevertheless keep both
terms when discussing the anisotropy or the gap in the spin--wave spectrum.
Two other conclusions of these authors were 1) the single--bond spin
Hamiltonian ${\cal H}_S(i,j)$ was rotationally invariant and 2) the
overall anisotropy of the Cu--O plane resulted from a frustration
between bonds with different values of ${\bf D}$.  In view of
the results of the present paper we are in a position to state clearly
the conditions under which the first conclusion is valid.  In
particular, our results show that rotational invariance
(when the Coulomb exchange ${\bf K}$ is zero) of the single--bond
spin Hamiltonian to all orders in $t$ is only to be expected when
hopping between excited states is ignored, as the SEA argument
does implicitly.  We remark that since the spin--orbit interaction
involves coupling to excited orbital states, it is only
nonnegligible when the energies of the excited states involved are
finite.  This being the case, strictly speaking, it is not
totally consistent to neglect hopping between such states, especially
since the associated hopping matrix elements are comparable to those
involving hopping to or from the orbital ground state.  Nonetheless,
as we have seen, the departures from the SEA rotational invariance
theorem (due to hopping between excited states) are numerically
quite small in most cases of physical interest.  In fact, for the case
of constant ${\bf U}$ considered by SEA, the deviations from
rotational invariance only enter at order $t^6$.

In Moriya's original work to order $t^2$ it was correct to ignore
hopping between excited states because he considered the case when
${\bf U}$ was a constant.  In this case, as our results in Sec. VI
show, rotational invariance only breaks down at order $t^6$,
because one has to go to that high order for hopping between excited
states to come into play.  When ${\bf U}$ is nonconstant, hopping
between excited states leads to anisotropy in ${\cal H}_S(i,j)$ at
order $t^2$, as our results in Sec. VI demonstrate.  These results
thus represent a generalization of those by Moriya and by most of
the literature which followed him and assumed constant ${\bf U}$.

>From this discussion one might now conclude that spin--orbit
interactions would lead to anisotropy for the Cu--O planes as long
as one includes hopping between excited Cu states.  However, the
theorem of Sec. IV shows that for the special case when the Cu sites
have tetragonal symmetry, the generic model of Eq. (\ref{HAM})
with only ${\bf t}$, $\lambda$,  and ${\bf U}$ nonzero does not yield
nonzero anisotropy.  The same result also applies to the ``real''
model including oxygen ions.  This theorem explains why most previous
calculations give no anisotropy for tetragonal site symmetry and it
emphasizes the importance of including Coulomb exchange terms,
${\bf K}$.  It is then clear why the exchange anisotropy is so
small, especially (as noted by BS) when compared to the anisotropy
in the $g$ tensor.  We thus find that for each bond the exchange
interaction has biaxial anisotropy ($J_\parallel$, $J_\perp$, and
$J_z$ are all different), where the anisotropy in $J$ is of order
$t_{\rm Cu-Cu}^2\lambda^2K$, or more correctly, it is of order
$t_{\rm Cu-Cu}^2\lambda^2B$ or $t_{\rm Cu-Cu}^2\lambda^2C$, where
$B$ and $C$ are the Racah parameters which represent deviations from
the simple constant ${\bf U}$ Hartree term.

Even though the single--bond exchange has biaxial anisotropy,
the classical ground state energy, because it is averaged over
bonds along [1,0,0] and [0,1,0], does not select an orientation
of the staggered magnetization within the easy plane.  As we have
shown, the anisotropy within the easy plane results from quantum
zero--point fluctuations.  In summary, a complete discussion of the
anisotropy of the Cu--O planes requires an interesting study of
several novel symmetries and the way they are broken by fluctuations.

\subsection{Conclusions}

We may summarize our conclusions as follows.

(1) For tetragonal site symmetry, with only Hartree--like direct
Coulomb terms, the effective spin Hamiltonian is isotropic at any order
in the parameters $t$ and $\lambda$.  Inclusion of Coulomb exchange
breaks this degeneracy at order $t^{2}_{\rm Cu-Cu}\lambda^{2} K$ for
our generic model and at order $t^{4}_{\rm Cu-O}\lambda^{2} K$ for
the cuprate system with an oxygen ion between the copper ions.

(2)  Since the easy--plane anisotropy (observed via the
``out--of--plane'' spin--wave gap at zero wave vector) has
comparable magnitudes for many orthorhombic and tetragonal cuprates,
it can not depend significantly on the orthorhombic distortion.
Our result, Eq. (\ref{JMU}), yields a biaxial anisotropy in the
exchange interaction of order $t_{\rm Cu-O}^4\lambda^2K$ which
can explain the observed [\onlinecite{CL}] out--of--plane spin--wave gap.

(3) In the tetragonal case, with the exchange interactions having
biaxial anisotropy given by Eq.
(\ref{JMU}), the ground state does not depend on the orientation of the
staggered magnetization within the easy plane.  [As shown in Ref.
\onlinecite{PRL1}, this remains true when dipolar interactions are
included.]  However, as we show,[\onlinecite{PRL1}]
quantum zero--point fluctuations
cause an anisotropy within the easy plane which leads to ordering
of the spins along the (1,0) axes, as indeed was claimed to be
observed in YBa$_2$Cu$_3$O$_6$.[\onlinecite{RM}]
A rough estimate yields a resulting ``in--plane'' spin--wave gap
of about 25 $\mu$eV.  An experimental measurement of this gap
would be very desirable.

(4)  In real crystals, the three--dimensional ordering of the
spins is determined by a competition between the anisotropies
treated in the present paper and several other mechanisms,
such as interplane hopping and interactions, as discussed
recently in Refs. \onlinecite{PRL1}, \onlinecite{JTLCOM},
and \onlinecite{REPLY}.  These may also affect the estimate
of the in--plane gap given in conclusion (3).

(5) For sufficiently low symmetry and  without exchange interactions,
the rotational invariance of the single--bond Hamiltonian is broken
at order $t^{6}\lambda^{2}$ for constant ${\bf U}$.
For arbitrary $U_{\alpha,\beta}$ and sufficiently low
symmetry, the single--bond Hamiltonian is not rotationally
invariant even at order $t^{2}\lambda^{2}$.

(6)  We have given results for arbitrary symmetry for the
effective spin Hamiltonian at order ${\bf t}^2$ including,
for the first time, the effects of realistic Coulomb
interactions.  These expressions are valid for the
orthorhombic phases of La$_2$CuO$_4$.

(7)  In view of the controversies in the literature concerning
the results which include spin--orbit interactions we have implemented
several checks of our perturbative results.  First of all,
we compared the results given in Eqs. (\ref{III26}) and
(\ref{III36}) with expressions obtained by treating both the
hopping {\it and} the spin--orbit interactions as perturbations.
In addition, we subjected our analytic results for the tetragonal
symmetry case to numerical
verification as follows.  We diagonalized exactly the Hamiltonian within
the basis of two holes on either a Cu--Cu cluster or a Cu--O--Cu cluster.
Then we compared the splittings of the ground state manifold (in this
case, the lowest four states) with those predicted on the basis of our
analytic evaluation of the perturbative contributions to the spin
Hamiltonian.  This comparison (see Fig. 3) was made with small
enough values of the perturbative parameters that we can easily
check how the results depend on the parameters.

ACKNOWLEDGEMENTS:
We acknowledge stimulating discussions with R. J. Birgeneau and
M. A. Kastner.  Work at the University of Pennsylvania was
partly supported by the National Science Foundation
MRL Program under Grant No. DMR--91--22784 and that at Tel Aviv by
the U. S.--Israel Binational Science Foundation and the USIEF.

\newpage
\appendix

\section{EXCHANGE AND 4 LOWEST ENERGY LEVELS}

In this Appendix, we explain what we mean by anisotropy and how
we study it by identifying the eigenvalues of ${\cal H}$ obtained
numerically with those of a general spin
Hamiltonian such as that given in Eq. (\ref{HJMN}).

The most general effective spin Hamiltonian for a single bond can
be written as
\begin{equation}
H_{\rm eff} = E_{0} + {\bf S}_{1} \left(
\begin{array}{ccc}
J_{11}&J_{12}&J_{13} \\
J_{21}&J_{22}&J_{23} \\
J_{31}&J_{32}&J_{33} \\
\end{array} \right)
{\bf S}_{2}
\end{equation}
where the matrix ${\bf J}$ is a $3\times 3$ real matrix.  There exist two
transformations ${\bf R}_{1}$ and ${\bf R}_{2}$ which transform ${\bf J}$
into diagonal form if we rotate the spins:
${\bf S}_{1} = {\bf R}_{1} {\bf S}^\prime_1$ and
${\bf S}_{2} = {\bf R}_{2} {\bf S}^\prime_{2}$.
To obtain ${\bf R}_1$ and ${\bf R}_2$ we first obtain the orthogonal
matrix ${\bf O}$ which diagonalizes ${\bf J}{\bf J}^t$, where
the subscript ``$t$'' indicates transpose:
\begin{equation}
{\bf O}^t ( {\bf J } {\bf J}^t ) {\bf O} = \tilde {\bf J}^2 \ ,
\end{equation}
where $\tilde {\bf J}^2$ is a diagonal matrix with nonegative entries.
For simplicity we assume that all its entries are actually positive.
Then we define $\tilde {\bf J}$ and $\tilde {\bf J}^{-1}$ to be the
corresponding diagonal matrices with positive entries.

Then we set
\begin{equation}
{\bf R}_{1}  =  {\bf O} \ , \ \ \ \ \
{\bf R}_{2}  =  {\bf J}^{t} {\bf O} \tilde {\bf J}^{-1} \sigma \ ,
\end{equation}
where $\sigma = {\rm Det} {\bf J} / | {\rm Det} {\bf J} |$.
In terms of the transformed spins the Hamiltonian is
\begin{equation}
H_{\rm eff}^\prime = E_{0}' + \sigma {\bf S}_{1}^\prime \left(
\begin{array}{ccc}
\tilde J_x&0&0 \\
0&\tilde J_y&0 \\
0&0&\tilde J_z \\
\end{array} \right)
{\bf S}_{2}^\prime \ ,
\end{equation}
where all the $\tilde J_\alpha$'s are positive.
Obviously, a further rotation could be made to change the sign of
any two components of ${\bf S}_1^\prime$ (or ${\bf S}_2^\prime$).
So the energy level scheme must be invariant under such a change of 
signs.  It also has to be invariant under permutations of the
$\tilde J_\alpha$.  We find the four energy levels to be
\begin{eqnarray}
\label{EIGVAL}
\lambda_S & = & E_0' - \sigma ( \tilde J_x +  \tilde J_y +  \tilde J_z
)/4 \nonumber \\ \lambda_\alpha & = & E_0' + 
\sigma ( \tilde J_x +  \tilde J_y +  \tilde J_z -  2 \tilde J_\alpha )/4
\ , \ \ \ \alpha = x, \ y, \ z .
\end{eqnarray}
This set of energies has the proper invariance under change of
the signs of any two $\tilde J_\alpha$'s.  For the case of arbitrary low
symmetry, we did not try to identify the principal axes, but
tabulated anisotropies, defined to be $\lambda_1 - \lambda_2$ and
$\lambda_1 - \lambda_3$, where $\lambda_1 < \lambda_2 < \lambda_3$
are the three eigenvalues of the set $\{ \lambda_\alpha \}$.

For D$_{2h}$ bond symmetry, ${\bf J}$ must be diagonal, so that
\begin{equation}
\label{D2H}
{\cal H}_{\rm eff} = E_0 + \sum_\alpha J_\alpha S_{1,\alpha}
S_{2,\alpha} \ .
\end{equation}
In this case the eigenvalues are given by Eq. (\ref{EIGVAL}) with
$\sigma=1$ and the $J_\alpha$ have whatever signs they
have in Eq. (\ref{D2H}).
The identification of the $J$'s from the set of eigenvalues of
Eq. (\ref{EIGVAL}) is not unique, because either permuting the
$J$'s or changing two of their signs leaves the set of eigenvalues
invariant.  Thus, identification
of the $J$'s with coordinate directions requires consideration of
the eigenfunctions.  For this purpose we write them explicitly:
\begin{eqnarray}
\psi_S & = & | \uparrow \downarrow \ - \downarrow \uparrow \rangle / \sqrt 2
\ , \ \ \ \ \ \ 
\psi_z = | \uparrow \downarrow \ + \downarrow \uparrow \rangle / \sqrt 2
\ , \nonumber \\
\psi_x & = & | \uparrow \uparrow \ - \downarrow \downarrow \rangle / \sqrt 2
\ , \ \ \ \ \ \ 
\psi_y = | \uparrow \uparrow \ + \downarrow \downarrow \rangle / \sqrt 2
\ .
\end{eqnarray}
Then the eigenfunctions are distinguished by their expectation values:
\begin{eqnarray}
\psi_S : &\ & \ \ \langle S_{1,z} S_{2,z} \rangle = \langle S_{1,x} S_{2,x}
\rangle = \langle S_{1,y} S_{2,y} \rangle = - 1/4 \nonumber \\
\psi_z : &\ & \ \ - \langle S_{1,z} S_{2,z} \rangle = \langle S_{1,x} S_{2,x}
\rangle = \langle S_{1,y} S_{2,y} \rangle = 1/4 \ ,
\end{eqnarray}
and so forth for the other $\psi_\alpha$.  Having identified which
wave functions (coming out of the diagonalization of the 190 $\times$ 190
matrix) are which, one can easily deduce the values of the $J_\alpha$.
For instance
\begin{equation}
J_z = - \lambda_S - \lambda_z + \lambda_x + \lambda_y \ .
\end{equation}

\section{COULOMB INTERACTION PARAMETERS IN TERMS OF THE RACAH
COEFFICIENTS}

Here we list the Coulomb interaction parameters for the tetragonal
symmetry crystal field states for a d$^8$ configuration
in terms of the Racah parameters. [\onlinecite{JSG}]
Here $\Delta U_{\alpha \beta} = -2K_{\alpha \beta}$,
where $U_{\alpha\alpha'}=U_0+\Delta U_{\alpha \alpha '}$, with
$U_0=A+4B+3C$.  In terms of the triplet and singlet energies
given in Ref. [\onlinecite{ESKES}] one has
${\bf U}^s={\bf U}+{\bf K}$ and ${\bf U}^t={\bf U}-{\bf K}$.

\begin{equation}
{\bf K} =
\begin{array}{c|ccccc}
& d_{x^2-y^2}& d_{3z^2-r^{2}} &  d_{xy} & d_{yz} & d_{zx} \\
\hline
d_{x^2-y^2} & 0  &4B+C      & C & 3B+C    &3B+C  \\
d_{3z^2-r^{2}} & 4B+C   & 0 &4B+C & B +C & B +C   \\
d_{xy} & C & 4B +C &  0 & 3B +C & 3B +C  \\
d_{yz} & 3B +C  & B +C &3B +C & 0 & 3B +C \\
d_{zx} & 3B +C  & B +C&3B +C & 3B +C & 0
\\
\end{array}
\end{equation}
In the numerical calculations we used (see Table I) $A=7.00$, $B=0.15$,
and $C=0.58$, so that $U_0=9.34$, all in eV.

\section{MATRIX ELEMENTS NEEDED FOR PERTURBATION THEORY}

Here we record some of the matrix elements needed to implement
perturbation theory.  If $|\psi_0 \rangle$ represents any state in the
ground state manifold (having one hole per site), then we may write
\begin{eqnarray}
\label{PSI1}
\mid\psi_{1} \rangle & \equiv & {1 \over {\cal H}_0 } T_{ji}
\mid\psi_{0} \rangle \equiv \sum_{\sigma_1 \sigma _2}
{1 \over {\cal H}_0 } T_{ji} c_{i 0 \sigma_1}^{\dag}
c_{j0\sigma_2}^{\dag} c_{j 0 \sigma_2} c_{i0\sigma_1}
\mid\psi_{0} \rangle \nonumber \\ & = &
\sum_{\sigma _1 \sigma_2 \sigma_3 b}
{ \Bigl( \tilde{t}_{b0}^{ji}\Bigr)_{\sigma_3 \sigma_1} \over
E_{jb} + U_0 } c_{jb\sigma_3}^{\dagger} c_{j0\sigma_2}^\dagger
c_{j0\sigma_2} c_{i0\sigma_1} \mid\psi_{0} \rangle \ ,
\end{eqnarray}
where ${\cal H}_0$ and $T_{ij}$ are defined in Eqs. (\ref{UNPERT})
and (\ref{I16b}), respectively, and where we set the ground state
energy of ${\cal H}_0$ to zero.
For results to second and fourth order in ${\bf t}$
with no Coulombic perturbations we need to generate the 
following matrix element:
\begin{eqnarray}
\label{PSI2}
\mid \psi_2 \rangle & \equiv &  T_{ij} \mid \psi_1 \rangle \nonumber \\ 
&=& \sum_{\stackrel{\sigma_1 \sigma_2 \sigma_3 \sigma_4}{bc}}
\Biggl[
\Bigl( \tilde{t}_{b0}^{ji} \Bigr)_{\sigma_3 \sigma_1}
\Bigl( \tilde{t}_{cb}^{ij} \Bigr)_{\sigma_4 \sigma_3}
c_{ic\sigma_4}^\dagger c_{j0\sigma_2}^\dagger
- \Bigl( \tilde{t}_{b0}^{ji} \Bigr)_{\sigma_3 \sigma_1}
\Bigl( \tilde{t}_{c0}^{ij} \Bigr)_{\sigma_4 \sigma_2}
c_{ic\sigma_4}^\dagger c_{jb\sigma_3}^\dagger \Biggr]
\nonumber \\ && \ \ \times \Biggl( E_{jb} + U_0 \Biggr)^{-1}
c_{j0 \sigma_2 } c_{i0 \sigma_1 } \mid \psi_0 \rangle \ .
\end{eqnarray}

We also need 
\begin{eqnarray}
\label{PSI3}
&\mid &\psi_{3} \rangle \equiv  {1 \over {\cal H}_0 }
\Delta{\cal H}_{c} {1 \over {\cal H}_0 }
T_{ji} \mid\psi_{0} \rangle \nonumber \\ 
&= & \sum_{\stackrel {\sigma_1 \sigma_2 \sigma_3} {s s' a_1 a_2 b} }
\Biggl\{ { \Bigl( \tilde{t}_{b0}^{ji}\Bigr)_{\sigma_3 \sigma_1} \over
( E_{jb} + U_0 )( E_{ja_1} + E_{j a_2} + U_0 ) } \Biggl(
\Delta \tilde U_{ss'\sigma_2 \sigma_3}(j;a_1 a_2 0b) + 
\tilde K_{ss'\sigma_2 \sigma_3}(j;a_1 a_2 0b) \nonumber \\ && \ \
- \Delta \tilde U_{ss'\sigma_3 \sigma_2}(j;a_1 a_2 b0)  -
\tilde K_{ss'\sigma_3 \sigma_2}(j;a_1 a_2 b0) \Biggr)
c_{ja_1 s}^{\dagger} c_{j a_2 s'}^\dagger 
\Biggr\} c_{j0\sigma_2} c_{i0\sigma_1} \mid \psi_0 \rangle \ .
\end{eqnarray}
Finally, to get the energy at order ${\bf t}^2 \Delta {\cal H}_c$ we
need
\begin{eqnarray}
\label{PSI4}
&& \mid \psi_4 \rangle \equiv T_{ij} \mid \psi_3 \rangle \nonumber 
\\ &&
= \sum_{\stackrel {\sigma_1 \sigma_2 \sigma_3 \sigma_4} {s s' a_1 a_2 b}}
\Biggl\{ { \Bigl( \tilde{t}_{b0}^{ji}\Bigr)_{\sigma_3 \sigma_1} \over
( E_{jb} + U_0 )( E_{ja_1} + E_{j a_2} +U_0 ) }
\nonumber \\ && \Biggl(
\Delta \tilde U_{ss'\sigma_2 \sigma_3}(j;a_1 a_2 0b) + 
\tilde K_{ss'\sigma_2 \sigma_3}(j;a_1 a_2 0b) -
\Delta \tilde U_{ss'\sigma_3 \sigma_2}(j;a_1 a_2 b0)  -
\tilde K_{ss'\sigma_3 \sigma_2}(j;a_1 a_2 b0) \Biggr)
\nonumber \\ &&
\Biggl( \Bigl( \tilde{t}_{0a_1}^{ij} \Bigr)_{\sigma_4 s}
c_{i0\sigma_4}^\dagger c_{j0s'}^\dagger \delta_{a_2,0} -
\Bigl( \tilde{t}_{0a_2}^{ij} \Bigr)_{\sigma_4 s' }
c_{i0\sigma_4}^\dagger c_{j0s}^\dagger \delta_{a_1,0}
\Biggr) \Biggr\} c_{j0\sigma_2} c_{i0\sigma_1} \mid \psi_0 \rangle \ .
\end{eqnarray}
In order to make sure this matrix element connects to the
ground state, we had to insert the factors $\delta_{a_2,0}$ and
$\delta_{a_1,0}$.

\section{SYMMETRY OF THE MAGNETIC HAMILTONIAN}

In this Appendix we analyze the eigenvalue spectrum of a system
of two spins $1/2$ with coupling which is arbitrary except that,
for simplicity, we consider the isotropic interaction to be
dominant.  In the presence of antisymmetric exchange interactions
one can always put the Hamiltonian into the following canonical
form:
\begin{equation}
\label{DMH1}
{\cal H}(i,j) = \alpha' {\bf S}(i) \cdot {\bf S}(j) + \beta \hat n \cdot
( {\bf S}(i) \times {\bf S}(j) ) + \gamma' {\bf S}(i) \cdot {\bf M}
\cdot {\bf S}(j) \ ,
\end{equation}
where $\hat n$ is a unit vector specifying the orientation of the
Dzyaloshinskii vector and ${\bf M}$ is a symmetric 
matrix.  Here we will show that the eigenvalue spectrum of
this Hamiltonian consists of a singlet and a triplet, if
and only if the matrix ${\bf M}$ is such that ${\cal H}(i,j)$
can written in the form
\begin{equation}
\label{DMH2}
{\cal H}(i,j) = \alpha {\bf S}(i) \cdot {\bf S}(j) + \beta \hat n \cdot
( {\bf S}(i) \times {\bf S}(j) ) + \gamma \Bigl( {\bf S}(i) \cdot
\hat n \Bigr) \Bigl( \hat n \cdot {\bf S}(j) \Bigr) \ ,
\end{equation}
where the coefficients obey the relation
\begin{equation}
\label{CONDIT}
\gamma = -\alpha + \alpha \mid 1 + (\beta /\alpha )^2 \mid^{1/2} \ ,
\end{equation}
for finite $\alpha/\beta$.  (The reason for phrasing the condition
in terms of Eq. (\ref{DMH2}) rather than Eq. (\ref{DMH1})
is that the former, unlike the latter, is a unique representation.)
In this case, as we shall see, the
spins can be rotated (about the same axis, but through opposite angles)
so that in terms of
the rotated spins the Hamiltonian looks isotropic.  This result
shows that for this relation between the parameters the Hamiltonian
is rotationally invariant, even if it is not isotropic.
(By isotropic, we mean $\beta=\gamma=0$.)

As seen in the text, many of the perturbative results have the form
\begin{equation}
{\cal H}(i,j)=\frac{2}{U_{0}}{\rm Tr} \Bigl\{\Bigl(A_{1}+
i{\bf B}_{1}\cdot\vec{\sigma}\Bigr)
\Bigl({\bf S}(j)\cdot\vec{\sigma}\Bigr)
\Bigl(A_{2}+i{\bf B}_{2}\cdot\vec{\sigma}\Bigr)
\Bigl({\bf S}(i) \cdot\vec{\sigma}\Bigr)\Bigr\}.\label{III9}
\end{equation}
Defining the vectors ${\bf D}$ and ${\bf D}_{2}$,
\begin{equation}
{\bf D}=A_{1}{\bf B}_{2}-A_{2}{\bf B}_{1},\ \
{\bf D}_{2}=A_{1}{\bf B}_{2}+A_{2}{\bf B}_{1},\label{III10}
\end{equation}
this becomes
\begin{eqnarray}
{\cal H}(i,j)&=&\frac{2}{U_{0}}A_{1}A_{2}{\rm Tr} \Bigl\{
\Bigl(1-i\frac{{\bf D}-{\bf D}_{2}}
{2A_{1}A_{2}}\cdot\vec{\sigma}\Bigr)
\Bigl({\bf S}(j) \cdot\vec{\sigma}\Bigr)
\Bigl(1+i\frac{{\bf D}
+{\bf D}_{2}}{2A_{1}A_{2}}\cdot\vec{\sigma}\Bigr)
\Bigl({\bf S}(i) \cdot\vec{\sigma}\Bigr)\Bigr\}\nonumber\\
&=&\frac{4A_{1}A_{2}}{U_{0}}\Biggl\{
\Bigl(1+\frac{D_{2}^{2}-D^{2}}{(2A_{1}A_{2})
^{2}}\Bigr){\bf S}(j) \cdot {\bf S}(i)+\frac{1}{A_{1}A_{2}}
{\bf D} \cdot{\bf S}(j) \times {\bf S}(i) \nonumber\\
&+&\frac{1}{2(A_{1}A_{2})^{2}}\Biggl[ \Bigl( {\bf S}(j)\cdot
{\bf D}\Bigr) \Bigl( {\bf S}(i) \cdot{\bf D} \Bigr)
- \Bigl( {\bf S}(j) \cdot {\bf D}_{2} \Bigr) \Bigl(
{\bf S}(i) \cdot {\bf D}_{2} \Bigr) \Biggr] \Biggr\} .\label{III11}
\end{eqnarray}
This is clearly of the general form (\ref{JDMEQ}), with the
Dzyaloshinskii vector $2{\bf D}/U_{0}$ and the symmetric
anisotropy matrix $2({\bf D}\otimes{\bf D}
-{\bf D}_{2}\otimes {\bf D}_{2})/( U_{0}A_{1}A_{2} )$.
The most general form for ${\bf M}$ would involve
introducing a third linearly independent vector ${\bf D}_3$.

We now show that the eigenvalues of the Hamiltonian of
Eq. (\ref{III11}) are a singlet and a triplet
if and only if ${\bf D}_{2}$ vanishes.
To see this we study its eigenvalue equation, which, after some
algebra, can be cast into the form

\begin{equation}
\label{III15}
(\lambda -x)^{2} \Bigl[ (\lambda -x)(\lambda +3x)
-4\mid {\bf V}_{2}\mid ^{2}\Bigr]=0\ ,
\end{equation}
where
\begin{equation}
x=\frac{4}{U_{0}}A_{1}A_{2}\Bigl(1+\frac{D^{2}-D_{2}^{2}}
{(2A_{1}A_{2})^{2}}\Bigr) ,\ \ \ \
{\bf V}_{2}=\frac{4}{U_{0}}\Bigl({\bf D}_{2}-\frac{1}{2A_{1}A_{2}}
{\bf D}\times{\bf D}_{2}\Bigr) \ .
\end{equation}
It is clear that a triplet occurs if and only if ${\bf V}_2=0$,
which, in turn, happens if and only if ${\bf D}_2 =0$. Q. E. D.
Furthermore, we see that in the presence of nonzero ${\bf D}_2$,
the triplet is split into a doublet and a singlet.  To
remove all degeneracy it is necessary to introduce a
third vector ${\bf D}_3$.

We make some further remarks about the case when ${\bf D}_2 =0$.
One can easily verify that the conditions of Eq. (\ref{CONDIT}) and
(\ref{DMH2}) are
equivalent to requiring that ${\bf D}_2$ in Eq. (\ref{III11})
vanish.  We further show now that when ${\bf D}_2=0$, the
Hamiltonian is rotationally invariant.  For this purpose note that
\begin{equation}
1+i\frac{{\bf D}}{2A_{1}A_{2}}\cdot\vec{\sigma}=\frac{1}{\cos\theta}
e^{i\theta \hat{\bf d}\cdot\vec{\sigma}},
\ \ \hat{\bf d}=\frac{{\bf D}}{\mid D\mid }
,\ \ \tan\theta =\frac{\mid D\mid }{2A_{1}A_{2}}.\label{III12}
\end{equation}
The Hamiltonian (\ref{III11}) with ${\bf D}_{2}=0$ then becomes 
\begin{eqnarray}
{\cal H}(i,j)&=&\frac{2}{U_{0}}A_{1}A_{2}\Bigl(1+\frac{D^{2}}
{(2A_{1}A_{2})^{2}}\Bigr){\rm Tr} \Bigl\{
\Bigl(\vec{\sigma}\cdot{\bf S}'(j)\Bigr)
\Bigl(\vec{\sigma}\cdot{\bf S}'(i)\Bigr)\Bigr\}\nonumber\\
&=&\frac{4}{U_{0}}A_{1}A_{2}\Bigl(
1+\frac{D^{2}}{(2A_{1}A_{2})^{2}}\Bigr)
{\bf S}'(i) \cdot{\bf S}'(j) \ ,\label{III13}
\end{eqnarray}
in terms of rotated variables (equivalent to those of SEA
[\onlinecite{SEA}]):
\begin{equation}
\vec{\sigma}\cdot{\bf S}(j)=
e^{i\frac{\theta}{2}\hat{\bf d}\cdot\vec{\sigma}}
\vec{\sigma}\cdot{\bf S}'(j)
e^{-i\frac{\theta}{2}\hat{\bf d}\cdot\vec{\sigma}},\ \ 
\vec{\sigma}\cdot{\bf S}(i)=
e^{-i\frac{\theta}{2}\hat{\bf d}\cdot\vec{\sigma}}
\vec{\sigma}\cdot{\bf S}'(i)
e^{i\frac{\theta}{2}\hat{\bf d}\cdot\vec{\sigma}}.
\label{III14}
\end{equation}

When ${\bf D}_{2}$ is finite, the triplet splits into a singlet
and a doublet.  One may ask whether it is possible to perform
rotations of the spins such that the antisymmetric Dzyaloshinskii
term will be eliminated and the Hamiltonian will contain only the
symmetric anisotropy. This is in general not the case. Returning
to the Hamiltonian (\ref{III9}), we put
\begin{equation}
\vec{\sigma}\cdot{\bf S}(j)=
e^{i\frac{\alpha}{2}\hat{\bf a}\cdot\vec{\sigma}}
\vec{\sigma}\cdot{\bf S}'(j)
e^{-i\frac{\alpha}{2}\hat{\bf a}\cdot\vec{\sigma}},\ \ 
\vec{\sigma}\cdot{\bf S}(i)=
e^{-i\frac{\alpha}{2}\hat{\bf a}\cdot\vec{\sigma}}
\vec{\sigma}\cdot{\bf S}'(i)
e^{i\frac{\alpha}{2}\hat{\bf a}\cdot\vec{\sigma}},
\label{III16}
\end{equation}
where the unit vector $\hat{\bf a}$ and the angle $\alpha$ are yet to
be determined.  The Hamiltonian then takes the form (\ref{III9}),
with the replacements ${\bf S}(i)\rightarrow {\bf S}'(i),$
${\bf S}(j)\rightarrow {\bf S}'(j) $,
and $A_{i}\rightarrow A_{i}' $,
${\bf B}_{i}\rightarrow {\bf B}_{i}' $, ($i=1,2$) with
\begin{equation}
A_{\stackrel{1}{2}}'=A_{\stackrel{1}{2}}\cos\alpha \mp\sin\alpha
\Bigl({\bf B}_{\stackrel{1}{2}}\cdot\hat{\bf a}\Bigr),\ \
{\bf B}_{\stackrel{1}{2}}'={\bf B}_{\stackrel{1}{2}}\pm A_{\stackrel{1}{2}}
\hat{\bf a}\sin\alpha-2\sin^{2}\frac{\alpha}{2}\Bigl(\hat{\bf a}
\cdot {\bf B}_{\stackrel{1}{2}}
\Bigr)\hat{\bf a}.\label{III17}
\end{equation}
The condition that the Dzyaloshinskii term vanish is therefore [cf. Eqs.
(\ref{III10}) and (\ref{III11})]
\begin{eqnarray}
\Bigl[&A&_{1}\cos\alpha-\sin\alpha({\bf B}_{1}\cdot\hat{\bf a})\Bigr]
\Bigl({\bf B}_{2} -\hat{\bf a}(\hat{\bf a}\cdot{\bf B}_{2})\Bigr)-
\Bigl[A_{2}\cos\alpha+\sin\alpha({\bf B}_{2}\cdot\hat{\bf a})\Bigr]
\Bigl({\bf B}_{1} -\hat{\bf a}(\hat{\bf a}\cdot{\bf B}_{1})\Bigr)\nonumber\\
&=&\hat{\bf a}\Bigl[\sin 2\alpha \Bigl(A_{1}A_{2}+(\hat{\bf a}\cdot{\bf B}_{1})
(\hat{\bf a} \cdot{\bf B}_{2})\Bigr)+\cos 2\alpha \Bigl(
(A_{2}{\bf B}_{1}-A_{1}{\bf B}_{2})\cdot\hat{\bf a}\Bigr)\Bigr].\label{III18}
\end{eqnarray}
Since the vector on the left--hand--side is orthogonal to $\hat{\bf a}$,
Eq. (\ref{III18}) yields
\begin{eqnarray}
\tan 2\alpha &=&\frac{(A_{1}{\bf B}_{2}-A_{2}{\bf B}_{1})\cdot\hat{\bf a}}
{A_{1}A_{2}+(\hat{\bf a}\cdot{\bf B}_{1})(\hat{\bf a}\cdot{\bf B}_{2})},\nonumber\\
\Bigl[A_{1}\cos\alpha - (\hat{\bf a}\cdot{\bf B}_{1})
\sin\alpha \Bigr]\Bigl({\bf B}_{2}
-\hat{\bf a}(\hat{\bf a}\cdot{\bf B}_{2})\Bigr) &=&
\Bigl[A_{2}\cos\alpha + (\hat{\bf a}\cdot{\bf B}_{2})
\sin\alpha \Bigr]\Bigl({\bf B}_{1}
-\hat{\bf a}(\hat{\bf a}\cdot{\bf B}_{1})\Bigr)\ . \nonumber \\
\label{III19}
\end{eqnarray}
It can be shown that these
two equations can be satisfied only when the vectors ${\bf B}_{1}$ and
${\bf B}_{2}$ are parallel.
Hence the antisymmetric anisotropy can be eliminated from
the Hamiltonian only for specific configurations. Moreover, the
criterion for complete rotational invariance of ${\cal H}(i,j)$ is that
${\bf D}_{2}=A_{1}{\bf B}_{2}+A_{2}{\bf B}_{1}=0$, or equivalently,
that ${\bf B}_{2}/A_{2}=-{\bf B}_{1}/A_{1}$, which is equivalent
to the condition that
\begin{equation}
\Bigl(A_{1}+i{\bf B}_{1}\cdot \vec{\sigma}\Bigr)^{\dagger}=\frac{A_{1}}{A_{2}}
\Bigl(A_{2}+i{\bf B}_{2}\cdot\vec{\sigma}\Bigr).\label{III20}
\end{equation}
Returning to Eq. (\ref{III7}) we note that
$(\tilde{\bf t}_{00}^{ij})^{\dagger}=\tilde{\bf t}_{00}^{ji}$, and
therefore that ${\cal H}^{(2)}(i,j)$ is indeed rotationally invariant.
This represents an alternative proof for the SEA result,
[\onlinecite{SEA}]
which holds to order $t^{2}$, to all orders in the spin--orbit coupling
$\lambda $ and for all site symmetries providing
$U_{\alpha \beta} =U$ and ${\bf K}=0$.

\section{PERTURBATION THEORY INCLUDING OXYGEN ORBITALS}

Here we show that the perturbation theory results for the Cu--O--Cu bond,
through the intermediate state in which the two holes are on the
copper ion (channel $a$) are obtained from those of the
Cu--Cu bond, with the replacement (\ref{III29}).

We start from the state $|\psi_1\rangle$ in Eq. (\ref{III30}).
As explained in the text,
the index $q$ that labels oxygen ions on the bonds along $x$ and
along $y$ may be omitted for simplicity.  Applying again the
hopping Hamiltonian yields
\begin{eqnarray}
| \psi_{2a} \rangle & = &
{1 \over {\cal H}_0} \left( T_j^\dagger {1 \over {\cal H}_0}T_i
+ T_i^\dagger {1 \over {\cal H}_0}T_j \right) c_{i 0 \sigma}^\dagger
c_{j0\sigma_1}^\dagger c_{j0\sigma_1} c_{i0\sigma} | \psi_0 \rangle
\nonumber \\ &=&
\sum_{n}\sum_{\stackrel{\sigma\sigma_{1}}{\sigma_{2}\sigma_{3}}}
\frac{1}{\epsilon_{n}} \Bigl[
\frac{\Bigl({\bar t}_{n0}^{i}\Bigr)_{\sigma_{2}\sigma}
\Bigl({\bar t}_{an}^{j}\Bigr)_{\sigma_{3}\sigma_{2}}}
{U_{0}+E_{ja}}c_{ja\sigma_{3}}^{\dagger}c_{j0\sigma_{1}}^{\dagger}
\nonumber \\ && \ \ \ +
\frac{\Bigl({\bar t}_{n0}^{j}\Bigr)_{\sigma_{2}\sigma_{1}}
\Bigl({\bar t}_{an}^{i}\Bigr)_{\sigma_{3}\sigma_{2}}}
{U_{0}+E_{ia}}c_{io\sigma}^{\dagger}c_{ia\sigma_{3}}^{\dagger}\Bigr]
c_{j0\sigma_{1}}c_{i0\sigma}\mid\psi_{0}> \ ,
\end{eqnarray}
where we have written the energy denominators explicitly. We now
concentrate on the terms that will eventually contribute to the
spin Hamiltonian. To order $t^{4}$, these are obtained by
applying two more factors of the hopping, that bring the holes back
to the ground state. The result is
\begin{eqnarray}
\Biggl( T_i^\dagger {1 \over {\cal H}_0} T_j &+&
T_j^\dagger {1 \over {\cal H}_0} T_i \Biggr) | \psi_{2a} \rangle
\nonumber \\ &=& - \sum_{nn'}\sum_{\sigma\sigma_{1}\sigma_{2}}
\sum_{\sigma_{3}\sigma_{4}\sigma_{5}}
\frac{1}{\epsilon_{n}\epsilon_{n'}U_{0}} \Bigl[
\Bigl({\bar t}_{n0}^{i}\Bigr)_{\sigma_{2}\sigma}
\Bigl({\bar t}_{0n}^{j}\Bigr)_{\sigma_{3}\sigma_{2}}
\Bigl({\bar t}_{n'0}^{j}\Bigr)_{\sigma_{4}\sigma_{1}}
\Bigl({\bar t}_{0n'}^{i}\Bigr)_{\sigma_{5}\sigma_{4}}
c_{j0\sigma_{3}}^{\dagger}c_{i0\sigma_{5}}^{\dagger} \ \nonumber\\
&& \ \  + \Bigl({\bar t}_{n0}^{j}\Bigr)_{\sigma_{2}\sigma_{1}}
\Bigl({\bar t}_{0n}^{i}\Bigr)_{\sigma_{3}\sigma_{2}}
\Bigl({\bar t}_{n'0}^{i}\Bigr)_{\sigma_{4}\sigma}
\Bigl({\bar t}_{0n'}^{j}\Bigr)_{\sigma_{5}\sigma_{4}}
c_{j0\sigma_{5}}^{\dagger}c_{i0\sigma_{3}}^{\dagger}\Bigr]
c_{j0\sigma_{1}} c_{i0\sigma} \mid \psi_{0}> \ ,
\end{eqnarray}
from which it is clear that using Eq. (\ref{III29}) one arrives at
Eq. (\ref{III7}).

\section{DIAGONALIZATION OF THE SPIN--ORBIT INTERACTION}

In this Appendix we show that the transformation of Eq. (\ref{CANON})
does indeed make the spin--orbit interaction diagonal in pseudospin
space.  We wish to show that
\begin{equation}
\langle \beta , \nu | {\bf L} \cdot \vec \sigma | \alpha , \mu \rangle
\end{equation}
vanishes unless $\mu=\nu$ in which case it is independent of $\mu$.
First of all, note that ${\bf L}$ has zero diagonal matrix elements,
i. e. the above matrix element vanishes when $\alpha =\beta$.
There are now three cases to consider: i) $\alpha=0$ and $\beta=1$
or $\alpha=1$ and $\beta=0$; ii) $\alpha=0,1$ and $\beta=x,y,z$;
and iii) $\alpha \not= \beta$ but both are $x$, $y$, or $z$.  In case i
the matrix element of ${\bf L}$ is again zero, so this case is 
as desired. In case ii with $\alpha=0$ ($\alpha=1$ is similar)
we express the above matrix element as
\begin{equation}
\sum_{\gamma, \rho} \langle \psi_\beta ({\bf r}) | L_\gamma |
\psi_0 ({\bf r}) \rangle \Bigl( \sigma_\beta \Bigr)_{\nu , \rho}
\langle \phi_\rho ( \sigma ) | \sigma_\gamma |\phi_\mu
(\sigma ) \rangle = 
\langle \psi_\beta ({\bf r}) | L_\beta | \psi_0 ({\bf r}) \rangle
\delta_{\mu , \nu} \ .
\end{equation}
In the last step we used the fact that the orbital matrix
element is only nonzero when $\gamma = \beta$.  So case ii is as
desired. In case iii we write the matrix element as
\begin{eqnarray}
&& \sum_{\gamma, \rho, \tau} \langle \psi_\beta ({\bf r}) | L_\gamma |
\psi_\alpha ({\bf r}) \rangle
\Bigl( \sigma_\alpha \Bigr)_{\mu , \tau}^*
\Bigl( \sigma_\beta \Bigr)_{\nu , \rho}
\langle \phi_\rho ( \sigma ) | \sigma_\gamma |\phi_\tau
(\sigma ) \rangle \nonumber \\ && =
\sum_{\gamma, \rho, \tau} \langle \psi_\beta ({\bf r}) | L_\gamma |
\psi_\alpha ({\bf r}) \rangle
\Bigl( \sigma_\alpha \Bigr)_{\tau , \mu}
\Bigl( \sigma_\beta \Bigr)_{\nu , \rho}
\Bigl( \sigma_\gamma \Bigr)_{\rho , \tau}
= \langle \psi_\beta ({\bf r}) | L_\gamma | \psi_\alpha ({\bf r}) \rangle
i \epsilon_{\beta \gamma \alpha } \delta_{\mu , \nu} \ ,
\end{eqnarray}
where $\epsilon_{\alpha \beta \gamma}$ is the totally
antisymmetric tensor.  In the last equality we used the fact
that $\alpha$, $\beta$ and $\gamma$ are Cartesian indices which
are all different.  Thus all the types of matrix elements
are diagonal and independent of pseudospin, as asserted.

\section{RESULTS FOR TETRAGONAL SYMMETRY}

The only nonzero matrix elements of the angular momentum within
the manifold of normalized tetragonal $d$ states,
$\mid 0>=d_{x^{2}-y^{2}}$, $\mid 1>=d_{3z^{2}-r^{2}},$
$\mid z>=d_{xy}$, $\mid x>=d_{yz}$, and $\mid y>=d_{xz}$ are
\begin{eqnarray}
&& \langle x^2-y^2 | L_x | yz \rangle = \langle zx | L_x | xy \rangle = -
\langle xy | L_x | zx \rangle = - \langle yz | L_x | x^2-y^2 \rangle = 
\langle xy | L_y | yz \rangle
\nonumber \\ &&
= \langle x^2-y^2 | L_y | zx \rangle = 
- \langle yz | L_y | xy \rangle = - \langle zx | L_y | x^2-y^2 \rangle = 
\langle yz | L_z | zx \rangle = - \langle zx | L_z | xy \rangle = i \ ,
\end{eqnarray}
\begin{eqnarray}
\langle 3z^2-r^2 | L_x | yz \rangle = - \langle yz | L_x | 3z^2-r^2 \rangle
= -
\langle 3z^2-r^2 | L_y | zx \rangle = \langle zx | L_y | 3z^2-r^2 \rangle
= i \sqrt 3  \ ,
\end{eqnarray}
\begin{equation}
\langle xy | L_z | x^2-y^2 \rangle = - \langle x^2-y^2 | L_z | xy \rangle
= 2i \ .
\end{equation}

>From Appendix F and the pseudospin transformation (\ref{CANON})
it follows that
\begin{equation}
\Bigl(\omega(\alpha ,\beta)\Bigr)_{\sigma\sigma '}
=\bar{\omega}(\alpha ,\beta) \Bigl( \sigma_\alpha \sigma_\beta
\Bigr)_{\sigma , \sigma'} \ ,
\end{equation}
where $\omega$ was defined in Eq. (\ref{I4}), and
$\bar \omega (\alpha, \beta)$ is a scalar.
Turning now to the diagonalization of the single--particle,
single--site Hamiltonian ${\cal H}_{x}+{\cal H}_{\rm so}$
of the Cu--Cu bond, one finds that this can be accomplished by putting 
\begin{equation}
\label{MMATEQ}
{\bf m}_{\alpha a}=\sigma_\alpha \bar{m}_{\alpha a} \ ,
\end{equation}
where $\bar{m}_{\alpha a}$ are scalars which are determined by
Eq. (\ref{I10}), which now becomes
\begin{equation}
\sum_{\alpha}\epsilon_{\alpha}\bar{m}_{\alpha a}^{\ast}
\bar{m}_{\alpha b} + \lambda\sum_{\alpha\beta}
\bar{\omega}(\alpha ,\beta)\bar{m}_{\alpha a}^{\ast}
\bar{m}_{\beta b}=\delta_{ab}E_{a}, \ \ 
\sum_{a}\bar{m}_{\alpha a}\bar{m}_{\beta a}^{\ast}=\delta_{\alpha\beta}.
\end{equation}
The solution of this equation, to second order in $\lambda $, reads
\begin{equation}
\bar{m}_{\alpha a}=\delta_{\alpha a}
\Bigl(1-\frac{\lambda^{2}}{2}
\sum_{\gamma}\frac{\bar{\omega}(\alpha,\gamma)
\bar{\omega}(\gamma, \alpha)}
{(\epsilon_{\alpha}-\epsilon_{\gamma})^{2}}\Bigr)
+(1-\delta_{\alpha a})\Bigl(
\lambda\frac{\bar{\omega}(\alpha, a)}
{\epsilon_{a}-\epsilon_{\alpha}}+ \lambda^{2}\sum_{\gamma}
\frac{\bar{\omega}(\alpha,\gamma)\bar{\omega}(\gamma, a)}
{(\epsilon_{\alpha}-\epsilon_{a})(\epsilon_{\gamma}-\epsilon_{a})}\Bigr).
\label{MEQ}
\end{equation}
The hopping matrix elements are diagonal in the tetragonal symmetry,
except for the states $\mid 0>$ and $\mid 1>$, and
are independent of the site indices $i$
and $j$. It follows that
$\tilde{t}_{ab}^{ij}$ are scalars independent of the site
indices as well, and that the hopping is not accompanied by a
pseudospin flip. This implies that ${\bf x}_{\alpha\alpha}$
[Eq. (\ref{III25a})] is diagonal in that space, and that
${\bf x}_{\alpha\alpha '}$ is independent of $i$ and $j$. Thus
the vector ${\bf D}_{ij}$, Eq. (\ref{III26b}), vanishes and
the matrix ${\bf M}(i,j)$, Eq. (\ref{III26c}), is given solely by the
$K_{\alpha\alpha '}$ terms, and is independent of $i$ and $j$.

>From these arguments and using Eqs. (\ref{III25}), one obtains that
the symmetric matrix ${\bf M}(i,j)$ (which is diagonal in tetragonal
symmetry) is given by
\begin{eqnarray}
\label{MIJ}
{\bf M}(i,j) &=&2\sum_{\alpha\alpha '}K_{\alpha\alpha '}
\Bigl({\rm Tr} \{ {\bf m}_{\alpha '0}^{\dagger}
{\bf m}_{\alpha 0}\vec{\sigma}\}\otimes
{\rm Tr} \{ {\bf w}_{\alpha\alpha '}^{ji}\vec{\sigma}\}\nonumber\\
&-&\frac{1}{2}{\rm Tr} \{
{\bf x}_{\alpha\alpha '}^{ji}\vec{\sigma}\}\otimes 
{\rm Tr} \{( {\bf x}_{\alpha\alpha '}^{ji})^{\dagger}\vec{\sigma}\}
-\frac{1}{2}{\rm Tr} \{( {\bf x}_{\alpha\alpha '}^{ji})^{\dagger}
\vec{\sigma}\}\otimes {\rm Tr} \{ {\bf x}_{\alpha\alpha '}^{ji}
\vec{\sigma}\}\Bigr)\nonumber\\
&=&2\sum_{ab}\frac{1}{(U_{0}+E_{a})}\frac{1}{(U_{0}+E_{b})}
\tilde{t}_{0a}^{ji} \tilde{t}_{b0}^{ij}
\sum_{\alpha\alpha '}K_{\alpha\alpha '}\Bigl(
{\rm Tr} \{ {\bf m}_{\alpha '0}^{\dagger}
{\bf m}_{\alpha 0}\vec{\sigma}\}\otimes
{\rm Tr} \{ {\bf m}_{\alpha a}^{\dagger}
{\bf m}_{\alpha 'b}\vec{\sigma}\}\nonumber\\
&-&\frac{1}{2}{\rm Tr} \{ {\bf m}_{\alpha a}^{\dagger}
{\bf m}_{\alpha '0} \vec{\sigma}\}\otimes
{\rm Tr} \{ {\bf m}_{\alpha '0}^{\dagger}
{\bf m}_{\alpha b}\vec{\sigma}\}-\frac{1}{2}
{\rm Tr} \{ {\bf m}_{\alpha '0}^{\dagger}
{\bf m}_{\alpha b}\vec{\sigma}\}\otimes
{\rm Tr} \{ {\bf m}_{\alpha a}^{\dagger}
{\bf m}_{\alpha '0}\vec{\sigma}\}\Bigr).
\end{eqnarray}
Performing the sums over $\alpha $ and $\alpha '$, this yields
\begin{eqnarray}
M_{xx}(i,j) &=& 8\sum_{ab}\frac{1}{(U_{0}+E_{a})}
\frac{1}{(U_{0}+E_{b})} \tilde{t}_{0a}^{ji}\tilde{t}_{b0}^{ij}
\Bigl[K_{0x} (\bar{m}_{x0}^{\ast}\bar{m}_{0a}^{\ast}
-\bar{m}_{00}^{\ast}\bar{m}_{xa}^{\ast})
(\bar{m}_{00}\bar{m}_{xb}-\bar{m}_{x0}\bar{m}_{0b})\nonumber\\
&+&K_{1x} (\bar{m}_{x0}^{\ast}\bar{m}_{1a}^{\ast}
-\bar{m}_{10}^{\ast}\bar{m}_{xa}^{\ast})
(\bar{m}_{10}\bar{m}_{xb}-\bar{m}_{x0}\bar{m}_{1b})\nonumber\\
&+&K_{yz} (\bar{m}_{z0}^{\ast}\bar{m}_{ya}^{\ast}
-\bar{m}_{y0}^{\ast}\bar{m}_{za}^{\ast})
(\bar{m}_{y0}\bar{m}_{zb}-\bar{m}_{z0}\bar{m}_{yb})\Bigr]\ ,
\end{eqnarray}
with analogous expressions for the $yy$ and $zz$ entries of ${\bf M}$.
The next step is to write $\bar{m}_{\alpha a}$ in terms of the
spin--orbit matrix elements, Eq.  (\ref{MEQ}). In doing this we keep
in mind that both states $a$ and $b$ cannot be the ground state $0$,
as they refer to intermediate states of the perturbation theory.
Therefore, it is sufficient to retain for the coefficient of $K_{0x}$
the terms
\begin{equation}
-\delta_{xa}\delta_{xb}-\lambda\delta_{xa}\frac{\bar{\omega}(x,b)}
{\epsilon_{b}-\epsilon_{x}}+\lambda\delta_{xb}\frac{\bar{\omega}(a,x)}
{\epsilon_{x}-\epsilon_{a}}-\lambda^{2}
\frac{\bar{\omega}(a,x)\bar{\omega}(x,b)}{(\epsilon_{x}-\epsilon_{b})
(\epsilon_{x}-\epsilon_{a})}\ .
\end{equation}
Similarly, the coefficient of $K_{1x}$ is
\begin{equation}
-\lambda^{2}\frac{\bar{\omega}(x,0)\bar{\omega}(0,x)}
{\epsilon_{x}^{2}}\delta_{1a} \delta_{1b}\ .
\end{equation}
The leading order of the coefficient of $K_{yz}$ is of order
$\lambda^{2}\delta_{yb}\delta_{za}$, etc. But then
$\tilde{t}_{0z}\tilde{t}_{y0}$ will be proportional to $\lambda^2$
too. Therefore, to order $\lambda^{2}$, the terms arising from
$K_{yz}$ do not contribute. Collecting terms we find the
contribution to the anisotropic exchange as
\begin{eqnarray}
J_{xx}^{\rm anis}(i,j) & = & -8K_{0x}
\Bigl[\frac{1}{U_{0}+\epsilon_{x})^{2}}
\tilde{t}_{0x}\tilde{t}_{x0}
+\lambda\frac{1}{U_{0}+\epsilon_{x}}\sum_{a}\Bigl(
\frac{\bar{\omega}(x,a)}{\epsilon_{a}-\epsilon_{x}}
\tilde{t}_{0x}\tilde{t}_{a0}
-\frac{\bar{\omega}(a,x)}{\epsilon_{x}-\epsilon_{a}}
\tilde{t}_{0a}\tilde{t}_{x0} \Bigr)\nonumber\\
&+&\lambda^{2}\sum_{ab}\frac{1}{U_{0}+\epsilon_{a}}
\frac{1}{U_{0}+\epsilon_{b}}
\frac{\bar{\omega}(a,x)\bar{\omega}(x,b)}
{(\epsilon_{x}-\epsilon_{b})(\epsilon_{x}
-\epsilon_{a})}\tilde{t}_{0a}\tilde{t}_{b0}\Bigr]\nonumber\\
&-&8K_{1x}\lambda^{2}\frac{1}{(U_{0}+\epsilon_{1})^{2}}
\tilde{t}_{01}\tilde {t}_{01}
\frac{\bar{\omega}(x,0)\bar{\omega}(0,x)}{\epsilon_{x}^{2}},
\label{C11}
\end{eqnarray}
where we have retained terms up to order $\lambda^{2}$. Finally
we write, using Eqs. (\ref{I16c}) and (\ref{MIJ})
\begin{equation}
\tilde{t}_{0x}=\lambda\frac{\bar{\omega}(0,x)}{\epsilon_{x}}
t_{00}- t_{xx})-\lambda t_{01}
\frac{\bar{\omega}(1,x)}{\epsilon_{1}- \epsilon_{x}}\ ,
\end{equation}
and put $a$ and $b$ in the sums of (\ref{C11}) equal to $1$, with
$\tilde{t}_{01}= t_{01}$. (These are the only possible
contributions up to order $\lambda^{2}$). It then follows that
\begin{eqnarray}
J_{xx}^{\rm anis}(i,j) &=&-8\lambda^{2}\Bigl[K_{0x}
\frac{1}{(U_{0}+\epsilon_{x})^{2}}
\left |\frac{\bar{\omega}(0,x)}{\epsilon_{x}}
(t_{00}-t_{xx})
-\frac{\bar{\omega}(1,x)}{U_{0}+\epsilon_{1}}
t_{01}\right |^{2}\nonumber\\
&+& K_{1x}\left |\frac{1}{U_{0}+\epsilon_{1}} t_{10}
\frac{\bar{\omega}(x,0)} {\epsilon_{x}}\right |^{2}\Bigr]\ .
\end{eqnarray}
This result reproduces that of Ref. \onlinecite{PRL2}. 
This result differs slightly from that given in Ref. 
\onlinecite{PRL2} and Eq. (\ref{JMU}) in the text, in that the
denominators include just the constant part $U_0$ of the
Coulomb interactions.  The corresponding expression in Ref. 
\onlinecite{PRL2} [and Eq. (\ref{JMU})] includes instead
$U_{0,\mu}$, $U_{0.1}$, and $U_{1,\mu}$ and thus represents
an expansion, described in Appendix H, in ${\bf K}$ but not
in $\Delta {\bf U}$.

\section{PERTURBATION THEORY FOR THE TETRAGONAL CASE}

In this Appendix we give an alternative derivation of Eq. (\ref{JMU})
based on conventional perturbation theory in which we treat
hopping, ${\cal H}_{\rm hop}$, spin--orbit, ${\cal H}_{\rm so}$,
and the Coulomb exchange interactions, ${\cal H}_{\rm ex}$,
as perturbations. 
For the ``generic'' model, anisotropic exchange appears at order
${\cal H}_{\rm hop}^2 {\cal H}_{\rm so}^2 {\cal H}_{\rm ex}$.
To perform this calculation we therefore need to work to fifth order
perturbation theory and will arbitrarily omit contributions to the
isotropic exchange.
Of course, ${\cal H}_{\rm ex}$ can only exist when there are two holes
on the same ion, so the five perturbations must be arranged
so that ${\cal H}_{\rm hop}$ and ${\cal H}_{\rm ex}$ occur in the
order ${\cal H}_{\rm hop} {\cal H}_{\rm ex} {\cal H}_{\rm hop}$.
In principle there are ten ways to insert the two factors
of ${\cal H}_{\rm so}$.  But some study shows that only if the
two powers of ${\cal H}_{\rm so}$ are
separated by ${\cal H}_{\rm ex}$ does the result lead to anisotropy.  So
the relevant fifth order terms in the effective Hamiltonian are
\begin{equation}
{\cal H} (i,j) = \left[ {\cal H}_{\rm so} {1 \over {\cal E} }
{\cal H}_{\rm hop} + {\cal H}_{\rm hop} {1 \over {\cal E} }
{\cal H}_{\rm so} \right]
{1 \over {\cal E} } {\cal H}_{\rm ex} {1 \over {\cal E} }
\left[ {\cal H}_{\rm so} {1 \over {\cal E} }
{\cal H}_{\rm hop} + {\cal H}_{\rm hop}
{1 \over {\cal E}} {\cal H}_{\rm so}  \right] \ ,
\end{equation}
where ${\cal E}$ is the appropriate energy denominator.
If we write the spin--orbit perturbation as
\begin{equation}
{\cal H}_{\rm so}= \lambda \sum_\alpha \biggl( \sum_{\rm holes, h}
L_\alpha (h) s_\alpha (h) \biggr) \equiv \sum_\alpha V_\alpha \ , 
\end{equation}
then it is easy to see that there are no cross terms, i. e.
terms involving $V_\alpha V_\beta$ with $\alpha \not= \beta$.
In addition, hopping from site $i$ to site $j$ and back will
give the same result as the reverse process.  So if hopping from
site $j$ to site $i$ is denoted $T_{ij}$, then we may write
\begin{eqnarray}
{\cal H} (i,j) &  = &  2 \sum_\alpha
\left[ V_\alpha {1 \over {\cal E} }
T_{ji} + T_{ji} {1 \over {\cal E} } V_\alpha \right]
{1 \over {\cal E} } {\cal H}_{\rm ex} {1 \over {\cal E} } 
\left[ V_\alpha {1 \over {\cal E} } T_{ij}
+ T_{ij} {1 \over {\cal E} } V_\alpha \right] \nonumber \\
& \equiv & 2 \sum_\alpha {\bf Q}_\alpha^\dagger {1 \over {\cal E} }
{\cal H}_{\rm ex} {1 \over {\cal E} } {\bf Q}_\alpha ,
\end{eqnarray}
where the operator ${\bf Q}_\alpha$ that we need to evaluate is simply
\begin{equation}
{\bf Q}_\alpha = \left[ V_\alpha {1 \over {\cal E} } T_{ij} +
T_{ij} {1 \over {\cal E} } V_\alpha \right] \ .
\end{equation}
There are two channels to be considered for the intermediate state
in which ${\cal H}_{\rm ex}$ operates.  Channel "0" is one in which site
$j$ has orbitals $|0 \rangle$ and $|\alpha \rangle$ occupied,
whereas channel "1" is one in which site
$j$ has orbitals $|1 \rangle$ and $|\alpha \rangle$ occupied.
Then we may define
\begin{equation}
\left[ Q_\alpha^{(\gamma)} \right]_{\sigma , \eta ; \sigma' , \eta' }
= \langle 0 | d_{i,\gamma ,\sigma} d_{i, \alpha , \eta}
\left[ V_\alpha {1 \over {\cal E} } T_{ij} +
T_{ij} {1 \over {\cal E} } V_\alpha \right]
d_{i, 0, \eta'}^{\dag} d_{j,0,\sigma'}^{\dag} | 0 \rangle \ ,
\end{equation}
where $\gamma=0$ or 1.  Then
\begin{equation}
{\cal H} (i,j)  =  2 \sum_\alpha \sum_\gamma 
\left[ {\bf Q}_\alpha^{(\gamma)} \right]^\dagger
{\cal H}_{{\rm ex},\alpha}^\gamma {\bf Q}_\alpha^{(\gamma)}
( \epsilon_\alpha + \epsilon_\gamma + U_{\alpha \gamma} )^{-2} \ ,
\end{equation}
where
\begin{equation}
{\cal H}_{{\rm ex},\alpha}^\gamma = - {1 \over 2 } {\cal K}_{\alpha \gamma}
\left[ {\cal I } {\cal I} + \vec \sigma \cdot \vec \sigma \right] \ .
\end{equation}
Here ${\cal I}$ is the identity operator and
$\sigma \cdot \sigma$ denotes the sum over direct products,
$\sum_\alpha \sigma_\alpha \sigma_\alpha$.
Also, each matrix [${\bf Q}^{(\gamma)}$ or ${\cal H}_{\rm ex}^{(\gamma)}$]
is a matrix in the direct product of the two
spin variables.  Any operator in this space can be written as a
linear combination of direct product operators.
We define ${\cal A}{\cal B}$ via
\begin{equation}
[ {\cal A} {\cal B}]_{\sigma , \eta; \sigma' , \eta'} =
A_{\sigma , \sigma'} B_{\eta , \eta'} \ .
\end{equation}

Explicit calculation of
the processes shown in Fig. 7 shows that
\begin{eqnarray}
\left[ Q_\alpha^{(0)} \right] _{\sigma , \eta ; \sigma' , \eta' }  & = &
{t_{00} \lambda \langle \alpha | L_x | 0 \rangle \over 2 }
\left( { \delta_{\sigma , \sigma'} [ \sigma_\alpha]_{\eta , \eta'}
\over \epsilon_\alpha } 
- { \delta_{\eta' , \sigma} [\sigma_\alpha]_{\eta , \sigma'}
\over U_{00} } \right)
+ { \delta_{\sigma , \sigma'} [\sigma_\alpha]_{\eta , \eta'}
\over U_{00} }
\nonumber \\
&-&  {\lambda \langle \alpha | L_\alpha | 0 \rangle t_{\alpha \alpha} \over
2 \epsilon_\alpha } \delta_{\eta' , \sigma}
[\sigma_\alpha]_{\eta , \sigma'}
-  {\lambda \langle \alpha | L_\alpha | 1 \rangle t_{01} \over
2 (\epsilon_1 + U_{10} ) } \delta_{\eta' , \sigma}
[\sigma_\alpha]_{\eta , \sigma'} \
\end{eqnarray}

To write this in operator form, note that
${1 \over 2} \Bigl[ {\cal I} {\cal I} + \vec \sigma \cdot \vec \sigma
\Bigr]_{ \sigma , \eta; \sigma' , \eta' }$ is unity if
$\sigma=\eta'$ and $\eta=\sigma'$ and is zero otherwise.  Thus
\begin{eqnarray}
\left[ {\bf Q}_\alpha^{(0)} \right]
& = & C_1 \left[ {\cal I } \sigma_\alpha \right] +
{1  \over 2 } C_2 \left[ {\cal I } \sigma_\alpha \right]
\left[ {\cal I } {\cal I} + \vec \sigma \cdot \vec \sigma \right]
\nonumber \\
& \equiv & (C_1 + C_2 ) \left[ {\cal I} \sigma_\alpha \right]
- C_2 \left[ {\cal I} \sigma_\alpha \right] \left[ {\cal O} \right] \ ,
\end{eqnarray}
where $\left[ {\cal O} \right] = 
[ {\cal I} {\cal I} - \vec \sigma \cdot \vec \sigma ]/2$ and
\begin{equation}
C_1 = {\lambda \over 2 } \Biggl\{ 
{t_{00} \langle \alpha | L_x | 0 \rangle \over \epsilon_\alpha } 
+ {t_{00} \langle \alpha | L_x | 0 \rangle \over U_{00} } \Biggr\} 
\end{equation}
\begin{equation}
C_2 = - { \lambda \over 2 } \Biggl\{
{t_{00} \langle \alpha | L_x | 0 \rangle \over U_{00} } 
+  {t_{\alpha \alpha} \langle \alpha | L_\alpha | 0 \rangle
\over \epsilon_\alpha }
+  {t_{01} \langle \alpha | L_\alpha | 1 \rangle 
\over (\epsilon_1 + U_{10} ) } \Biggr\}  \ .
\end{equation}
Also
\begin{equation}
\left[ Q_\alpha^{(1)} \right] _{\sigma , \eta ; \sigma' , \eta' }  =
{ \lambda t_{01} \langle \alpha | L_\alpha | 0 \rangle
\over 2 } \left( {1 \over \epsilon_\alpha }
+ {1 \over (\epsilon_1 + U_{10}) } \right)
\delta_{\sigma , \sigma'} [\sigma_\alpha]_{\eta, \eta'}
\end{equation}
so that
\begin{equation}
\left[ {\bf Q}_\alpha^{(1)} \right] 
= { \lambda t_{01} \langle \alpha | L_\alpha
| 0 \rangle \over 2 } \left( {1 \over \epsilon_\alpha }
+ {1 \over (\epsilon_1 + U_{10}) } \right)
\left[ {\cal I } \sigma_\alpha \right]
\equiv C_3 \left[ {\cal I} \sigma_\alpha \right]  \ .
\end{equation}

Thus we have the result
\begin{eqnarray}
\label{5RES}
{\cal H}(i,j) & = & - \sum_\alpha \Biggl\{
{K_{0 \alpha} \over ( \epsilon_\alpha + U_{0\alpha })^2 }
\left[ C_1^* {\cal I} {\cal I}  + C_2^*  {\cal I} {\cal I}
- C_2^* {\cal O} \right] \left[ {\cal I} \sigma_\alpha \right]
\left[ {\cal I } { \cal I} + \vec \sigma \cdot \vec \sigma \right]
\left[ {\cal I} \sigma_\alpha \right]
\nonumber \\ && \times \left[
C_1  {\cal I} {\cal I} + C_2 {\cal I} {\cal I}  - C_2 {\cal O} \right]
+ { K_{1 \alpha} | C_3 |^2 \over ( \epsilon_\alpha
+ \epsilon_1 + U_{1\alpha} )^2 } \left[ {\cal I} \sigma_\alpha \right]
\left[ {\cal I } {\cal I} + \vec \sigma \cdot \vec \sigma \right]
\left[ {\cal I } \sigma_\alpha \right] \Biggr\} \ .
\end{eqnarray}
To simplify the above result we use the identity for Pauli matrices,
\begin{equation}
\label{SIGMA}
\left[ {\cal I } \sigma_\alpha \right]
\left[ {\cal I } {\cal I} + \vec \sigma \cdot \vec \sigma \right]
\left[ {\cal I } \sigma_\alpha \right] = \left[ {\cal I} {\cal I} 
+ 2 \sigma_\alpha \sigma_\alpha - \vec \sigma \cdot \vec \sigma \right] \ .
\end{equation}
>From the form of Eq. (\ref{5RES}) we see that
all the anisotropic contributions come from the term 
$\sigma_\alpha \sigma_\alpha$ in Eq. (\ref{SIGMA}).
Keeping only such terms we have
\begin{eqnarray}
{\cal H}(i,j) & = & - \sum_\alpha \Biggl\{
{2K_{0 \alpha} \over ( \epsilon_\alpha + U_{0\alpha })^2 }
\left[ C_1^* {\cal I} {\cal I}  + C_2^* {\cal I} {\cal I}
- C_2^* {\cal O} \right] \left[ \sigma_\alpha \sigma_\alpha \right]
\left[ C_1 {\cal I} {\cal I}  + C_2 {\cal I} {\cal I} - C_2 {\cal O} \right]
\nonumber \\ &&
\ \ + { 2K_{1 \alpha} | C_3 |^2 \left[ \sigma_\alpha \sigma_\alpha \right]
\over ( \epsilon_\alpha + \epsilon_1 + U_{1\alpha} )^2 } \Biggr\} \ .
\end{eqnarray}
The terms involving the operator ${\cal O}$ give only isotropic terms.
This can be seen by using the equality
\begin{equation}
2 \left[ \sigma_\alpha \sigma_\alpha \right] \left[ {\cal O} \right] =
\left[ \sigma_\alpha \sigma_\alpha \right] \left[[ {\cal I} {\cal I} -
\vec \sigma \cdot \vec \sigma \right] =
\left[ \vec \sigma \cdot \vec \sigma - {\cal I} {\cal I} \right] ,
\end{equation}
which is isotropic.  Thus the anisotropic exchange terms are correctly
given by ${\cal H}(i,j) = (1/4) \sum_\mu J_{\mu\mu}^{\rm anis} \sigma_\mu
\sigma_\mu = \sum_\mu J_{\mu \mu}^{\rm anis} S_\mu(i) S_\mu(j)$, with
\begin{eqnarray}
J_{\mu\mu}^{\rm anis} & = & - 2  \lambda^{2}  \left\{
 \frac{  \mid L_{0,\mu }^{\mu }\mid^{2}
t_{0,1}^{2}K_{1,\mu }} {  (\epsilon_{\mu } + \epsilon_{1}+U_{1,\mu })^{2}}
\left[ \frac{1}{\epsilon_{\mu }}
+ \frac{1}{\epsilon_{1}+U_{0,1}} \right]^{2} \right.  \nonumber \\
&+& \left.  \frac{ K_{0,\mu } }{(\epsilon_{\mu }+U_{0,\mu })^{2}}
\left | \frac{(t_{\mu ,\mu } - t_{0,0}) L_{0,\mu }^{\mu }}{\epsilon_{\mu }}
+ \frac{t_{0,1}  L_{1,\mu }^{\mu }}{\epsilon_{1}+U_{0,1}} \right |^{2}
\right\} ,
\end{eqnarray}
where $L_{\alpha \beta}^\mu$ denotes the orbital angular momentum matrix
element, $\langle \alpha | L^\mu | \beta \rangle$.

\section{EXPRESSIONS FOR $J_{\mu\mu}^{\rm anis}$}

In this Appendix we give expressions for $J_{\mu \mu}^{\rm anis}$
[see Eq. (\ref{JMU})] assuming the relations for the hopping matrix
elements implied by Eq. (\ref{IV4}) and the relations involving
$(pd\sigma)$ and $(pd\pi)$ listed in Sec. IV.D:
\begin{equation}
t_{xx}=0\ , \ \ \ \ t_{yy}=t_{zz}=t_{00}/3, \ \ \ \ 
t_{01}=-t_{00}/\sqrt 3 \ .
\end{equation}
Also, we use the identifications of the Racah coefficients given
in Appendix B.  Thereby we obtain
\begin{equation}
J_{xx}^{\rm anis} = - 2 \lambda^2 t_{00}^2 \left[ 
{B+C \over 3( \epsilon_x+ \epsilon_1 + A + 2B +C)^2 } +
{3B+C \over (\epsilon_x +A-2B+C)^2 } \right]
\left[ {1 \over \epsilon_x} + {1 \over \epsilon_1 + A -4B+C} \right]^2
\ , \end{equation}
\begin{eqnarray}
J_{yy}^{\rm anis} = - 2 \lambda^2 t_{00}^2 && \left\{
{B+C \over 3( \epsilon_x+ \epsilon_1 + A + 2B +C)^2 }
\left[ {1 \over \epsilon_x} + {1 \over \epsilon_1 + A -4B+C} \right]^2
\right. \nonumber \\ && \left. + {3B+C \over (\epsilon_x +A-2B+C)^2 }
\left[ - {2 \over 3\epsilon_x} + {1 \over \epsilon_1 + A -4B+C} \right]^2
\right\}
\ , \end{eqnarray}
\begin{eqnarray}
J_{zz}= -8 \lambda^2 t_{00}^2 && \left\{ {4B+C \over 3(\epsilon_z +
\epsilon_1 +A -4B+C)^2 } \left[ {1 \over \epsilon_z} + {1 \over
\epsilon_1+A-4B+C } \right]^2  \right. \nonumber \\ && + \left.
{C \over (\epsilon_z +A+4B+C)^2 } \left[
{2 \over 3 \epsilon_z } \right]^2 \right\} \ ,
\end{eqnarray}
where we set $\epsilon_y=\epsilon_x$ for tetragonal symmetry.

We now compare our results with those of Eq. (14) of BS.  In that
equation the only hopping matrix element that was included was that
between the ground state orbitals of the Cu ions.  If we keep only
such terms in Eq. (\ref{JMU}), we obtain
\begin{equation}
J_{xx}^{\rm anis}=J_{yy}^{\rm anis} =
-2 \lambda^2 t_{00}^2 {3B+C \over \epsilon_x^2 (\epsilon_x+A
-2B+C)^2} \approx - { \lambda^2 t_{0p_x}^4 \over \epsilon_{p_x}^2
\epsilon_x^2} \left[ {1 \over \epsilon_x + A -5B } - {1 \over
\epsilon_x +A +B +2C } \right] \ ,
\end{equation}
\begin{equation}
J_{zz}^{\rm anis} = -8 \lambda^2 t_{00}^2  {C \over \epsilon_z^2
(\epsilon_z+A +4B +C )^2 } \approx -4 { \lambda^2
t_{0,p_x}^4 \over \epsilon_{p_x}^2
\epsilon_z^2 } \left[ {1 \over \epsilon_z +A +4B} - 
{1 \over \epsilon_z +A +4B +2C } \right] \ ,
\end{equation}
where we used Eq. (\ref{IV4}) to set
$t_{00}=t_{0,p_x}^2/\epsilon_{p_x}$ and wrote our expression in
terms of singlet and triplet energy denominators
to facilitate comparison with BS. We note the
following differences between their results and ours.  1)  Our results
are smaller by an overall factor of 2.  2)  Instead of our evaluation
in which $t_{00}=t_{pd}^2/\Delta$ (to use their notation), they
use $t_{00}=t_{pd}^2/(\Delta+E_{xy})$.  3) The last energy denominator in
their $J_{yy}$ is wrong: (their $4B$ should be replaced by $B$).  With
respect to the first difference we would note that, as described in
the text, we did compare results from the full diagonalization with
those using the spin Hamiltonian.  Thus an error by a factor of two
in our calculations is extremely unlikely.  Difference 2 comes about
because BS do
not sum over all processes.  In particular consider process 1 of
Fig. 7.  When the hop actually consists of two hops, one from a
Cu to an O and another from an O to a Cu, this process corresponds
to two orderings of the three perturbations, one in which the spin--orbit
interactions comes first and one in which it comes second.  (The case
when it comes third should be identified with processes 2 or 3 of
Fig. 7.)  Summing over these two orderings converts the denominator
$(\Delta+E_{xy})$ of BS into ours.  The correct energy denominator
can also be obtained from Eqs. (E1) and (E2).

In principle, we ought also to compare with Eq. (20) of BS, where
contributions involving $\psi_1({\bf r})$ are claimed to be
included.  Here their results are so different from ours that we
can not identify their terms with ours.  In particular, we note
the following.  1)  Although they claim to include the effects of
$\psi_1({\bf r})$, their expressions do not include any
energy denominators which depend on the associated crystal field
energy $\epsilon_1$.  Obviously, when $\epsilon_1=0$, the
ground manifold would be described by a totally different spin
Hamiltonian to remove the spin {\it and} orbital degeneracy.  2)
In our Eq. (\ref{JMU}) [\onlinecite{PRL2}] $t_{01}$ enters in several
places and thus gives rise to many more terms than appear in BS.
3)  When $t_{01} \not=0$, as we have noted, each bond has biaxial
anisotropy in contrast to the axial anisotropy they implicitly
assume.  4)  We include hopping between excited crystal field
states.  These give contributions which are of the same order of
magnitude as those involving hopping into the ground state.
5)  We have not included covalency corrections.  It may indeed be
a good idea to include such corrections, but at present the
parameters are themselves so uncertain that we regard this
correction as a refinement.

\begin{table}
\caption{Values in eV of Parameters Used}
\begin{tabular} {c c c c c c c c c c c }
\hline
$U_0^{\rm a}$ &
$A^{\rm b}$ & $B^{\rm b,c}$ & $C^{\rm b,c}$ & $(pd\sigma)^{\rm d}$ &
$\epsilon_1^{\rm e}$ & $\epsilon_x\equiv \epsilon_y^{\rm e}$ &
$\epsilon_z^{\rm e}$ & $\epsilon_{p_x}^{\rm f}$ & $\epsilon_{p_y}^{\rm f}$
& $\epsilon_{p_z}^{\rm f}$ \\
9.34 & 7.00 & 0.15 & 0.58 & 1.5 & 1.8 & 1.8 & 1.8 & 3.25 & 3.25 & 3.25 \\
\end{tabular}

$^{\rm a}$ Refs. \onlinecite{ESKES} and \onlinecite{EANDS} use
$U_0=8.8$.  Local density calculations of Refs. \onlinecite{HSC}
and \onlinecite{MAM} give 10.5 and 9.4, respectively.

$^{\rm b}$ The Racah coefficients, $A$, $B$, and $C$ are defined in
Ref. \onlinecite{JSG}.

$^{\rm c}$  For the solid, the values of $B$ and $C$ are appropriately
taken from the free--ion optical values of Ref. \onlinecite{CEM},
as is discussed by Eskes et al.. [\onlinecite{ETS}]

$^{\rm d}$ See Ref. \onlinecite{ESF}. 

$^{\rm e}$ See Ref. \onlinecite{ETS}.

$^{\rm f}$ Ref. \onlinecite{ESF} gives $\epsilon_p=3.5$, but
smaller values of $\epsilon_p$ are plausible. [\onlinecite{GS}]

\end{table}

\begin{figure}
\caption{A CuO plaquette.  Here
we distinguish between ``y'' oxygen ions (on $y$--directed bonds) and
``x'' oxygen ions (on $x$--directed bonds.)  In each case we show a p
orbital on the oxygen ion to which we give the symmetry label, $z$,
since these orbitals can only hop to an orbital on a copper ion with
that same symmetry label, i.e., to $\psi_z \sim xy$, which is
also shown.}
\end{figure}

\vspace{0.5cm}
\begin{figure}
\caption{
Schematic view of $d_{xz}$ and $d_{yz}$ of two Cu ions
when they are on the $x$--axis. Note that while $d_{xz}$
orbitals are in the same plane, the plane of  $d_{yz}$
orbitals are parallel to each other. Hence there is no
reason that $t_{x,x} $ should be equal to $t_{y,y}$.}
\end{figure}

\begin{figure}
\caption{ Comparison of perturbation results (dotted line)
with the exact results (solid line).  Here $J_\parallel$,
$J_\perp$, and $J_z$ correspond to $J_{xx}$, $J_{yy}$,
and $J_{zz}$ of Eq. (\protect\ref{JMU}).
The hopping matrix elements $t_{\alpha,\beta}$ are estimated
from Eq. (\protect\ref{IV4}) as explained in the text.  In the
left and right panel
the $t_{\alpha,\beta} $ and $K_{\alpha,\beta}$
are replaced by  $ t \; t_{\alpha,\beta}$ and
$ K \; K_{\alpha,\beta}$ respectively.  The values of $\lambda$
(in eV), $t$, and $K$  are given in the panels.
Preprint version: labels for curves didn't print well.
In each panel the upper curves are $J_\perp-J_z$ and the lower
curves are $J_\perp$-$J_\parallel$.}
\end{figure}

\begin{figure}
\caption{Noninteracting spin--wave spectrum along different high
symmetry lines in the Brillouin zone according to Eq.
(\protect\ref{SW12}) for $\theta=0$.  For this plot the lattice constant
$a$ is set equal to unity. The $J$'s and $\omega$ are all
in the same arbitrary units.}
\end{figure}

\begin{figure}
\caption{ Variation of the zero--point energy as a function of
the angle $\theta$ between the staggered magnetization and a $[1,0]$
direction in the easy plane. Dotted and solid lines are from
the approximate [Eq. (\protect\ref{SW21})] and exact
[Eq. (\protect\ref{SW14})] expressions for $E_{Z}(\theta)$,
respectively.  The $J$'s and $E_Z(\theta)$ are all in the same
arbitrary units.}
\end{figure}

\medskip
\begin{figure}
\caption{ Anisotropies (energy differences between triplet states)
as a function of $t$ and $\lambda$ for non--constant (left) and
constant (right) on site Coulomb interaction.  If the energies of
the triplet states are $\lambda_1 < \lambda_2 < \lambda_3$,
the data points are $\lambda_1-\lambda_2$ (circles) and
$\lambda_1 - \lambda_3$ (diamonds). The solid and dashed lines
are power--law fits, as indicated.}
\end{figure}

\medskip
\begin{figure}
\caption{Processes (1--5) which contribute to $Q_\alpha^{(0)}$
[in the order written in Eq. (H9)]
and those (6--7) which contribute to $Q_\alpha^{(1)}$
[in the order written in (H13)].  Here the left site is
the site $i$ and the right site is site $j$.
The dashed line depicts the first matrix element and
the full line the second. The orbitals are the ground
state lowest, the state $|1\rangle$ next, and the state
$|\alpha \rangle$ highest. In term 3 the second process
promotes the left--hand hole to an excited state, whereas in
term 2 the left--hand hole remains in the ground state.}
\end{figure}


\begin{references}
\bibitem{ERRATUM}
In Ref. \onlinecite{PRL2} there are some misprints in the
Hamiltonian of Eq. (3).  The last two lines there are rewritten
correctly here in Eq. (\ref{HAM}).
\bibitem{WRITE}
In Eq. (\ref{ROTINV}) when
$\alpha \not= \alpha'$ $f^\dagger_{k\alpha \mu}$
can be anticommuted to the right of $f_{k\alpha' \mu'}$.  When
$\alpha=\alpha'$ the Fermion operators can be written as
$(n_{k\alpha}^2-n_{k\alpha})/2$, where
$n_{k\alpha}=Q_{\alpha \alpha} (k,k)$.
\bibitem{SPLIT}
   This degeneracy is also removed by second--neighbor interactions.
\end{references}
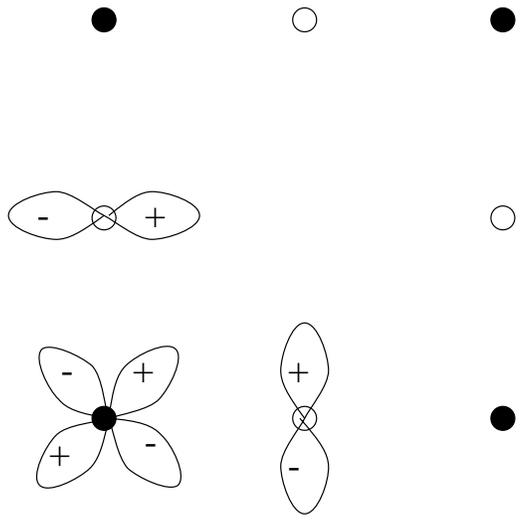
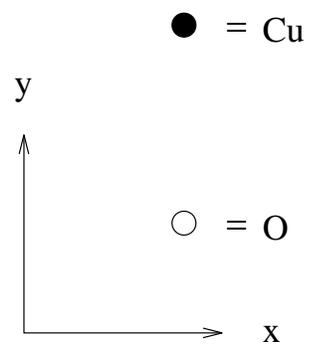
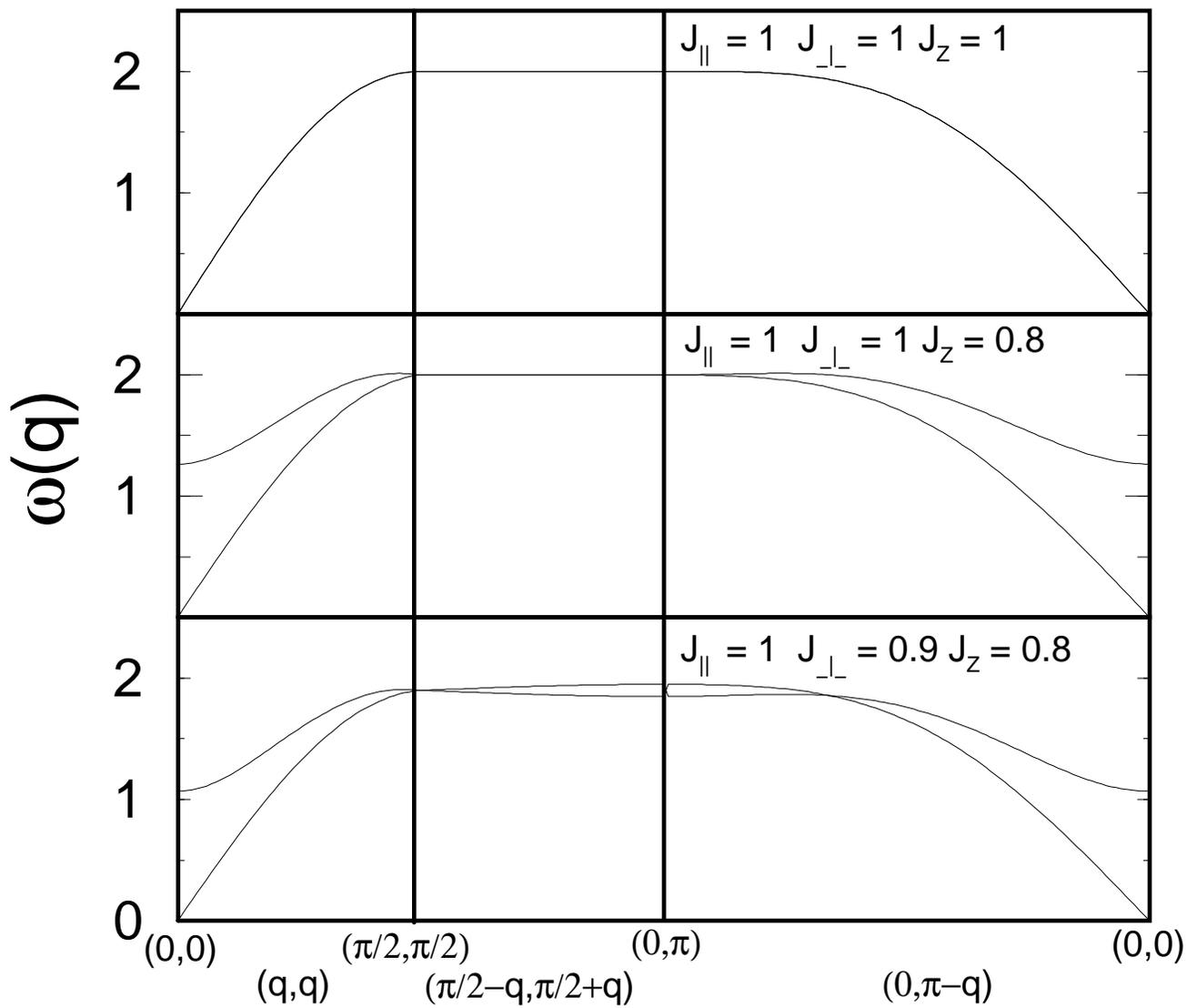
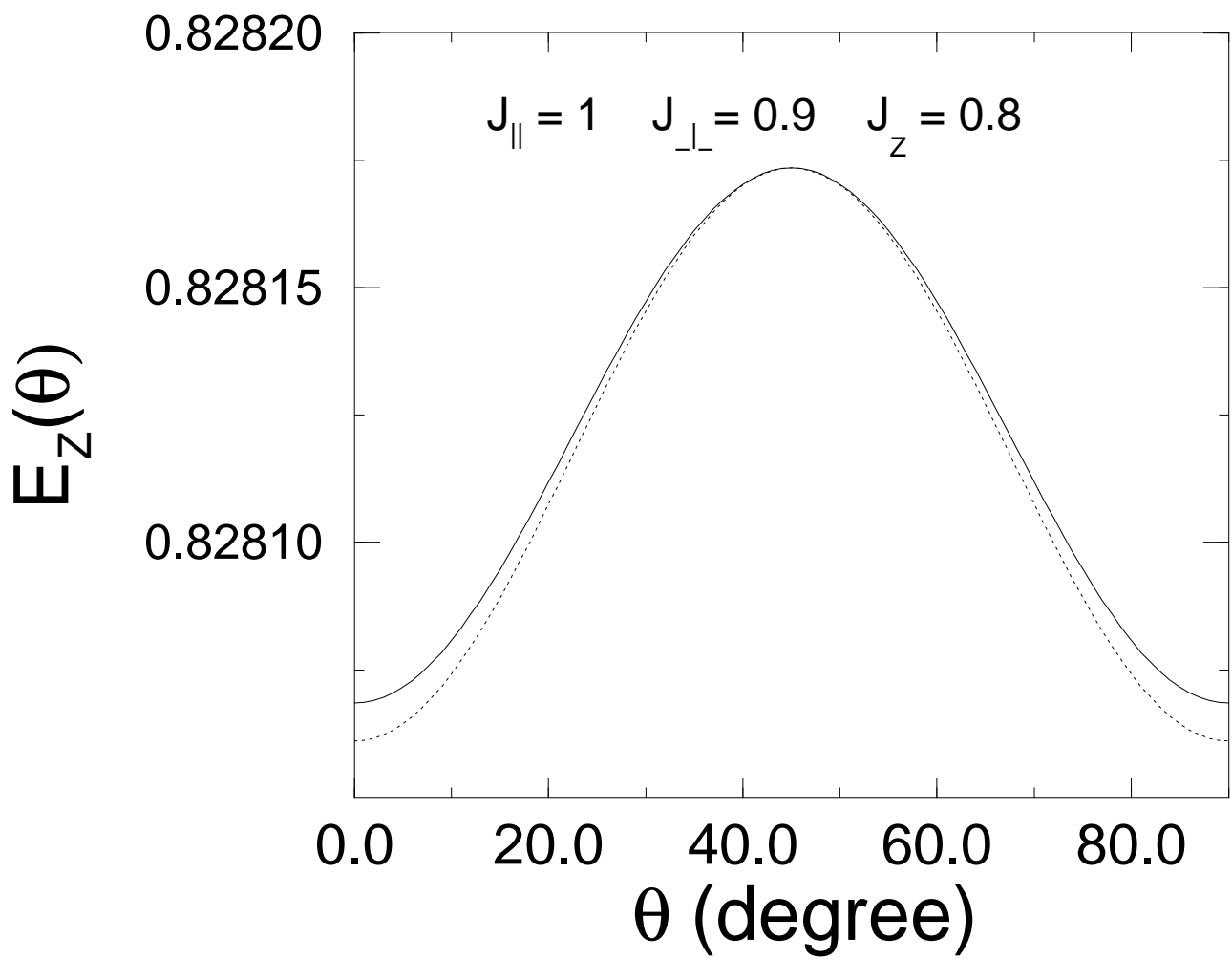
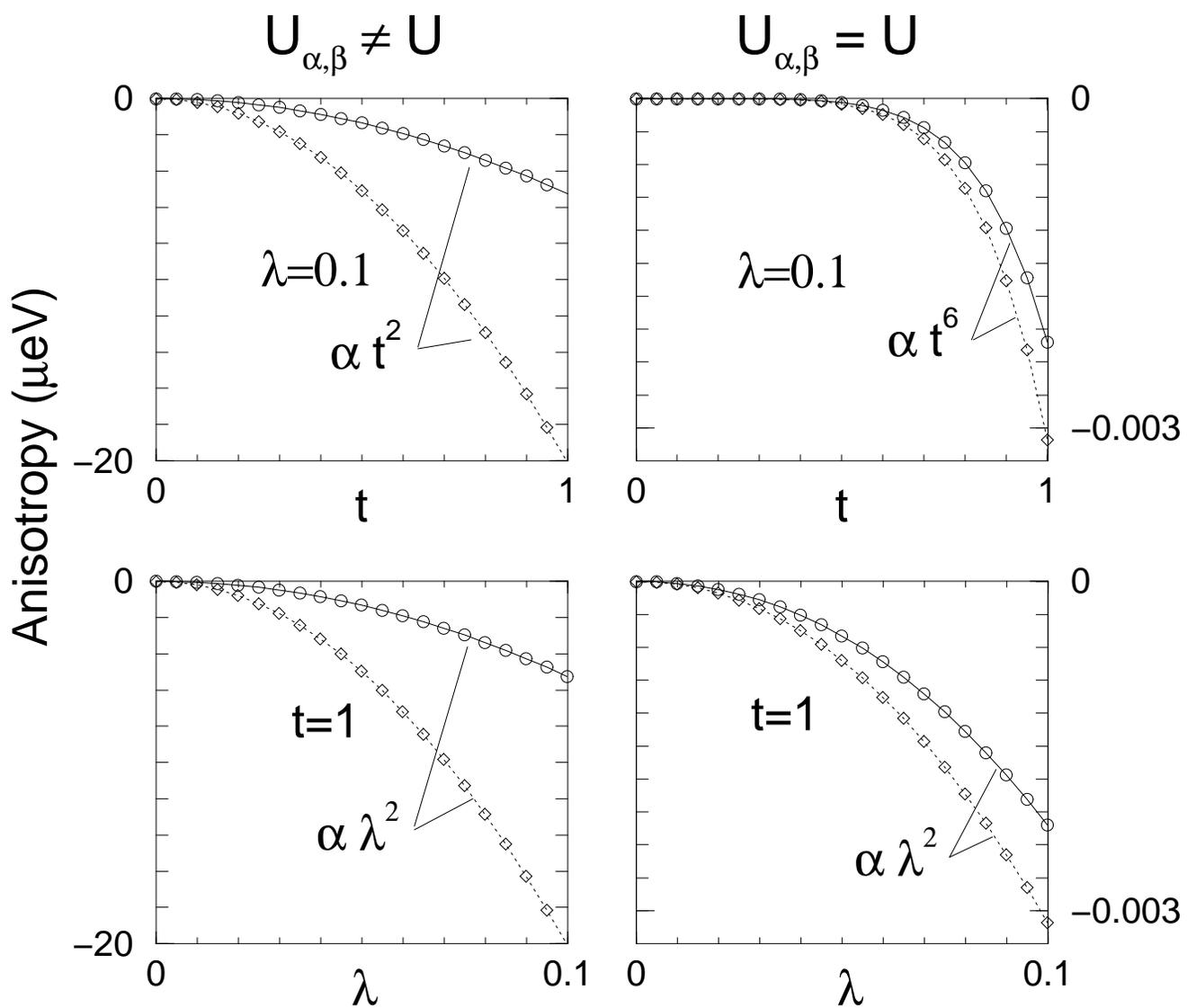
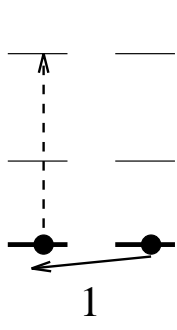
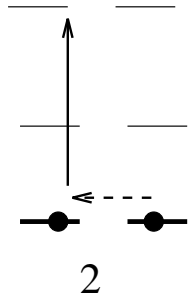
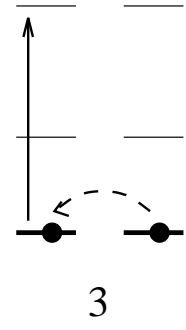
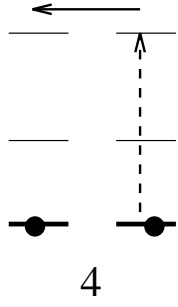
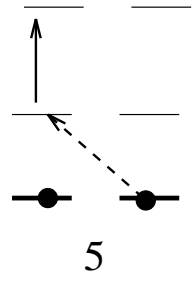
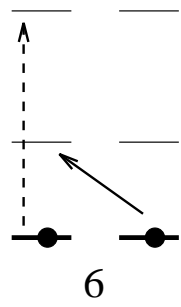
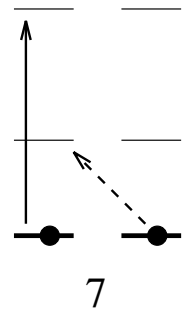
\end{document}